%% file: WP001.tex
\documentclass[12pt]{article} %[headings = normal, titlepage=true, toc=listof, toc=indent, listof=indent, 12pt]

\usepackage[usenames, dvipsnames]{xcolor}
\usepackage[right=20mm,top=15mm, bottom=25mm, left=20mm, includeheadfoot]{geometry}
\usepackage{microtype} %
\usepackage[utf8]{inputenc}
\usepackage[babel,german=quotes]{csquotes}
\usepackage{booktabs}
\usepackage{listings}
\usepackage{texnames}
\usepackage{paralist}
\usepackage{float}
\usepackage{amsmath}
\usepackage{amsthm}
\usepackage{textcomp}
\usepackage{lmodern}
\usepackage[nodisplayskipstretch]{setspace} %
\usepackage{bbm}
\usepackage{blindtext}
\usepackage{layout}
\usepackage{multirow}
\usepackage{amssymb}
\usepackage{pdflscape}
\usepackage[inline]{enumitem}
\usepackage{graphicx}
\usepackage{tikz}
\usetikzlibrary{patterns}
\usepackage{pgfplots}
\usepackage{mathrsfs}
\usepackage[english]{babel}
\usepackage[T1]{fontenc} %
\usepackage[utf8]{inputenc} %
\usepackage{lmodern} %
\usepackage{comment}
\usepackage{verbatim}
\usepackage{booktabs} % lininien in Tabelle
\usepackage{caption}

\usepackage{subfig}
\usepackage{threeparttable}

\pgfplotsset{compat=1.6}
\usepackage[longnamesfirst, round, sort&compress,  authoryear]{natbib}

\usepackage{footmisc}
\usepackage{subfig}

\makeatletter
\def\@xfootnote[#1]{%
	\protected@xdef\@thefnmark{#1}%
	\@footnotemark\@footnotetext}

\makeatother  % To definde stars as footnotes \footnote[*] gibt star, [$\dagger$] kreuz und [$\star$] anderen stern

\theoremstyle{plain}
\newtheorem{theorem}{Theorem}

\newtheorem{lemma}{Lemma}
\newtheorem{assumption}{Assumption}
\newtheorem{corollary}{Corollary}
\newtheorem{proposition}{Proposition}

\theoremstyle{definition}

\newtheorem{remark}{Remark}
\newtheorem{definitions}{Definition}[section]

\usepackage{chngcntr}
\usepackage{apptools}
\AtAppendix{\counterwithin{lemma}{section}}
\AtAppendix{\counterwithin{assumption}{section}}
\AtAppendix{\counterwithin{corollary}{section}}
\AtAppendix{\counterwithin{theorem}{section}}
\AtAppendix{\counterwithin{proposition}{section}}
\newcommand{\captionfonts}{\small}

\makeatletter  % Allow the use of @ in command names
\long\def\@makecaption#1#2{%
	\vskip\abovecaptionskip
	\sbox\@tempboxa{{\captionfonts #1: #2}}%
	\ifdim \wd\@tempboxa >\hsize
	{\captionfonts #1: #2\par}
	\else
	\hbox to\hsize{\hfil\box\@tempboxa\hfil}%
	\fi
	\vskip\belowcaptionskip}
\makeatother   % Cancel the effect of \makeatletter
\usepackage{titlesec}
\titleformat{\section}[block]{\centering\normalfont}{\thesection.}{0.5em plus .1em minus .1em}{\uppercase }
\titleformat{\subsection}[runin]{\normalfont}{\thesubsection.}{0.3em plus .1em minus .1em}{\bfseries}[.]
\titleformat{\subsubsection}[runin]{\normalfont}{\thesubsubsection.}{0.3em plus .1em minus .1em}{\it}[.]
\titlespacing*\section{0pt}{14pt plus 2pt minus 2pt}{5pt plus 2pt minus 2pt}
\titlespacing*\subsection{0pt}{4pt plus 1pt minus 1pt}{5pt plus 2pt minus 2pt }
\titlespacing*\subsubsection{0pt}{4pt plus 1pt minus 1pt}{5pt plus 2pt minus 2pt}
\usepackage{chngcntr}
\usepackage{apptools}
\makeatletter
\def\mythanks#1{%
	\protected@xdef \@thanks {\@thanks \protect \footnotetext [\the \c@footnote ]{#1}}%
}
\makeatother

\def\boxit#1{\vbox{\hrule\hbox{\vrule\kern6pt
			\vbox{\kern6pt#1\kern6pt}\kern6pt\vrule}\hrule}}

\usepackage{pgfplots}
\usepgfplotslibrary{dateplot,fillbetween}
\definecolor{darkbrown}{RGB}{139, 69, 19}
\definecolor{darkred}{RGB}{180,0, 10}
\definecolor{darkorange}{RGB}{255,140,0}

\newcommand{\IR}{\mb{R}}

\newcommand{\IP}{\mb{P}}

\newcommand{\1}[1]{\mathbf{1}\{#1\}}

\newcommand{\Cov}{\textrm{Cov}}

\newcommand{\wh}{\widehat}
\newcommand{\wt}{\widetilde}
\newcommand{\mc}{\mathcal}
\newcommand{\mb}{\mathbb}
\newcommand{\A}{\textnormal{SR}}

\newcommand{\pdf}{\pi_{\textnormal{DF}}}
\newcommand{\pco}{\pi_{\textnormal{CO}}}
\newcommand{\pat}{\pi_{\textnormal{AT}}}
\newcommand{\pnt}{\pi_{\textnormal{NT}}}
\newcommand{\pd}{\pi_{\textnormal{d}}}

\newcommand{\updf}{\underline{\pi}_{DF}}
\newcommand{\opdf}{\overline{\pi}_{DF}}

\newcommand{\Fcod}{F_{Y_d^{CO}}}

\newcommand{\Fdfd}{F_{Y_d^{DF}}}

\newcommand{\Qdd}{Q_{\textnormal{dd}}}

\newcommand{\Fucop}{\overline{F}_{Y_1^{CO}}}
\newcommand{\Fucoa}{\overline{F}_{Y_0^{CO}}}
\newcommand{\Flcop}{\underline{F}_{Y_1^{CO}}}
\newcommand{\Flcoa}{\underline{F}_{Y_0^{CO}}}
\newcommand{\Fucod}{\overline{F}_{Y_d^{CO}}}
\newcommand{\Flcod}{\underline{F}_{Y_d^{CO}}}

\newcommand{\Fudfd}{\overline{F}_{Y_d^{DF}}}
\newcommand{\Fldfd}{\underline{F}_{Y_d^{DF}}}

\newcommand{\yld}{y, \pdf, \delta}
\newcommand{\Pddd}{P_{dd}}
\newcommand{\Pdds}{P_{d(1-d)}}
\newcommand{\Gdsup}{G_d^{\sup}}
\newcommand{\Gdinf}{G_d^{\inf}}

\newcommand{\deltamin}{\underline{\delta}}
\newcommand{\deltamax}{\overline{\delta}}

\pgfmathdeclarefunction{gauss}{2}{%
	\pgfmathparse{1/(#2*sqrt(2*pi))*exp(-((x-#1)^2)/(2*#2^2))}%
}
\def\cdf(#1)(#2)(#3){0.5*(1+(erf((#1-#2)/(#3*sqrt(2)))))}%
\tikzset{
	declare function={
		normcdf(\x,\m,\s)=1/(1 + exp(-0.07056*((\x-\m)/\s)^3 - 1.5976*(\x-\m)/\s));}
}

\begin{document} 
\selectlanguage{english}

\begin{titlepage}
	
	\newpage
	\thispagestyle{empty}
	\vspace{-2cm}
	\begin{verbatim}
	\end{verbatim}
%	\vspace{0.5cm}
	\begin{center}
		\Large \scshape{ Sensitivity of LATE Estimates to}\\
	\scshape{Violations of the Monotonicity Assumption}
	\end{center}
	\vspace{0.5cm}
	\begin{flushleft}
		\begin{center}
			\large Claudia Noack\footnote[$\star$]{This version: 	June 10, 2021. Department of Economics, University of Mannheim;  E-mail: claudia.noack@gess.uni-mannheim.de. Website: claudianoack.github.io.  I am grateful to my advisor Christoph Rothe for invaluable support on this project. I am thankful for insightful discussions with  Tim Armstrong, Matthew Masten, Yoshiyasu Rai, and Ed Vytlacil.  Furthermore, I thank  Michael Dobrew, Jasmin Fliegner, Martin Huber, Paul Goldsmith-Pinkham, Sacha Kapoor, Lukas Laffers,  Tomasz Olma, Vitor Possebom, Alexandre Poirier,  Jonathan Roth,   Pedro S'Antanna, Konrad Stahl, Matthias Stelter,  Jörg Stoye,  Philipp Wangner, and  seminar and conference participants at the University of Mannheim, University of Heidelberg, University of Bonn, University of Oxford, Yale University,   ESEM  2019, ESWC 2020, IAAE 2019,  and 1st International Ph.D. Conference at Erasmus University Rotterdam  for  helpful comments and suggestions. I gratefully acknowledge financial support by the European Research Council (ERC) through grant SH1-77202.} \\
			\vspace{0.75cm}
			\end{center}
	\end{flushleft}

	\begin{abstract}
		\noindent
		In this paper, we develop a method to assess the sensitivity of local average treatment effect estimates to  potential violations of the monotonicity assumption of Imbens and Angris~(1994). We parameterize the degree to which monotonicity is violated using two sensitivity parameters: the first one determines  the share of defiers in the population,  and the second one measures differences in the distributions of outcomes between compliers and defiers. For each pair of values of these  sensitivity parameters, we derive sharp bounds on the  outcome distributions of compliers in the first-order stochastic dominance sense.   We identify the robust region that is the set of all values of  sensitivity parameters for which a given empirical conclusion, e.g. that the local average treatment effect is positive, is valid. 
		Researchers can assess the credibility of their conclusion by evaluating whether all the plausible sensitivity parameters lie in the robust region. We  obtain confidence sets for the robust region through a bootstrap procedure and illustrate the sensitivity analysis in an empirical application. We also extend this framework to analyze treatment effects of the entire population.
	
	\end{abstract}
	\setcounter{page}{0}\clearpage
\end{titlepage}

\pagenumbering{arabic}
\onehalfspacing

\newcommand{\md}{\mathbb}
\section{Introduction} \label{sectionintroduction}
The local average treatment effect framework (LATE) is used  for instrumental variable analysis in setups of  heterogeneous treatment effects \citep{imbens1994identification}. We consider settings of a binary instrumental variable and a binary treatment variable. The Wald estimand then equals the treatment effect of \textit{compliers}, individuals for which the instrument influences the treatment status, given the  well-known classical LATE assumptions: monotonicity, independence, and relevance. Monotonicity states that the   effect of the instrument on the treatment decision is monotone across all units. In the canonical example, in which the instrument encourages units to take up the treatment, monotonicity  rules out the existence of \textit{defiers}, i.e., units  that receive the treatment only if the instrument discourages them. Researchers might  question the validity of this assumption in empirical applications. In these settings, the  local treatment effect estimates  might be biased and might lead the researchers to draw incorrect conclusions about the true treatment effect.

As  an example of a setup in which  monotonicity could plausibly be violated,  consider the study of \cite{angristevens1998}, who analyze the effect of having a third child on the labor market outcomes of mothers.  As the decision to have a third child is endogenous, the authors use a dummy for whether the first two children are of the same sex as an instrument. The underlying reasoning is that some parents would only decide to have a third child if their first two children were of the same sex; these parents are compliers.  The monotonicity assumption seems questionable in this setting as parents, who have a strong preference for one specific sex, might act as a defier in this setup. Consider, for example parents who want to have at least two boys and their first child is a boy.
Contrary to the incentive given by  instrument, they have two children if their second child is a boy, and three children if their second child is a girl. As the monotonicity assumption might  be questionable in this example, one can question the validity of empirical conclusions drawn from the classical LATE analysis.\footnote{The other LATE assumptions seem to be plausible here. As the sex of a child is determined by nature and as only the number of and not the sex of the child arguably influences the labor market outcome of mothers, the independence assumption seems to be satisfied. The relevance assumption is testable.}

In this paper, we provide a framework to evaluate the sensitivity of treatment effect estimates to a potential violation of the monotonicity assumption.  As noted in \cite{AngristImbensRubin1996}, a violation of the monotonicity assumption always has two dimensions: The first dimension is the heterogeneous effect of the instrumental variable on the treatment variable,  the presence of defiers. The second dimension  is the heterogeneous effect of the treatment variable on the outcome variable,   the outcome  heterogeneity between defiers and compliers. We derive the degree to which monotonicity is violated by parameterizing these two dimensions. 

We parameterize the existence of defiers by their population size and the outcome  heterogeneity  by the Kolmogorov-Smirnov norm, which bounds the difference of the cumulative distribution functions of compliers and defiers.  For each of these two sensitivity parameters, we identify sharp bounds of the outcome distribution of compliers  in a first-order stochastic dominance sense. These bounds also imply sharp bounds on various treatment effects, e.g., the average treatment effect or quantile treatment effects of compliers.

Our analysis proceeds in two steps. 
In a first step, we identify the \textit{sensitivity region}. The sensitivity region defines the set of sensitivity parameters for which a data generating process exists, that is consistent with our model assumptions and implies both the observed probabilities and the sensitivity parameters. Since sensitivity parameters lying in the complement of the sensitivity region are not compatible with our model, we do not analyze them further. For the derivation of the sensitivity region, we also derive sharp bounds of the population size of defiers.
In a second step, we identify the \textit{robust region}, which is the set of sensitivity parameters that imply treatment effects  that are consistent with a particular empirical conclusion; for instance, the treatment effect of compliers has a specific sign or a particular order of magnitude.\footnote{See \cite{masten2020} for a detailed exposition of this approach.}  Parameters lying in the complement of the robust region, the \textit{nonrobust region}, imply treatment effects that are not, or may not be, consistent with the given empirical conclusion. The robust region and the nonrobust region are separated from each other by the \textit{breakdown frontier}, following the terminology of \cite{masten2020}. For each population size of defiers, the breakdown frontier identifies  the weakest assumption about outcome heterogeneity, which is necessary to be imposed to imply treatment effects being consistent with the particular empirical conclusion under consideration. 

This framework can be used in the following ways. First, by evaluating the size of the sensitivity region, one can determine the plausibility of the model.  If this set is empty, the model is refuted, which implies that even if one would allow for an arbitrary violation of the monotonicity assumption, at least one of the  model assumptions has to be violated. Second, researchers can analyze the sensitivity of their estimates with respect to the degree to which the monotonicity assumption  is violated by varying the sensitivity parameters within the sensitivity region. Third, by evaluating the plausibility of the parameters within the robust region, researchers can assess the sign or the order of magnitude of the treatment effect. 
While being transparent about the imposed assumptions, they might still arrive at a particular empirical conclusion of interest in a credible way. 
Fourth, one can assess to which degree monotonicity has to be violated to overturn a particular empirical conclusion. 
Within our framework, researchers can use their economic insights about the analyzed situation to judge the severity of a violation  monotonicity. 

While the main focus of this paper lies on treatment effects of compliers, we also show how this framework can be exploited to analyze treatment effects of the entire population. Under further support assumptions of the outcome variable and for given sensitivity parameters, the average treatment effect of the entire population  is partially identified, which complements known results in the literature \citep[see][]{kitagawa2021identification, Balke97, machado2019instrumental, kamat2018identifying}.
Since the analytic expressions of the sensitivity and robust regions are rather complicated and difficult to interpret, we provide simplified analytical expressions of these regions for a binary outcome.  

To construct confidence sets for both the sensitivity and the robust region, we show that both regions are determined through mappings of some underlying parameters. These mappings are not Hadamard-differentiable, and inference methods relying on standard  Delta method  arguments are therefore not applicable. We  show how to construct smooth mappings that bound the parameters of interest. This construction leads to mappings for which standard Delta method arguments are applicable, and  we use the nonparametric bootstrap to construct  valid confidence sets for the parameters of interest.  With a binary outcome variable, the mappings resulting in the sensitivity and robust region are considerably simpler. Therefore, we can use a generalized Delta method to show  asymptotic distributional results and apply a bootstrap procedure to construct asymptotically valid confidence sets. 

We show in a Monte Carlo study that our proposed inference method  has good finite sample properties. We further apply our method to the setup studied by  \cite{angristevens1998} introduced above. We show that relatively strong assumptions on either the population size of the defiers or the outcome heterogeneity have to be imposed to preserve the sign of the estimated treatment effect.
This result demonstrates that the monotonicity assumption is key in the local treatment effect framework.

The remainder of this paper is structured as follows: A literature review follows, and Section~\ref{sectionsetup} illustrates the setup in a simplified setting.  Section~\ref{sectionsensitivitypara} introduces the sensitivity parameters and Section~\ref{sec_identification_bounds} derives sharp bounds on the distribution functions of compliers.
The   main sensitivity analysis is presented in Section~\ref{sectionsensitivtyanalysis}. Section~\ref{sectionextension}  discusses extensions and Section~\ref{sectioninference} derives estimation and inference results. Section~\ref{sectionsimulations} contains a simulation study and  Section~\ref{sectionempirical} an empirical example. Section~\ref{sectionconclusion} concludes. All proofs and additional materials are deferred to the appendix.

\subsection*{Literature} This paper relates to several strands of the literature. First, this paper contributes to the growing strand of the literature, which considers sensitivity analysis in various applications. These applications include, among many others,  violations of parametric assumptions, violations of moment conditions, and multiple examples within the treatment effect literature 
\citep[see, among others,][]{armstrong2021sensitivity, mukhin2018sensitivity, christensen2019counterfactual, kitamura2013robustness, bonhomme2018minimizing, bonhomme2019posterior, andrews2017measuring, andrews2020informativeness, andrews2020model, andrews2020transparency, rambachan2020honest, conley2012plausibly, imbens2003, chen2015sens}.
This paper is closely related to the literature  about  breakdown points  of \cite{HorowitzManski1995, imbens2003, KleinSantos2013, stoye2005, stoye2010partial}, and especially closely related to \cite{masten2020, masten2021salvaging}.  These papers consider several assumptions in the treatment effect literature, but not the monotonicity assumption. 

Second, it is related to the local average treatment effect framework literature, which is formally introduced in \cite{imbens1994identification} and further in \cite{vytlacil2002independence}. Several papers consider violations of the monotonicity assumption through different types of assumptions. 
\cite{Balke97, machado2019instrumental, hubermellace2010, manski1990, huber2015testing, huber2015testmonotonicity} consider a binary and \cite{kitagawa2021identification} a continuous outcome variable and  partially identify the average treatment effect.   
\cite{small2017instrumental,  manskipepper2000,dahl2017s, hubermellace2010} propose alternative assumptions on the data generating process, which are strictly weaker than monotonicity and obtain bounds on various treatment effects. \cite{richardson2010analysis} consider a binary outcome variable and derive bounds on outcome distributions for a given population size of always takers. We consider not only a binary outcome variable and we introduce a second parameter to also bound outcome heterogeneity. We then consider these sensitivity parameters within the framework of a  breakdown frontier. 

\cite{de2017tolerating} shows that in the presence of defiers, under certain assumptions, the Wald estimand still identifies a convex combination of causal treatment effects  of only a subpopulation of compliers. In a policy context, the  treatment effect of compliers might be of particular interest because the  treatment status of compliers  is most likely to change with  a small policy change. However, the same reasoning does not apply to the subpopulation of compliers.
\cite{Klein2010} evaluates the sensitivity of the  treatment effect of compliers to random departures from monotonicity. \cite{FioriniStevens2014} give examples of analyzing the sensitivity of the monotonicity, and \cite{Huber2014} considers a violation of monotonicity in a specific example. They do not provide sharp identification results of the  treatment effect of compliers in the presence of defiers, nor do they derive the robust region. A violation of the monotonicity assumption with a non-binary instrumental variable is considered, and alternative assumptions and testing procedures are proposed in \cite{mogstad2019identification, frandsen2019judging, norris2020effects}.
This paper contributes to this literature by presenting an effective tool to analyze the severity of a potential violation of  the monotonicity assumption. It thus gives applied researchers a new tool to evaluate the robustness of their estimates to a violation of the monotonicity assumption, and their estimates may thereby gain credibility.

Our proposed inference procedure builds on seminal work about Delta methods for non-differentiable mappings by  \cite{Shapiro1991, fansantos2016, dumbgen1993nondifferentiable, Hong2016}, and it further exploits ideas of smoothing population parameters by \cite{masten2020, chernozhukov2010quantile, haile2003}.

\section{Setup}\label{sectionsetup} 
\subsection{Model of the Local Average Treatment Effect}
We observe the distribution of the random variables $(Y, D, Z)$,  where $Y$ is the outcome of interest; $D$ is the actual treatment status, with $D=1$ if the person is treated and $D=0$ otherwise; and $Z$ is the instrument, with $Z=1$ if the person is assigned to treatment and $Z=0$ otherwise. We assume that each unit has potential outcomes $Y_0$  in the absence and $Y_1$ in the presence of treatment, and potential treatment status  $D_1$ when assigned to treatment and $D_0$  when not assigned to treatment.  The observed and potential outcomes are related by $Y=D Y_1 + (1-D)Y_0$, and observed and potential treatment status by $D=Z D_1 + (1-Z)D_0$.

Based on the effect of the instrument on the treatment status, we distinguish four different groups: compliers that are only treated if they are assigned to treatment   (CO); defiers that are only treated if they are not assigned to treatment (DF); always takers that are independently of the instrument always treated (AT),  and    never takers that are never treated (NT).
We denote the population sizes of the respective group by $\pat$, $\pnt$, $\pco$, and $\pdf$. We denote by $Y_d^{T}$ the potential outcome variable of group $T \in \{AT, NT, CO, DF\}$ under treatment status $d$. To simplify the notation, we write $Y_d^{dT}$  for the potential outcome variable  of always takers if $d=1$ and otherwise of never takers, and similarly $\pi_{dT}$ for the respective population size.  
We denote the outcome distribution of a variable $Y$ by $F_Y$, its density function, if it exists, by $f_Y$, and its support by $\mb{Y}$.\footnote{Throughout the paper, we implicitly assume  that all necessary moments of all random variables for the parameter of interest exist; for instance,  if we consider the local average treatment effect, we assume $Y_d^{T}$ has first moments for all $d\in\{0,1\}$ and $T \in \{C, DF, AT, NT\}$.}

The key parameters of interest in this analysis are   treatment effects of compliers. We denote the average treatment effect of compliers  by\footnote{Similarly, the  average treatment effect of defiers is denoted by $\Delta_{DF}=\mb{E}[Y_1-Y_0 |D_0=1, \; D_1=0]$.} $$\Delta_{CO}=\mb{E}[Y_1-Y_0 | D_0=0, \; D_1=1]. $$
  
Throughout the paper, we assume that  $\mb P(D=1|Z=1) \geq \mb P (D=1|Z=0)$ without loss of generality, and we impose the following identifying assumptions.
\begin{assumption} 
	\label{assumptionLATE} The instrument satisfies   $(Y_{1}, Y_{0}, D_1, D_0) \perp Z$ (Independence), and  $\mb P(D=1|Z=1)>\mb P(D=1|Z=0)$ (Relevance).
\end{assumption}
We refer to \cite{AngristImbensRubin1996} for an extensive discussion of these assumptions.
\subsection{Illustration of the Sensitivity Analysis}
In this section, we illustrate the sensitivity analysis in a very simplified framework, where we introduce the sensitivity parameters, the sensitivity and the robust region. In contrast to our main sensitivity analysis in Section~\ref{sectionsensitivitypara}-\ref{sectionsensitivtyanalysis}, we do not consider any sharp identification results in this illustration.
\subsubsection{Sensitivity Parameter Space} 
In the presence of defiers, \cite{AngristImbensRubin1996} show that the average treatment effect of compliers is not  point identified. The Wald estimand, $\beta^{IV} =\Cov(Y,Z) / \Cov(D,Z)$,  equals a weighted difference of the average treatment effect of compliers and defiers:	
\begin{align}
	\beta^{IV} = \frac{1}{\pco -\pdf}  \left(\pco \Delta_{CO} - \pdf \Delta_{DF} \right).\label{equationidentificationaverage}
\end{align}
Three parameters in equation~\eqref{equationidentificationaverage} are in general not identified:  the population size of defiers $\pdf$, the treatment effect of compliers $\Delta_{CO}$ and of defiers $\Delta_{DF}$.\footnote{Clearly, if either $\pdf=0$, implying the absence of defiers, or $\Delta_{CO}= \Delta_{DF}$, implying that compliers and defiers have the same average treatment effect, the treatment effect $\Delta_{CO}$ is still point identified.}
To bound the average treatment effect of compliers, we introduce two sensitivity parameters.  The first one determines the population size of defiers, and the second one outcome heterogeneity between compliers and defiers. These two parameters measure the degree to which monotonicity is violated and represent the two dimensions of heterogeneity: (i) heterogeneous effects of the instrument on the treatment status and (ii) heterogeneous effects  of the treatment on the outcome.

The heterogeneous impact of the instrument on the treatment status, is  parameterized, in the most simplest ways,  by the population size of  defiers 
\begin{align}
	\pdf =\mb P(D_0=1 \text{ and } D_1=0  ). \label{equ_pdf}
\end{align} 
A larger sensitivity parameter $\pdf$  implies a more severe violation of monotonicity.  It is clear that, for a given population size of defiers, $\pdf$,  the population sizes of the other groups are point identified. In our analysis, these population sizes are, therefore,  functions of the sensitivity parameter $\pdf$, but we leave this dependence implicit.\footnote{It follows from the definitions of the groups and our assumptions that  ${\pat=\mb{P}(D=1|Z=0)-\pdf}$,  $\pnt=\mb{P}(D=0|Z=1)-\pdf $ and  $\pco=\mb{P}(D=1|Z=1)-\mb{P}(D=1|Z=0)+\pdf$.} 

We parameterize the second dimension of heterogeneity by the  sensitivity parameter $\delta_a$ which equals the absolute differences in treatment effects of both groups
\begin{equation*}
	\delta_a= | \Delta_{CO} - \Delta_{DF}|.
\end{equation*}
A larger sensitivity parameter $\delta_a$ implies a more severe violation of monotonicity. 

\subsubsection{Sensitivity Region and Robust Region}
The \textit{sensitivity region} is the set of sensitivity parameters which do not violate our model assumptions. For instance, a sensitivity parameter ${\pdf\geq0.5}$ would violate our model assumptions as the relevance assumption implies that ${\pco > \pdf}$. Therefore, such a sensitivity parameter does not lie within our sensitivity region, which is identified without imposing any additional assumptions.  
In this illustrative example, we simplify the derivation and say that the sensitivity region is trivially given by $$\A_a=  [0,0.5) \times \IR_+.$$
In our main sensitivity analysis, this set, however, is nontrivial and can be empty. In this case,  the model is rejected, implying that even though the monotonicity assumption may be violated, at least one of the other model assumptions has to be violated as well.  

Even though the treatment effect of compliers is generally not point identified if $\pdf>0$, using~\eqref{equationidentificationaverage}, it~is  partially identified  for  any given pair of sensitivity parameters $(\pdf, \delta_a)$~by  $$ \Delta_{CO} \in   \left[ \beta^{IV}- \frac{\pdf}{\pco-\pdf}  \delta_a, \; \beta^{IV}+\frac{\pdf}{\pco - \pdf}  \delta_a \right].$$ 
In a typical sensitivity analysis, researchers now consider different values of the sensitivity parameters to evaluate the identified sets of the parameter of interest and to evaluate the robustness of the LATE estimates to a potential violation of monotonicity.
However, in many empirical applications, the interest does not lie in the precise treatment effect but in its sign or in its order of magnitude. It is, therefore, natural to start with the empirical conclusion of interest and to ask which sensitivity parameters imply treatment effects that are consistent with this conclusion. This approach is formalized by the breakdown frontier  \citep[see, e.g.,][]{KleinSantos2013, masten2020}. 

We now consider the empirical conclusion that $\Delta{CO} \geq \mu$,  and we assume that ${\beta^{IV} \geq \mu}$.\footnote{If $\beta^{IV} \leq \mu$, the robust region for the conclusion  that $\Delta{CO} \geq \mu$ is empty.}
Under our model assumptions and for a given value of the population size of defiers $\pdf$, the breakdown point determines the largest value of outcome heterogeneity $\delta_a$ that implies treatment effects that are consistent with our empirical conclusion of interest. Specifically, for any $\pdf \in [0,0.5]$, the breakdown point is given by
$$BP_a(\pdf) =\frac{\pco-\pdf}{\pdf} (\beta^{IV} - \mu).$$

The breakdown frontier (BF) is the set of all breakdown points  and the robust region (RR) is the set of all sensitivity parameters that are consistent with the empirical conclusion of interest. They  are respectively given by
\begin{align*}
	BF_a= \left\lbrace  (\pdf , BP_a(\pdf)) \in \A_a \right\rbrace  \quad \text{and} \quad RR_a= \left\lbrace  (\pdf , \delta_a) \in  \A_a: \delta_a \leq BP_a(\pdf) \right\rbrace.
\end{align*} 
\begin{figure}
	\centering
	\resizebox{0.45\linewidth}{!}{\input{Graphics/average_identification}}
	\caption[Illustration of sensitivity and robust region I.]{Illustration of Sensitivity and Robust Region. Non-shaded area represents sensitivity region. $[\updf, \opdf]$ represent some bounds on the population size of defiers.}\label{figurelates}
\end{figure}
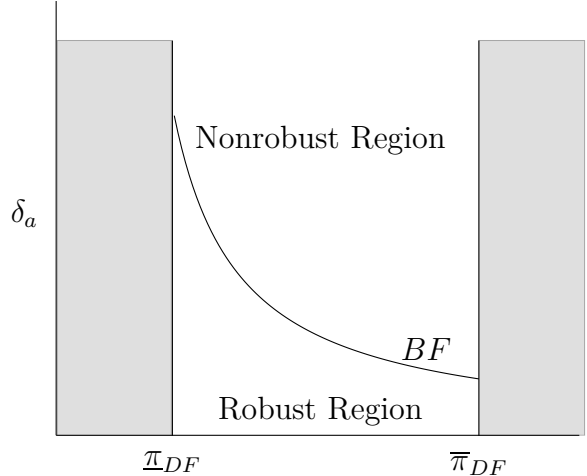
The nonrobust region is the complement of the robust region within the sensitivity region. It contains sensitivity parameters that may or may  not be consistent with the empirical conclusion. Due to the functional form of the breakdown frontier, the nonrobust region is a convex set in this example.  An illustrative example of this setup is shown in Figure~\ref{figurelates}. 

In this simple example, neither the sensitivity region nor the robust regions are sharp. 
For example, if the outcome is binary than the difference between compliers and defiers treatment effects is clearly bounded. Similarly, the robust region might also be substantially reduced by taking into account the actually observed outcomes.\footnote{To give a concrete example, assume that all treated units have a realized outcome of 1 and all nontreated units have a realized outcome of 0. Then it is clear, that the treatment effect of compliers is point identified to be one.}
This reasoning means that even though a parameter pair may lie within the sensitivity region, it might not imply a well-defined data generating process that is consistent with the model assumptions and the observed probabilities. Similarly, even though a parameter pair may lie within the nonrobust region, it might be robust.  Empirical conclusions that can be drawn from this analysis might, therefore,  not be very informative. Consequently, we improve upon this framework in the remainder of this paper.

\section{Sensitivity Parameters}\label{sectionsensitivitypara}
In this section, we introduce two sensitivity parameters  that are interpretable and imply bounds on the outcome distributions of compliers so that the parameter of interest is partially identified.   They allow us to consider a trade-off between the strength of the imposed assumption and the size of the identified set. 
To derive the sensitivity parameters, we consider  the following function:
\begin{gather*}
	G_d(y) = \frac{\Cov(\1{Y\leq y}, \1{D=d})}{\Cov(Z,\1{D=d})},
\end{gather*} 
for $d \in \{0,1\}$. In the absence of defiers, $G_d(y)$  is the cumulative distribution function of  compliers under treatment status $d$. In the presence of defiers, it  holds analogously to  \eqref{equationidentificationaverage} that    \begin{align}\label{equation_gd}
	G_d(y)&= \frac{1}{\pco-\pdf} \left(\pco F_{Y_d^{CO}}(y) - \pdf  F_{Y_d^{DF}}(y)\right).
\end{align}
The outcome distributions of compliers are thus identified up to the population size of defiers and the heterogeneity between the outcome distributions of compliers and defiers. 
We introduce two sensitivity parameters to parameterize these two dimensions. First, the presence of defiers is parameterized by the population size of defiers $\pdf$ \eqref{equ_pdf}. Second,  outcome heterogeneity is represented by $\delta$, which bounds  the maximal difference  between  cumulative distribution functions of the outcome of compliers and defiers by the Kolmogorov-Smirnov (KS) norm
\begin{equation*}
	\max_{d \in \{0, 1 \}} \; \underset{y \in \mb{Y}}{\sup} \{|F_{Y_d^{CO}}(y) - F_{Y_d^{DF}}(y)|\} = \delta,
\end{equation*}
where $\delta \in [0,1]$. Without a restriction on $\delta$, the outcome distributions can be arbitrarily different.  If $\delta=0$, the outcome distributions are restricted the most as both distribution functions coincide. A larger value of the parameter $\delta$ implies a more severe violation of monotonicity. 

There are clearly many different possibilities for how heterogeneity between distribution functions can be specified. In this paper, we choose the Kolmogorov-Smirnov norm, as it leads to tractable analytical solutions of the bounds on the compliers outcome distribution. More importantly, this parameterization is simple enough to be interpretable in an empirical conclusion.

A similar parameterization is chosen in \cite{KleinSantos2013} in a different context.\footnote{Since the parameterization of $\delta$  is  weak on the tails of the distributions, the bounds on the tails are likely to be uninformative. Imposing a \textit{weighted} KS assumption, that penalizes deviations at the tails of the two distributions more, would overcome this issue but would also lead to less tractable results.}

\section{Partial Identification of Distribution Functions}\label{sec_identification_bounds}
Since our main sensitivity analysis exploits bounds on parameters defined by the distribution function $\Fcod$ for $d  \in \{0, 1\}$, we  bound this distribution function  for a fixed given sensitivity parameter pair $(\pdf, \delta)$ in this section. We illustrate the derivation of the bounds in the subsequent sections, and the main result  is stated in Section~\ref{sec_mainresult_sharpbounds}. 

\subsection{Preliminaries} 
\subsubsection{Identification Strategy}
Our goal is to obtain  sharp lower and upper bounds of the distribution function $\Fcod$ in a first-order stochastic dominance sense. That is, we derive analytical characterizations of the distribution functions $\Flcod$  and $	\Fucod$ that are feasible candidates
for   $\Fcod$, in the sense that they are compatible with the imposed sensitivity parameters, our assumptions, and the population distributions of observable probabilities.  They are further such that $	\Flcod(y) \leq  \Fcod(y) \leq \Fucod(y), $ for all $y \in \mathbb Y$.
The identification strategy for deriving such sharp bounds $\Flcod$ and $\Fucod$ is based on the  premise that any candidate distribution function of $\Fcod$ then also implies  distribution functions of $F_{Y_d^{dT}}$ and $F_{Y_d^{DF}}$. Our candidate function $F_{Y_d^{CO}}$ is  therefore feasible, only if the implied functions of  $F_{Y_d^{dT}}$ and $F_{Y_d^{DF}}$  are indeed distribution functions.

The explicit analytical characterization of these sharp bounds  illustrates the effect of the sensitivity parameters on the bounds, and more importantly,  it implies  sharp bounds on a variety of treatment effects  of interest, e.g., the average treatment effect of compliers \citep[][Lemma 1]{stoye2010partial}.\footnote{The explicit characterization  also allows the inference procedure to be  based on $\Flcod$ and $\Fucod$.}

\subsubsection{Notation} We here collect the notation used in the following subsections.  Let $d,s \in \{0,1\}$  and  $y \in \mb Y$. Let the differences in population sizes of compliers and defiers be denoted by $\pi_{\Delta}=\pco-\pdf$. 
Let $Q_{ds}(y) \equiv \mb{P}(Y\leq y, D=d|Z=s)$ be  the observed joint distribution of $Y$ and $D$. We further let, for $\mathscr{B}$ denoting the Borel $\sigma$-algebra,
\begin{align*}
	&\wt G_d^+(y)=\sup_{B \in \mathscr{B}} \{ \mb{P}(Y \in B, Y\leq y,  D=d|Z=d) - \mb{P}(Y \in B,  Y\leq y,  D=d |Z=1-d) \}.
\end{align*} 
and $G_d^+=\frac{1}{\pco}  \wt G_d^+(y)$.
Our sensitivity analysis is based on the following observed underlying parameters
\begin{align}
	\theta= \left(Q_{11},Q_{10}, Q_{01},Q_{00}, \wt G_1^+,  \wt G_0^+\right).\label{eq_underlying_parameters}
\end{align}  

\subsection{Preliminary Bounds}\label{sec_simple_bounds_maintext} To illustrate the identification argument, we first derive preliminaries bounds on the distribution function $\Fcod$, which are not necessarily sharp in general.
Based on the law of total probability and our assumptions, the probability function $Q_{dd}$ is a weighted average of the distribution functions $\Fcod$ and $F_{Y_d^{dT}}$, specifically $Q_{dd}(y)=\pco \Fcod(y)+ \pd F_{Y_d^{dT}}(y) $.
Any feasible distribution function of $\Fcod$ has to imply a  function $F_{Y_d^{dT}}$ that is a distribution function. Exploiting this argument and   using our sensitivity parameter $\pdf$, it follows that
\begin{align}
		\frac{1}{\pco} Q_{dd}(y) \leq 	F_{Y_d^{CO}}(y) & \leq   \frac{1}{\pco} \left( Q_{dd}(y)- \pd \right). \label{equ_otheroutcomes_compliers_always_takers}
\end{align}
These bounds correspond to the extreme scenarios where compliers have the highest or the lowest outcomes compared to always and never takers.

Using the same argument for defiers and the definition of $G_d(y)$ in \eqref{equation_gd}, it  further follows that
\begin{align}
	\frac{ \pi_{\Delta}}{\pco}   G_d(y)   \leq 	F_{Y_d^{CO}}(y) & \leq  \frac{1}{\pco} \left(   \pi_{\Delta} G_d(y) + \pdf  \right). \label{equ_otheroutcomes_compliers_defiers}
\end{align}
We now consider the second sensitivity parameter $\delta$. Based on the definition of $G_d(y)$ in \eqref{equation_gd}, we conclude that any feasible candidate of $\Fcod$ also has to satisfy  that
\begin{align}
	G_d(y)-  \frac{\pdf}{\pi_{\Delta}} \delta  \leq 	\Fcod(y) \leq G_d(y) +  \frac{\pdf}{\pi_{\Delta}} \delta. \label{equ_restriction based_on_senstivityparamter}
\end{align}
Since the function $G_d$ is not necessarily increasing in $y$ for all $y \in \mathbb Y$,  bounds  on the distribution function $\Fcod$  based on \eqref{equ_otheroutcomes_compliers_defiers} and \eqref{equ_restriction based_on_senstivityparamter} have to take this into account. 
We therefore directly consider bounds  on $\Fcod$ that employ this information.
To be precise, for the lower bound,   we consider equation \eqref{equ_otheroutcomes_compliers_defiers} and \eqref{equ_restriction based_on_senstivityparamter}, where  we  replace $G_d$ by its smallest, nondecreasing  upper envelope; vice versa, for the upper bound, where  we  replace $G_d$ by its greatest, nondecreasing lower envelope.\footnote{We give an illustration of this derivation in Appendix~\ref{sec_illustraion_bounds}.}
Following this reasoning and taking \eqref{equ_otheroutcomes_compliers_always_takers}-\eqref{equ_restriction based_on_senstivityparamter} into account,  the lower bound is given by  
\begin{align} 
	\underline{H}_{Y_d^{CO}}(y,\pdf,\delta)=   \max\{0, \frac{1}{ \pco } (\Qdd (y)- \pi_d) ,\; \frac{\pi_{\Delta}}{\pco}  \sup_{\tilde y \leq y} G_d (\tilde y), \;  \sup_{\tilde y \leq y} G_d(\tilde y) -   \frac{\pdf}{\pi_{\Delta}} \delta \}, \label{simpleboundslower}
\end{align}
and  the upper bound  by
\begin{align}
	\overline{H}_{Y_d^{CO}}(y,\pdf,\delta)= \min\{ 1, \frac{1}{\pco} Q_{dd}(y), \frac{\pi_{\Delta}}{\pco} (\inf_{\tilde y \geq y} G_d(\tilde y)+\pdf),    \inf_{\tilde y \geq y} G_d(\tilde y)+   \frac{\pdf}{\pi_{\Delta}} \delta \}.\label{simpleboundsupper}
\end{align}
Any value outside of these bounds is clearly incompatible with the distribution of $(Y,D,Z)$ and our assumptions. 
To illustrate the effect of our sensitivity parameters, we consider the  width of these bounds for any fixed $y \in \mb Y$ as a function of $(\pdf, \delta)$, that is\footnote{This comparison is helpful as the qualitative size of the width of the bounds on the distribution functions is related to the width of the identified set of many parameters of interest, e.g., the LATE.}
$
\overline{H}_{Y_d^{CO}}(y,\pdf,\delta) - \underline{H}_{Y_d^{CO}}(y,\pdf,\delta).
$
The width is weakly increasing in the sensitivity parameter $\delta$, which implies that a larger violation of monotonicity leads to a larger identified set. 
However, the effect of the sensitivity parameter $\pdf$ on this width can be both negative and positive depending on the specific underlying parameters $\theta$. 
For example, we note that $\Fcod$  is point identified either if $\pdf=0$ or  $\pi_d=0$, which denotes the absence of always or never takers.
Heuristically speaking, the parameter $\pdf$, therefore,   trades off the identification power gained from the non-existence of defiers and the non-existence of always or never takers.

The  functions $\underline{H}_{Y_d^{CO}}$ and $\overline{H}_{Y_d^{CO}}$  clearly bound $\Fcod$ in a first-order stochastic dominance sense. However, since they do not imply that the implied functions of $F_{Y_d^{dT}}$ and $F_{Y_d^{DF}}$ are nondecreasing, 
they are not necessarily a feasible candidate of $\Fcod$. 
To give an intuition for this result and for the sake of argument, we now assume that all outcome variables are continuously distributed. We consider  $\underline{H}_{Y_d^{CO}}$, and we assume that  the bound on  the outcome heterogeneity $\delta$ determines the bound, i.e.,
$
\underline{H}_{Y_d^{CO}}(y)=  G_d(y) -   \frac{\pdf}{\pi_{\Delta}} \delta.
$
This bound does not necessarily imply that the always takers  have a positive density. 
Specifically, the density of the lower bound is $
g_d(y) = (q_{dd}-q_{d(1-d)}(y))/\pi_{\Delta},
$ whereas to guarantee that the density function $f_{Y_d^{dT}}$ does not take any negative value, any feasible candidate  of $f_{Y_d^{CO}}$ has to  satisfy that 
\begin{align}
	f_{Y_d^{CO}}(y)  \leq  \frac{q_{dd}(y)}{\pco} \label{equ_at_nondecreasing}
\end{align} 
for all $y \in \mathbb Y$.\footnote{To be precise, one can assume that $q_{d(1-d)}(y)=0$ and as $\pi_\Delta =\pco- \pdf  \leq \pco$ the claim follows.}
A similar restriction as \eqref{equ_at_nondecreasing} can be derived for defiers such that any feasible candidate of the density 	$f_{Y_d^{CO}}(y)$  has to  also satisfy that, for all $y \in \mathbb Y$,
\begin{align}
	f_{Y_d^{CO}}(y) \geq \frac{\pi_\Delta}{\pco} \max \{ g_d(y), 0\}. \label{equ_df_nondecreasing}
\end{align}
Based on this argument, we construct
our final bounds, $\Flcod$ and $\Fucod$.  Specifically,  the distribution function $\Flcod$ is dominated by $\underline{H}_{Y_d^{CO}}$ in a first-order stochastic dominance sense, and the distribution function $\Fucod$  dominates $\overline{H}_{Y_d^{CO}}$ in a first-order stochastic dominance sense, and they both carefully take into account the reasoning of \eqref{equ_df_nondecreasing}~and~\eqref{equ_at_nondecreasing}.
In Appendix~\ref{sec_proof_theoremdist}, we show that these distribution functions both bound the distribution function $\Fcod$ and  are feasible candidates.

\subsection{Identification Result}\label{sec_mainresult_sharpbounds} We first provide the analytical expressions of the bounds in the following. The lower bound of the distribution functions $\Fcod$ is given by
\begin{align}
	\Flcod(y,&  \pdf,\delta)	= \frac{1}{\pco} \Qdd(y) \label{boundslower}    \\
	& \hfill - \frac{1}{\pco}  \inf_{\tilde{y} \geq  y} \left( Q_{dd}(\tilde{y})  - \left(\pi_{\Delta} G_d^+(\tilde{y})  -  \inf_{\wh{y} \leq \tilde{y}  } \left( \pi_{\Delta} G_d^+(\wh{y}) -  \pco  \underline{H}_{Y_d^{CO}}(\wh y, \pdf,\delta)  \right) \right) \right),\notag
\end{align}
and similarly the upper bound  by
\begin{align}
	\overline{F}_{Y_d^{CO}} (y,	& \pdf,\delta) = \frac{ \pi_{\Delta} }{\pco}    G_d^+(y)  \label{boundsupper}\\
	& \hfill  -\frac{1}{\pco} \sup_{\tilde{y}\geq y} \left(   \pi_{\Delta}  G_d^+(\tilde{y}) - \left( Q_{dd}(\tilde{y})-  \sup_{\wh{y} \leq \tilde{y}} \left(  Q_{dd}(\wh{y})- \pco \overline{H}_{Y_d^{CO}}(\wh y, \pdf,\delta)   \right)\right) \right). \notag
\end{align}
Based on the derivation above, Theorem~\ref{theoremdist} summarizes the result. 
\begin{theorem} \label{theoremdist}
	Suppose that Assumption~\ref{assumptionLATE} holds,  and the data generating process is compatible with the sensitivity parameters $(\pdf, \delta)$. Then, it holds that 
	\begin{align*}
		\underline{F}_{Y_d^{CO}}(y,\pdf,\delta) \leq F_{Y^{CO}_d}(y)  \leq \overline{F}_{Y_d^{CO}}(y,\pdf,\delta),
	\end{align*}
	for $d \in \{0,1\}$ and for all  $y \in \mb Y$.
	Moreover, there exist data generating processes which are compatible with the above assumptions such that the outcome distribution of compliers equals either $\underline{F}_{Y_d^{CO}}(y,\pdf,\delta)$, $\overline{F}_{Y_d^{CO}}(y,\pdf,\delta)$,  or  any convex combination of these bounds. 
\end{theorem}
Theorem~\ref{theoremdist} shows not only that the proposed bounds are valid but also that without imposing further assumptions, the bounds cannot be tightened in a first-order stochastic dominance sense.\footnote{As the derived bounds are rather complicated, we propose simpler bounds for each of our sensitivity parameters in Appendix~\ref{sec_simplified_bounds}. These bounds are possibly conservative.  We explain how to evaluate in an empirical setting whether they are close to the sharp bounds derived in this section.}

\begin{remark}
	Theorem~\ref{theoremdist} does clearly not imply that all distribution functions that are bounded by the distribution functions 	$\underline{F}_{Y_d^{CO}}$ and  $\overline{F}_{Y_d^{CO}}$ are feasible candidates of the distribution function of $F_{Y^{CO}_d}$. The reason for that is that these functions do not necessarily imply nondecreasing distribution functions of the other groups. Since we are not  interested  in the distributions functions themselves but in parameters defined through the bounds, this result is sufficient to derive sharp bounds on  the  sensitivity and robust region for empirical conclusions about these parameters. 
\end{remark}

\begin{remark}
The parameter of interest  is often not only the average treatment effect but also, e.g., quantile and distribution treatment effects.  As Theorem~\ref{theoremdist} bounds the entire outcome distribution functions of compliers, in a first-order stochastic dominance sense,  these treatment effects are identified as well and are sharp for many relevant parameters. 
	We present them in Appendix~\ref{sec_additional_treatment_effects_appendix}.
\end{remark}
\begin{remark}
	In empirical applications, researchers also often have access to pre-intervention covariates. In Appendix~\ref{sec_additional_covariates}, we show how these covariates can be exploited to reduce the size of the identified set of the distribution function $\Fcod$. These covariates can then be used to tighten the sensitivity and to enlarge the robust regions. 
\end{remark}

\section{Sensitivity Analysis}\label{sectionsensitivtyanalysis}
We present our main sensitivity analysis in this section.
\subsection{Sensitivity Region}\label{sub_sec::sr}
We derive the sensitivity region, which is the set of sensitivity parameter pairs for which a feasible candidate of the distribution function $\Fcod$ exists.  Sensitivity parameters that lie in the complement of this set refute the model, and we, therefore, do not consider them further.\footnote{\cite{masten2021salvaging} denote the complement of the sensitivity region the falsification region.}  
\subsubsection{Population Size of Defiers}
We show that the population size of defiers is partially identified.  We denote an upper bound  by
\begin{align}
	\opdf = \min\{\mb{P} (D=1|Z=0), \;\mb{P} (D=0|Z=1) \}.\label{equ_maximal_pdf}
\end{align}  
The first element of the minimum represents the sum of the population size of always takers and defiers, whereas the second one of never takers and defiers. The population size of defiers is  clearly smaller than both of these quantities.

The lower bound on the population size of defiers is denoted by 
\begin{align}
	\updf= &  \max_{s \in \{0,1\}} \{ \sup_{B \in \mathscr{B}} \{\mb{P}(Y \in B, D=s|Z=1-s) - \mb{P}(Y  \in B, D=s |Z=s) \} \}.\label{equa::minimalpdf}
\end{align}
The supremum is taken  over the differences in population distributions of defiers and compliers, which bounds the population size of defiers from below. The Proposition~\ref{theoremlambda} shows that these bounds are sharp.\footnote{\cite{richardson2010analysis} present sharp bounds on $\pdf$ for a binary outcome variable.}
\begin{proposition} \label{theoremlambda}
	Suppose Assumption~\ref{assumptionLATE}  holds.  Then the population size of defiers, $\pdf$, is bounded by $[\updf, \opdf]$. Moreover, there exist data generating processes which are compatible with the above assumptions such that the population size of defiers equals any value within these bounds. Thus, the bounds are sharp.	
\end{proposition}
If the lower bound on population size of defiers is greater than zero, $\updf > 0$, at least one of the classical LATE assumptions, including monotonicity, is violated.

 This reasoning is align with the result of \cite{Kitagawa2015}, who shows  that $\updf > 0$ is sufficient and necessary such that the LATE assumptions are valid.

However, if the above inequalities contradict, i.e., $\updf>\opdf$, the sensitivity region is empty. This implies that even if one allows for a violation of monotonicity, our model assumptions must be violated as well. 

\subsubsection{Outcome Heterogeneity}
We now consider the sensitivity parameter $\delta$. 
Based on Theorem~\ref{theoremdist},  we can bound the sensitivity parameter $\delta$ from below and from above for a given value of the sensitivity parameter $\pdf$.

A given pair fo sensitivity parameters  $(\pdf, \delta)$ is refuted if the implied lower and upper bounds, $\Flcod$ and $\Fucod$, intersect, so that there does not exists a feasible candidate of the distribution function $\Fcod$ which is compatible with these sensitivity parameters. The domain of the sensitivity parameter $\delta$ is bounded from below by
\begin{align}
	\deltamin(\pdf) =\min_{d \in\{0,1\}} \inf \{\delta:  \inf_y \Flcod(\yld)-\Fucod(\yld)\geq 0 \}.\label{equation_delta_min}
\end{align}
The feasible set of the sensitivity parameter $\delta$ is further bounded from above. The bounds $\Flcod$ and $\Fucod$ imply bounds on the distribution function of $\Fdfd$, where the largest value of the Kolmogorov-Smirnov norm between the distributions of $\Fcod$ and $\Fdfd$ is achieved when $\delta=1$. It follows that there does not exists a feasible candidate function of $\Fcod$ such that the implied outcome heterogeneity parameter exceeds this value. 
We denote the upper bounds by
\begin{align} 
	\overline{\delta}(\pdf) = \max_{d \in \{0,1\} } \sup_{y \in \mb Y}  & \left\lbrace | \Fucod(y, \pdf, 1) - \Fudfd(y,\pdf, 1) |, \right. 	\notag\\
	& \qquad  \qquad  \left. | \Flcod(y,\pdf, 1) - \Fldfd(y,\pdf, 1) | \right\rbrace. \label{equation_delta_max}
\end{align}
By the reasoning of  Theorem~\ref{theoremdist}, these bounds are sharp, and any convex combination of these bounds is feasible as well. 
It follows that our sensitivity region is given by
\begin{align}
	SR= \{(\pdf, \delta): \pi \in [\updf, \opdf]  \text{ and } \underline{\delta}(\pdf) \leq \delta \leq  \overline{\delta}(\pdf)    \}. \label{equ_sensitvityregion_main}
\end{align}

\subsection{Robust Region}\label{sub_sec::rr}
We now derive the robust region for the empirical conclusion that $\Delta_{CO}\geq \mu$.\footnote{In Appendix~\ref{sec_additional_treatment_effects_appendix}, we also consider other treatment effects than the average treatment effect of compliers. Sensitivity and robust regions for empirical conclusions about these parameters can then also be  derived based on the reasoning of this section.}  
To simplify the presentation, we assume in the following that the sensitivity region is nonempty and that $\Delta_{CO}(\updf, \deltamin(\updf))\geq \mu$.\footnote{If $\Delta_{CO}(\updf, \deltamin(\updf))< \mu$, the robust region is empty.}

By first-order stochastic dominance of the distribution functions $\Flcod$ and $\Fucod$,  we can construct sharp bounds on many  treatment effect parameters,  that depend  on these bounds \citep[see Lemma 1 in ][]{stoye2010partial}. Specifically, let
\begin{align}
	\underline{\Delta}_{CO}(\pdf, \delta)& =  \int_{\mb Y} y \; d\Fucop(y,\pdf,\delta) - \int_{\mb Y} y\; d\Flcoa(y,\pdf,\delta) \label{equ_LATE_lowerbound} \\
	\overline{\Delta}_{CO}(\pdf, \delta) & =   \int_{\mb Y} y \;  d\Flcop(y,\pdf,\delta) - \int_{\mb Y} y \;  d\Fucoa(y,\pdf,\delta). \label{equ_LATE_upperbound}
\end{align}

\begin{figure}[!h]
	\centering
	\resizebox{0.4\linewidth}{!}{\input{Graphics/quantile_identification}}
	\caption[Illustration of sensitivity and  robust region II.]{Sensitivity and  Robust Region. Non-shaded region represents sensitivity region.}\label{iden_figurelqtes}
\end{figure}
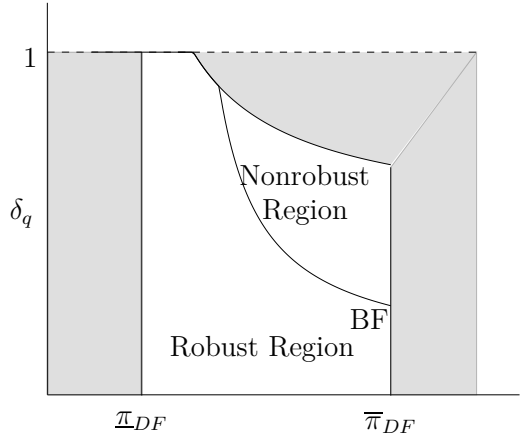

\begin{corollary}\label{coro_treatment_effect}	Suppose that Assumption~\ref{assumptionLATE} holds,  and the data generating process is compatible with the sensitivity parameters $(\pdf, \delta)$. Then, the average treatment effect of compliers, $\Delta_{CO}$, is bounded by $	[\underline{\Delta}_{CO}(\pdf, \delta) , 	\overline{\Delta}_{CO}(\pdf, \delta) ]$. 
	Moreover, there exist data generating processes which are compatible with the above assumptions such that the average treatment effect of compliers equals any value within these bounds. Thus, the bounds are sharp.
\end{corollary}
For a given sensitivity parameter $\pco$, we now consider the breakdown point given by
$$BP(\pdf )=\sup \{\delta: (\pdf, \delta) \in \A \text{ and }\underline{ \Delta}_{CO}(\pdf,\delta) \geq\mu\}. $$
For a given sensitivity parameter $\pdf$, it identifies the weakest assumption on outcome heterogeneity between compliers and defiers such that the empirical conclusion holds. The breakdown point, as a function of the sensitivity parameter $\pdf$, is not necessarily decreasing in the population size of defiers as the bounds on the outcome distribution of compliers can become tighter if the value of $\pdf$ increases (see the discussion in Section~\ref{sec_simple_bounds_maintext}).
The breakdown frontier of the average treatment effect is the boundary of the robust region and given by the set of all breakdown points
\begin{align}
	BF &  =  \{ (\pdf, \delta) \in \A:\delta=BP(\pdf)\}.  \label{equation_BF}
\end{align}
The robust region of the empirical conclusion that $\Delta_{CO}\geq \mu$ is  characterized by 
\begin{align}
	RR & =\{(\pdf, \delta) \in \A: \delta \leq  BP(\pdf) \}. \label{equation_robustregion}
\end{align}
The nonrobust region, that is the complement of the robust region within the sensitivity region,  contains pairs of sensitivity parameters which only may not imply treatment effects being consistent with the empirical conclusion.
Figure~\ref{iden_figurelqtes} illustrates one example of the sensitivity and robust region.\footnote{We refer to a discussion on how these sets can be used in an empirical setting to Section~\ref{sectionsetup}~and~\ref{sectionempirical}}

\section{Extensions}\label{sectionextension}
In this section, we show how our framework can be exploited to draw empirical conclusions about other population parameters, and how it simplifies if the outcome variable is binary.
\subsection{Treatment Effects for other Populations}\label{sec_additional_treatment_effects}
To show how  empirical questions about treatment effects of the entire population can be analyzed, we exploit that the proof of Theorem~\ref{theoremdist}  presents sharp bounds on all groups in a first-order stochastic dominance sense. For $d \in \{0,1\}$, let the lower bound be denoted by 
\begin{align*}
	\underline F_{Y_d} (y, \pdf, \delta) & =  \pco \cdot \underline F_{Y_0^{CO}} (y,\pdf , \delta) + Q_{d(1-d)}(y), 
\end{align*}
and the upper bound by
\begin{align*}
	\overline F_{Y_d} (y, \pdf, \delta) & =\pi_d + \pco \cdot \overline  F_{Y_0^{CO}} (y,\pdf , \delta) + Q_{d(1-d)}(y). 
\end{align*}
\begin{proposition}\label{prop_extensions_average}
	Suppose the instrument satisfies Assumption~\ref{assumptionLATE},  and the data generating process is compatible with the sensitivity parameters $(\pdf, \delta)$.  Then, it holds  that 
	$$ \underline F_{Y_d} (y, \pdf, \delta) \leq   F_{Y_d} (y, \pdf, \delta) \leq  \overline  F_{Y_d} (y, \pdf, \delta)$$
	for $d \in \{0,1\}$ and for all  $y \in \mb Y$.
	Moreover, there exist data generating processes which are compatible with the above assumptions such that the potential outcome distributions equal either $\overline{F}_{Y_d}(y,\pdf,\delta)$, $\underline{F}_{Y_d}(y,\pdf,\delta)$, or  any convex combination of these bounds. 
\end{proposition}
As the data do not contain any information about the distribution functions $F_{Y_0^{AT}}$ and $F_{Y_1^{NT}}$, the bounds $ \underline F_{Y_d}$ and  $\overline F_{Y_d}$ are such that their respective probability mass is shifted to the extreme of the support $\mb Y$.
To interpret these bounds, for any $y$ in the interior of $\mathbb Y$, we consider the difference 
$$\overline F_{Y_d} (y, \pdf, \delta) - \underline F_{Y_d} (y, \pdf, \delta)
= \pi_d.$$
The sensitivity parameter $\delta$  does not influence the distribution of the outcome of the entire population, as it only influences how the observed outcome probability mass is distributed between the groups.  
However, the size of the bounds decreases  with the population size of defiers,  $\pdf$,  as the population size of always and never takers $\pd$ decreases with  $\pdf$. This reasoning is intuitive as if $\pd$ decreases, the observed outcome probability mass  represent more of the population under consideration.  
This result aligns with results of the literature \cite{kitagawa2021identification,kamat2018identifying}, who shows that imposing monotonicity (e.g., $\pdf=0$) does not imply a smaller identified set of the average treatment effect of the entire population  if the LATE assumptions are not violated. 

Based on the bounds presented in Proposition~\ref{prop_extensions_average}, we can now derive a sensitivity analysis similar to the one presented in Section~\ref{sectionsensitivtyanalysis}. However, to derive informative results about the average treatment effect of the entire population, we would have to impose that the outcome is bounded as otherwise the average treatment effect is not identified in general. 

The sensitivity analysis of this paper is based on the premise that the treatment effect of compliers is the object of interest. However, if the parameter of interest is the treatment effect of the entire population, one might then be willing to impose assumptions not only on outcome heterogeneity between compliers and defiers but also between other groups. To be precise, we can replace the sensitivity parameter $\delta $ by $\delta_p\in [0,1]$ such that
$$ \max_d \sup_y \{ | F_{Y_d^{T}} (y) -F_{Y_d^{T'}} (y)| \}   | \leq  \delta_p \quad \forall \;  T,T' \in \{ AT, NT, CO, DF\}. $$
Using similar arguments as in the proof of Theorem~\ref{theoremdist}, one can then derive sharp bounds on the outcome distribution functions of the entire population and then conduct a sensitivity analysis similar to the one described in  Section~\ref{sectionsensitivtyanalysis}.
Empirical conclusions drawn  on this parameterization might be substantially more informative. 

\subsection{Binary Outcome Variable}\label{sec:binaryoutcomemodel}
In many empirical applications, the outcome of interest is binary. The results of Section~\ref{sec_identification_bounds}~and~\ref{sectionsensitivtyanalysis} are still valid in this case, but we show in this section that the bounds substantially simplify so that  they are easier applicable. Let $P_d^{T}= \mathbb P (Y_d^T=1)$ denote the probability that the random variable $Y_d^T$ equals one, and let the conditional joint probability of the outcome and the treatment status be given by $P_{ds}=\mb P(Y=1, D=d|Z=s)$. We denote the underlying parameters by  $\theta_b=(P_{11},P_{10},P_{01}, P_{00}, P_0, P_1) \in [0,1]^6$.

Following the same arguments as above, 
the sensitivity and robust region depend on the marginal outcome distributions of the compliers.
The presence of defiers is also bounded by  $\pdf $, and the parameter of outcome heterogeneity  simplifies  to   
$$ \delta_b= \max_{d \in \{0,1\}} |P_d^{CO}-P_d^{DF}|. $$
The outcome probabilities of compliers are bounded from below  by 
\begin{align}
	\underline{P}_d^{CO}( \pdf,\delta) & = \max \left\lbrace  0, \frac{\Pddd -  \pd    }{\pco},  \frac{\Pddd -  \Pdds    }{\pco},  \frac{\Pddd -  \Pdds - \pdf  \delta_b }{\pi_{\Delta} }\right\rbrace, \label{equ_lower_bound_binary}
\end{align}
and from above by
\begin{align}		
	\overline{P}_d^{CO}(\pdf,\delta)&= \min \left\lbrace 1, \frac{\Pddd }{\pco}, 
	\frac{\Pddd -  \Pdds +\pdf    }{\pco },  \frac{\Pddd -  \Pdds + \pdf  \delta_b }{\pi_{\Delta} } \right\rbrace . \label{equ_upper_bound_binary}
\end{align}
 
\begin{corollary}\label{extensions_corollarybinary}
	Suppose Assumption~\ref{assumptionLATE}  holds,  and the data generating process is compatible with the sensitivity parameters $(\pdf, \delta)$. The outcome probabilities of compliers are  bounded by
	$ \underline P_d^{CO} \leq P_d^{CO} \leq  \overline P_d^{CO}$. 
	Moreover, there exist data generating processes which are compatible with the above assumptions such that the population size of defiers equals any value within these bounds. Thus, the bounds are sharp.	
\end{corollary}
The interpretation of the width of these bounds follows the same reasoning as in Section~\ref{sec_simple_bounds_maintext}.
The lower bound of the population size of defiers simplifies to
\begin{align*}
	\updf= \max_{d \in \{0,1\}}  &   \left\lbrace  \sum_{y=0}^1 \max \{0, \mb P (Y=y, D=d|Z=1-d) -   P (Y=y, D=d|Z=d ) \}  \right\rbrace.
\end{align*} 
The upper bound on $\pdf$ cannot be simplified further and is  given by \eqref{equ_maximal_pdf}.
The lower bound on outcome heterogeneity is  given by
\begin{align*}
	\underline{\delta}_b(\pdf)  = \frac{\updf}{\pdf }.
\end{align*}
The lower bound on the sensitivity parameter $\delta$  decreases with the population size of defiers. 
The upper bound on the sensitivity parameter $\delta$ is given by the maximal difference between the  outcome probabilities of compliers and defiers 
\begin{gather*}\overline{\delta}_b(\pdf)  = \max_{d \in \{0,1\}}  \max \{ |\underline P_d^{CO}(\pdf , 1) - \underline P_d^{DF}(\pdf , 1)| ,|\overline P_d^{CO}(\pdf , 1) - \overline P_d^{DF}(\pdf , 1)|   \}.
\end{gather*}
The sensitivity parameter space is given by 
$$ 
\A_b=\{ (\pdf , \delta_b) \in [\updf, \opdf] \times [0,1]: \underline{\delta}_b(\pdf) \leq \delta_b \leq \overline{\delta}_b(\pdf) \},
$$
and the robust region for the claim $\Delta_{CO}\geq \mu$, if $\underline P_1^{CO}(\updf , \deltamin_b)  -  \overline P_0^{CO}(\updf , \deltamin_b) \geq \mu$, is given by 
\begin{align*}
	RR_b= \{ (\pdf , \delta_b) \in \A_b: \underline P_1^{CO}(\pdf , \delta_b)  -  \overline P_0^{CO}(\pdf , \delta_b) \geq \mu\}. 
\end{align*}
Using the simple algebra structure of the bounds of the outcome probabilities, a closed-form expression for both the robust and the sensitivity region can be derived. As this expression is rather lengthy without providing much intuition, we state it in Appendix~\ref{appendixproofextensions_corollarybinary}.

\section{Estimation and Inference} \label{sectioninference}
Even though the contribution of this paper is the derivation of the sensitivity and robust region for a particular empirical conclusion,  we consider some methods for estimation and inference of these two regions. While the technical details are deferred to  Appendix~\ref{appendixestimationandinference}, in this section, we sketch the main issues of conducting inference in this setting and our proposed solutions.  To simplify the exposition, we consider the setting of a continuous and a binary outcome variable, but our method is not restricted to these distributions.

Throughout this section, we assume that we have access to the data $\{(Y^z_{i},D^z_{i})\}_{i=1}^{n_z}$  for $z\in \{0,1\}$  that   are independent and identically distributed according to the distribution of $(Y,D)$  conditionally on $Z=z$ with support $\mb{Y}\times \{0,1\}$. We denote this distribution by $(Y^z, D^z)$ and we let $n=n_0+n_1$, where $n_0/n$ converges to a nonzero constant as $n\rightarrow \infty$.\footnote{We discuss this assumption in  Assumption~\ref{assumption_inference_sampling_distribution}.} 
\subsection{Estimation} 
To construct estimators of the sensitivity and robust region for a particular   empirical conclusion, we note that the identification argument of these regions are constructive. It follows from Section~\ref{sectionsensitivtyanalysis} that the boundaries of both  regions are identified   by the following  mapping,\footnote{The signs of  the components of  the mapping~$\phi(\theta, \pdf)$   simplify the subsequent analysis.}
\begin{align}
	\phi(\theta, \pdf) = (\updf, \, - \opdf, \, \deltamin(\pdf), \, - \deltamax( \pdf), \, BP(\pdf)), \label{equ_mapping}
\end{align} which is   evaluated at  the sensitivity parameter $\pdf \in \left[0,0.5\right)$ and the underlying parameters  $\theta$, that is defined in \eqref{eq_underlying_parameters}. 
Estimating the sensitivity and robust region is then equivalent to estimating this mapping.
To  do so, we consider
estimates of the underlying parameters $\theta$ that are simply obtained by replacing unknown population quantities by their corresponding nonparametric sample counterparts and by standard nonparametric kernel methods.   
We denote the estimates of $\theta$ by $\wh \theta$.	Point estimates of the mapping $\phi(\theta, \pdf)$ can then be derived by simple plug-in methods. We defer a detailed description to Appendix~\ref{appendixestimation}.

\subsection{Goal of Inference}\label{sec:goalofinference}
We propose to construct confidence sets for the sensitivity and robust region such that the confidence set for the sensitivity region is an outer confidence set  and for the robust region is an inner confidence set.\footnote{Considering inner confidence set for the robust region follows from \cite{masten2020}.} These confidence sets should therefore 
jointly satisfy with probability approaching the confidence level, $1-\alpha$, that  (i) any sensitivity parameter pair of the sensitivity region lies within  the confidence set for the sensitivity region and (ii) not any single parameter pair of the nonrobust region  lies within the confidence set for the robust region.\footnote{To give one more interpretation of the confidence sets and using the language of hypothesis testing,  a sensitivity parameter pair, $(\pdf, \delta)$,  does not lie in the sensitivity region only if we can  reject such a hypothesis with  confidence level $1-\alpha$. Contrary, $(\pdf, \delta)$ lies in the robust region, only if we can reject that it is nonrobust with confidence level $1-\alpha$. The confidence sets are constructed so that the hypothesis tests are valid uniformly in the sensitivity parameter space. }
Let  $\widehat \A_{L}$ and  $\widehat{RR}_{L}$ denote two sets of the sensitivity parameters. They satisfy the described condition if
\begin{align}
	{\underset{n \rightarrow \infty}{\lim} \; \mb{P}( \A \subseteq \widehat \A_{L}  \text{ and } \widehat{RR}_{L}(\A) \subseteq RR(\A) ) \geq 1-\alpha}\label{equ_goal_of_inference}.
\end{align} 
Based on the definition of the mapping  $\phi (\theta, \pdf) $, it therefore suffices to construct  a  lower confidence band for each component of the estimator  $\phi (\widehat \theta, \pdf) $ as a function of $\pdf$ that are jointly valid.\footnote{Throughout this section,  we consider confidence sets that are uniformly valid in the sensitivity parameter space, but not necessarily in the distribution of the underlying parameters $\theta$} That is, we need to find a function that is componentwise a uniformly   lower bound  $\phi_L (\widehat \theta, \pdf)$ in $\pdf$ of $\phi ( \theta, \pdf) $ so that\footnote{We verify this equivalence in Appendix~\ref{sec_preliminariesforconfidencesets}.}
\begin{align}
	\lim_{n \rightarrow \infty} \mb P \left(\min_{ 1 \leq l \leq 5 } \inf_{ \pdf \in [0,0.5) }   e_l^\top (\phi_L (\widehat \theta, \pdf) - \phi ( \theta, \pdf)) \leq  0\right) \geq 1- \alpha, \label{equ_goal_of_indeference_2}
\end{align} 
where $e_l$ is the l-th unit vector.\footnote{Conservative confidence sets for only the average treatment effect of compliers for specific values of $(\pdf, \delta)$  directly follow from the presented procedure. To obtain nonconservative confidence sets, one can follow the literature on partially identified parameters  \cite[see, e.g., ][]{imbens2004confidence}}

\subsection{Inference for a Continuous Outcome Variable}
We analyze the distribution of  $\phi(\wh \theta, \pdf)$  in order to construct confidence sets for the mapping $\phi(\theta, \pdf)$.\footnote{We want to emphasize that this procedure is valid for a fixed distribution. In particular, we do not consider settings of weak instruments or data generating processes which are such that the  robust region becomes empty.}
Under  regularity assumptions presented in Appendix~\ref{sec_appendix_assumption}, the estimators of the underlying parameters $\widehat  \theta$ converge in $\sqrt{n}$ to a tight Gaussian process.
Since the mapping $\phi$ is not Hadamard-differentiable, as it depends on minimum, maximum, supremum, and infimum  of random functions, standard Delta method arguments do  not apply in this setup  \citep[see][]{fansantos2016}. We propose a method to construct confidence sets that are asymptotically conservative but valid in the sense of \eqref{equ_goal_of_inference}. It is  based on ideas of population smoothing that have  been suggested by, e.g.,  \cite{haile2003, chernozhukov2010quantile, masten2020}.

In contrast to considering the mapping  $\phi$, which identifies the sensitivity and robust region, we  construct a smooth mapping, $\phi_\kappa$,  which yields valid bounds of both regions. The smoothed mapping $\phi_\kappa$ is indexed by a fixed smoothing parameter $\kappa\in \mb N$. The mapping $\phi_\kappa$ is  differentiable such that the standard functional Delta method can be applied to $\phi_\kappa$ and we can study its asymptotic distribution by standard methods.
The mapping $\phi_\kappa$ is further  such that it  yields an   outer set of the sensitivity region and an inner set of the robust region. This reasoning implies  that confidence sets of the smooth mappings $\phi_\kappa$,  which are valid in the sense of \eqref{equ_goal_of_inference}, are also valid for the  mapping $\phi$.

In finite samples, the choice of the smoothing parameter $\kappa$ comprises  the trade-off of constructing conservative confidence sets and better finite sample approximations of the underlying distributions.   
Suppose the smoothing parameter $\kappa$ is small. In that case, the smoothed sensitivity and robust region are very similar to the original regions. However, the finite-sample distribution of $\phi_\kappa(\wh \theta \pdf)$ might not be well-approximated by its asymptotic distribution. 
Vice versa, suppose the smoothing parameter $\kappa$ is large. The  finite-sample distribution of $\phi_\kappa(\wh \theta \pdf)$ might be well-approximated by its asymptotic distribution.  However, the smoothed sensitivity and robust region are conservative to the original regions.

In Appendix~\ref{sec_population_smoothing},  we show how the smoothed mappings can be constructed. It then follows that plug-in estimators of the smoothed mappings converge in  $\sqrt{n}$ to a Gaussian process by standard functional Delta method arguments. 
The covariance structure of this process is, in general, rather complicated and tedious to estimate. We, therefore, apply the nonparametric bootstrap to simulate its distribution. Consistency of this bootstrap procedure follows from arguments of \cite{fansantos2016}. 
In  Appendix~\ref{appendixestimationandinference}, we show how to construct the confidence sets based on the described procedure and that they achieve the outlined goal \eqref{equ_goal_of_inference}.

\subsection{Inference for a Binary Outcome Variable}
Following the discusssion about a bianry outcome model in Section~\ref{sec:binaryoutcomemodel}, the mapping yielding the sensitivity and the robust region for a particular   conclusion for a binary outcome variable is  given by\footnote{where its precise definition follows from Section~\ref{sec:binaryoutcomemodel} and  Appendix~\ref{sec:appendix:inference_binary}.} $$\phi_b(\theta_b, \pdf)= (\underline{\pi}_{\text{DF,b}}, -\overline{\pi}_{\text{DF,b}},  - \overline{\delta}_b( \pdf), BP_b(\pdf)),$$  The interpretation of $\phi_b$ follows the one for a continuously distributed outcome variable, and in principle, we could apply the same inference procedure  as described above. However, the mapping $\phi_b(\theta_b, \pdf)$  is substantially simpler than the mapping $\phi(\theta, \pdf)$ so that, in this section,  we can apply more classical inference procedure to obtain confidence sets in the sense of  \eqref{equ_goal_of_inference}; in particular, we follow ideas of \cite{masten2020} and the literature about moment inequalities \citep[see, e.g., ][]{andrews2010inference}. 

Under standard sampling assumptions, it follows that the estimators of the underlying parameters are jointly $\sqrt{n}$ normally distributed  (see Appendix~\ref{sec:appendix:inference_binary}). The  mapping  $\phi_b(\theta_b, \pdf)$ is clearly not Hadamard-differentiable, as it consist of  minimum and maximum of random functions. Standard Delta method arguments are therefore not applicable here as well.  Valid confidence sets could be obtained by projection arguments, which, however, are known to be conservative in general.  

We show instead that the mapping $\phi_b$ is  Hadamard directionally differentiable in the direction of $\theta$ when evaluated at finitely many  $\{ \pdf^k\}_{k=1}^K$.  Using generalized Delta method arguments, the estimator of the mapping $\phi_b$  converges to some tight random process, which is a continuous transformation of a  Gaussian process,  indexed at the finite set  $\{\pdf^k\}_{k=1}^K$.  As this limiting distribution is rather complicated, we do not construct our inference procedure directly on its limiting distribution, but one can choose various modified bootstrap methods to simulate this distribution,  e.g.,  subsampling or numerical-Delta method \citep[see][]{ dumbgen1993nondifferentiable, Hong2016}. In this paper, we follow a  bootstrap method which relies on ideas based on the moment inequality literature \citep[see, e.g., ][]{andrews2010inference, bugni2010bootstrap} and we explain the procedure in detail in Appendix~\ref{sec:appendix:inference_binary}. 
Based on this bootstrap procedure, we can construct valid lower confidence sets for $\phi_b$ indexed at the finite set of sensitivity parameters  $\{\pdf^k\}_{k=1}^K$. Using these confidence sets and exploiting the functional form of $\phi_b$, we then obtain lower confidence sets for the estimator of the mapping $\phi_b$, which are uniformly valid in $\pdf$. We state these arguments precisely in Appendix~\ref{sec:appendix:inference_binary} and show that these confidence sets are asymptotically valid in the sense of our goal of inference  \eqref{equ_goal_of_inference}.

\section{Simulations}\label{sectionsimulations}
\subsection{Setup}
We  study the finite sample performance of the proposed estimators of the sensitivity and robust regions through a Monte Carlo study. We consider  different data generating processes with varying degrees of violations of monotonicity, implying different sizes and shapes of both the sensitivity and robust regions.   Specifically, we consider the following population sizes $(\pco, \pdf) \in \{ (0.35, 0.05) , (0.25,0.15) \}$, where $\pdf=\pat =0.3.$
We set $\mb{P}(Z=1)=0.5$ and we generate the  outcome by
\begin{align*}
	Y^{CO}_1& \sim \mathcal{B}( 1,0.5+\Delta_{CO})  & Y^{DF}_1& \sim \mathcal{B}( 1,0.5+\Delta_{DF})  &
	Y^{AT}_1, Y^{NT}_0 ,   Y^{DF}_0, Y^{CO}_0& \sim \mathcal{B}(1, 0.5),
\end{align*}
where $\Delta_{CO} \in \{0.2,0.1\}$,   and $\mathcal{B}( 1,p)$ denotes the Bernoulli distribution with parameter $p$. 
The sensitivity region is nonempty as the data generating process satisfies our model assumptions.
We consider the empirical conclusion of a positive  treatment effect of compliers, so that  the robust region is nonempty in each of the data generating processes as the Wald estimand is positive.   The bootstrap procedure requires to choose the tuning parameter $\eta$, which is explained in Appendix~\ref{sec:appendix:inference_binary}.  We consider different values of $\eta$  given by $\{0.2, 0.5,1,1.5,2\}/\sqrt{N}$. The  results are based on 10,000 Monte Carlo draws. 

\subsection{Simulation Results}
Table~\ref{table_mc_sim} shows the results of the simulated coverage rates at which the confidence sets  cover the  population sensitivity and nonrobust region for the different data generating processes and choices of tuning parameters. Our considered choice of tuning parameters implies that  the simulated coverage of our confidence sets is close to the nominal one in most data generating processes.
These results illustrate that the confidence method performs reasonably well in finite samples.
\begin{table}
	\centering
	\caption{Simulated coverage rates of the sensitivity and robust region for  a positive treatment effect. }\label{table_mc_sim}
	\begin{tabular}{cccrrrrr}
		\toprule
		$\pco$ & $\Delta_{CO}$ & $\Delta_{TE}$ & $\eta=0.2$ & $ \eta=0.5$ & $ \eta=1$  & $\eta=1.5$ & $\eta=2$ \ \\
		\midrule
		\rule{0pt}{3ex} 
		\multirow{4}{*}{0.35} & 0.3 & 0 & 99.1 & 97.8 & 95.1 & 91.3 & 90.8 \\ 
		& 0.3 & -0.3 & 96.9 & 94.3 & 91.2 & 92.5 & 91.6 \\ 
		& 0.1 & 0 & 99.3 & 98.5 & 96.0 & 92.9 & 91.2 \\ 
		& 0.1 & -0.3 & 99.3 & 98.5 & 95.6 & 93.1 & 91.3 \\ 
		\rule{0pt}{3ex} 
		\multirow{4}{*}{0.25} & 0.3 & 0 & 98.9 & 98.1 & 95.2 & 92.4 & 91.2 \\ 
		& 0.3 & -0.3 & 99.1 & 98.0 & 94.3 & 92.9 & 91.4 \\ 
		& 0.1 & 0 & 99.3 & 98.6 & 96.0 & 93.5 & 91.2 \\ 
		& 0.1 & -0.3 & 99.0 & 97.7 & 94.3 & 92.9 & 90.9 \\ 
		\bottomrule
	\end{tabular}
	\begin{flushleft}
		\footnotesize{The data generating process and the expressions follow the description of the text. Results are based on 10,000 Monte Carlo draws.}
	\end{flushleft}
\end{table}

\section{Empirical Application}\label{sectionempirical}
To illustrate our proposed framework, we apply this sensitivity analysis to data from  \cite{angristevens1998}, who analyze the effect of having a third child on the labor market outcomes of mothers. It is shown that even small violations of the monotonicity assumption may have a large impact on the robustness of the estimated treatment effects such that even the sign of the treatment effects may be indeterminate. The same-sex instrument in \cite{angristevens1998} arguably satisfies Assumption~\ref{assumptionLATE}: The independence assumption seems to be plausible by the following reasoning: The sex of a child is determined by nature, and only the number of and not the sex of the child arguably influences the labor market outcome. The relevance assumption is testable. However, monotonicity might be violated. We apply the proposed sensitivity analysis to evaluate the robustness of the estimated treatment effects to a potential violation of monotonicity in this setting.  For simplicity, we focus on two outcome variables: the labor market participation of mothers and their annual wage.\footnote{The annual wage is a continuously distributed running variable with a point mass at zero.} The binary decision to treat represents the extensive margin and the continuous outcome variable a mix of extensive and intensive variables. 
We use the same data as \cite{angristevens1998}.\footnote{Data are taken from the website Joshua D. Angrist website www.economics.mit.edu/faculty/angrist from 1980. The sample is restricted to women at the age of 20-36, having at least two children, being white, and having their first child at the age of 19-25.} The sample size is 211,983. The point estimated difference of the population sizes of compliers and defiers is given by 0.06.

\subsection{Sensitivity Analysis for Binary Outcome Variable }
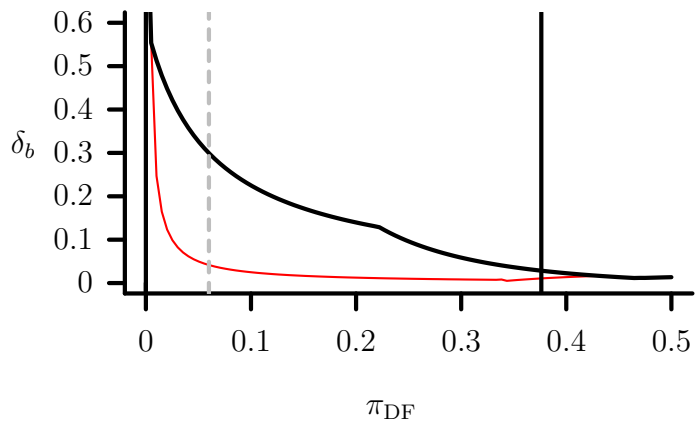
\begin{figure} 
	\centering
	\input{Graphics/binary_outcome_angrist_evans_res}
	\caption[Application - Confidence sets for the sensitivity and robust region I.]{
		Confidence sets for the sensitivity and robust region for a negative treatment effect of compliers. % is negative. %
		The confidence level is 95 \%. The treatment effect of compliers is the effect of having a third child on the labor market participation of mothers complying with the same-sex instrument. The black lines bound the sensitivity region, and the red line indicates the boundary of the robust region. The population size of defiers is on the horizontal axis, and outcome heterogeneity between compliers and defiers on the vertical axis. 	\label{figurelates_example}}
\end{figure}

We consider the labor market participation of mothers as the outcome variable. The Wald estimate is given by $-0.13$. Figure~\ref{figurelates_example}  illustrates the 95\% confidence set for the sensitivity and the robust region for the claim that the treatment effect of compliers is negative. The formal definition of these confidence sets is given in Section~\ref{sec:goalofinference}.  
In this example, a (conservative) 95\% confidence set for $\pdf$ is given by $[0, 0.37]$. Following the literature, one can therefore not conclude that monotonicity is violated in this example \citep[see for a comparison, e.g., ][]{small2017instrumental}.
The sensitivity parameter pairs below the red line represent the  robust region, which is the estimated set of sensitivity parameters implying a negative  treatment effect.
This figure shows that concerns about the validity of the monotonicity assumption have to be taken seriously.  Since  $BP(0.37)$ is almost zero, 
the hypothesis that the treatment effect is negative cannot be rejected without imposing any assumptions on the data generating process,
If the population size of defiers increases,  the breakdown frontier is relatively steeply declining, and thus the robust region is rather small. This implies that relatively strong assumptions on the outcome distributions of compliers and defiers have to be imposed to conclude that the treatment effect is negative in the presence of defiers. In contrast, if the population size of defiers is small, it is not necessary to impose strong assumptions about heterogeneity in the outcome variables to imply a negative effect.

This example shows that without imposing any assumptions on the data generating process, only non-informative conclusions can be drawn in this example, which is the case as the population size of defiers is not much restricted and is arguably implausible high.\footnote{ 
	To interpret these numbers, we note that the upper bound is a rather conservative estimate. If roughly  37\% of the population were a defier, then approximately 43\% of the population would have been a complier. This reasoning implies that roughly 90\% of the population would base their decision to have a third child on the sex composition of the first two children.}
One, therefore, might be willing to impose further assumptions to arrive at more interesting results, and we show how one could plausibly proceed. These assumptions should only serve as an example, and obviously, they have to be always adapted to the analyzed situation. We adopt the approach of \cite{de2017tolerating}. One of the most essential inherently unknown quantities of interest is the population size of defiers. Imposing a smaller upper bound of this quantity based on economic reasoning allows us to derive sharper results. Based on a survey conducted in the US, \cite{de2017tolerating} states that it seems reasonable that  5\% of defiers is a conservative upper bound of the population size of defiers in this setting.
If one is willing to impose this assumption, one would still have to assume that the differences in the Kolmogorov-Smirnov norm are less than 0.05, which is a  quite strong assumption. Therefore, we would conclude that the treatment effect is not robust to a potential violation in this specific example.  

\subsection{Sensitivity Analysis for Continuous Outcome Variable}
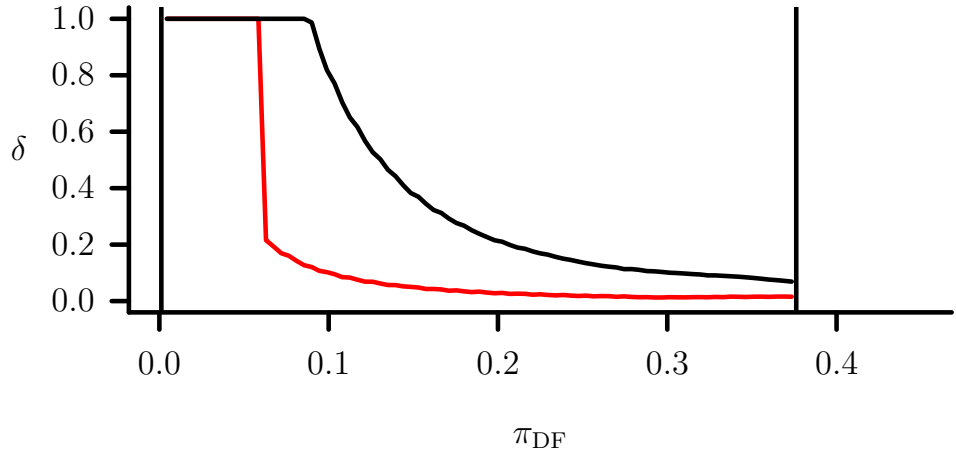
\begin{figure}
	\centering
	\resizebox{0.8\textwidth}{!}{	\input{Graphics/application_pic_continuous}}
	\caption[Application - Confidence Sets for the Sensitivity and Robust Region II]{Confidence Sets for the Sensitivity and Robust Region for a Negative Treatment Effect of Compliers. The confidence level is 95 \%. The compliers treatment effect is the effect of getting a third child on the annual log wage of mothers complying with the same-sex instrument. The black lines bound the sensitivity region, and the red line indicates the boundary of the robust region.
		\label{figureapplication:cont}}
\end{figure}

We now consider the annual log income of the mother. This variable has a point mass at zero, representing all women who do not work but is otherwise continuously distributed. The Wald estimate is given by $-1.23$. Figure~\ref{figureapplication:cont} shows the corresponding 95\% confidence sets for the robust and sensitivity region. If the monotonicity assumption were not violated, this estimate would imply that women who get a third child have an annual log wage reduced by $1.23.$

Figure~\ref{figureapplication:cont} shows 95 \% confidence sets for both the sensitivity and robust region. The same line of interpretation applies as in the case of a binary outcome variable. One can see that without imposing any assumption about the population size of defiers, the empirical conclusion of a negative treatment effect is not robust to a potential violation of monotonicity. However, applying the same reasoning as above and imposing a maximal population size of 5\% as an upper bound of the population size of defiers, one can see that the empirical conclusion is now robust to a potential violation of monotonicity.

To conclude,  this sensitivity analysis is of interest, as one can identify the sign and the order of magnitude of the treatment effects by imposing further assumptions. These imposed assumptions are substantially weaker than the monotonicity assumption so that the estimates gain credibility.

\section{Conclusion} \label{sectionconclusion}
The local average treatment effect framework is popular to evaluate heterogeneous treatment effects in settings of endogenous treatment decisions and instrumental variables. In some empirical settings, one might doubt the validity of one of its key identifying assumptions, the monotonicity assumption. Conducting a sensitivity analysis of the estimates in these settings improves the reliability of the results. This paper, therefore, proposes a new framework, which allows researchers to assess the robustness of the treatment effect estimates to a potential violation of monotonicity.  It parameterizes a violation of monotonicity by two parameters, the presence of defiers and heterogeneity of defiers and compliers. The former parameter is represented by the population size of defiers and the latter by the Kolmogorov-Smirnov norm bounding the outcome distributions of both groups. Based on these two parameters, we derive sharp identified sets for the average treatment effect of compliers and for any other group under further mild support assumption on the outcome variable. These identification results allow us to identify the sensitivity parameters that imply conclusions of treatment effect being consistent with the empirical conclusion. 
The empirical example of \cite{angristevens1998} same-sex instruments underlines the importance of the validity of the monotonicity assumptions as small violations of monotonicity may already lead to uninformative results.

\appendix
\renewcommand\thefigure{\thesection.\arabic{figure}} \setcounter{figure}{0}  
\newpage
 \begin{center}
 	\Large{\scshape Appendix}
 \end{center}

\section{Additional Materials for the Sensitivity Analysis}\label{appendixsensitivityregion}
In this section, we collect additional materials on identification of the sensitivity region.  In Section~\ref{sec_simplified_bounds}, we present simplified bounds, and we consider additional treatment effects in Section~\ref{sec_additional_treatment_effects_appendix}. We explain how covariates can be used to tighten the bounds in Section~\ref{sec_additional_covariates} and  we  give further results on a binary outcome variable in Section~\ref{sec_appendix_more_details_bianry_outcome}.

\subsection{Simple Bounds}\label{sec_simplified_bounds}
The bounds presented in Theorem~\ref{theoremdist} are rather tedious, and conducting inference for the sensitivity and robust regions based on these bounds is complicated as it depends on many tuning parameters to choose. In this section, we, therefore, present simpler bounds, which might be more easily applicable in an empirical context, and confidence sets of these regions might be more reliable.
The proposed simple bounds are especially suited for settings in which the empirical researcher does not have evidence for the existence of defiers and expects that the number of defiers is, if any, smaller than of defiers. In this case, the simplified bounds on the distribution function $\Fcod$ are similar to the sharp bounds of Theorem~\ref{theoremdist}. 

The main reason for the rather complicated expression of the sharp bounds $\Fucod$ and $\Flcod$ is heuristically that these bounds   exploit all the information contained about the defiers included  in the function $G_d(y)$.  However, if we forgo the aim of constructing sharp bounds, we can simply look at the functions  
\begin{align*}
	\Flcod^{S}(y, \pdf,\delta)& =  \max \left\lbrace 0,  \; \frac{1}{\pco}  (\Qdd (y)- \pi_d) , \; \frac{\pi_{\Delta}}{\pco}   G_d (y), \;  G_d(y) -  \frac{\pdf}{\pi_{\Delta}} \delta \right\rbrace  ,
\end{align*}
and
\begin{align*}
	\Fucod^{S}(y, \pdf,\delta) & =  \min\left\lbrace 1, \frac{1}{\pco}  Q_{dd}(y), \; \frac{\pi_\Delta}{\pco}  (G_d(y)+\pdf), \;     G_d(y)+   \frac{\pdf}{\pi_{\Delta}} \delta \right\rbrace.	
\end{align*}
It follows from the reasoning of the main text that these functions satisfy that any value outside of these bounds is incompatible with the distribution of $(Y,D,Z)$, and our model assumptions.\footnote{This reasoning directly follows from the discussion of Section~\ref{sec_identification_bounds}.} They are therefore always valid bounds of the distribution function $\Fcod$.  

Our simplified sensitivity analysis is then based on the simplified bounds, and we further ignore the lower bound on the sensitivity parameters  $\updf$ and $\deltamin$ and set both to zero.
These bounds are still valid for both the sensitivity and robust region. Estimation and conducting inference within this sensitivity analysis is substantially more straightforward.

In an empirical setting, it remains the question under which conditions these are actually "good" bounds in the sense that they are close to the sharp bounds and do not lead to a substantial loss in information. The most important difference between the sharp and these simple bounds is that we do not exploit the information about the defiers, which is contained in the function $G_d$ for $d \in \{0,1\}$ and the information that the distribution function is not allowed to increase too much. So if $G_d$ is indeed decreasing, we expect that the sharp bounds are substantially more informative than the simple bounds. 
To assess how conservative these bounds might be, a researcher could also  test whether $G_d(y)$ is non-decreasing for all  $y \in \mathbb Y$   \citep{Kitagawa2015}. 

\subsection{Additional  Treatment Effects}\label{sec_additional_treatment_effects_appendix}
The sharp bounds on the distribution function, $\Fcod$, in a first-order stochastic dominance sense, allows us to consider various other treatment effects as well. In this section, we consider quantile treatment effects and define the $\tau$-th quantile effect of the compliers by $\Delta_{CO}(\tau)$ and we consider empirical conclusions of the form  $\Delta_{CO}(\tau)\geq \mu $.

We define the lower and upper bounds of the quantile functions by the respectively left and right inverse of the bounds of the outcome distributions
\begin{align*}
	\underline{Q}_{Y_d^{CO}}(\tau,\pdf , \delta ) &=  \inf \{ y \in \mb Y: \Fudfd(\yld)\geq \tau\}  \\
	\overline{Q}_{Y_d^{CO}}(\tau,\pdf ,\delta) &=  \sup \{y \in \mb Y:  \Fldfd(\yld)\leq \tau\}. 
\end{align*} 

The quantile treatment effect of a quantile $\tau$ is then  given by 
\begin{align*}
	&	[\underline{\Delta}_{CO}(\tau,\pdf , \delta) , 	\overline{\Delta}_{CO}(\tau,\pdf , \delta) ] \\ 
	& \hspace*{0.5cm} = 	[ \underline{Q}_{Y_1^{CO}}(\tau,\pdf , \delta)  - 	\overline{Q}_{Y_0^{CO}}(\tau,\pdf ,\delta), \overline{Q}_{Y_1^{CO}}(\tau,\pdf ,\delta)  -\underline{Q}_{Y_0^{CO}}(\tau,\pdf ,\delta) ].  
\end{align*}
It follows from the reasoning of Lemma~1 in \cite{stoye2010partial} that these bounds are indeed sharp as well, and there exist feasible candidate distribution functions of $\Fcod$, which also imply any value between these bounds.

The sensitivity region is defined independently of the particular empirical conclusion under consideration and is therefore given by the expression of the main text \eqref{equ_sensitvityregion_main}.
It follows that the breakdown point for the conclusion that $\Delta_{CO}(\tau) \geq \mu $  is given by 
$$BP_{\tau}(\pdf )=\sup \{\delta: (\pdf, \delta) \in \A \text{ and }\underline{\Delta}_{CO}(\tau,\pdf , \delta)  \geq\mu\}. $$
The breakdown frontier of the quantile  treatment effect is given by
\begin{align*}
	BF_{\tau} &  =  \{ (\pdf, \delta) \in :\delta=BP_{ \tau}(\pdf)\}.  
\end{align*}
and the robust region by
\begin{align*}
	RR_{\tau} & =\{(\pdf, \delta) \in \A: \delta \leq  BP_{\tau}(\pdf) \}. 
\end{align*}

\subsection{Additional Covariates}\label{sec_additional_covariates}
Additional covariates, which are measured prior to treatment assignment, can be used to tighten the bounds on the identified set of treatment effects of compliers and can thus lead to greater sets of the robust region; the arguments are similar to those in \cite{lee2009training}. It further holds that conditioning on pretreatment covariates can imply that the identified set of the sensitivity parameters can be reduced such that the analysis becomes more informative. 
We, therefore, assume that the covariates are discrete and given by $\mc X=\{x_1, \dots, x_K\}$, which splits the population into non-overlapping groups. We further impose the following assumptions.
\begin{assumption}\label{assumptions_late_cond}
	(i)  Conditional independence assignment: $(Y_{1}, Y_{0}, D) \perp Z | X=x$, (ii)  Conditional relevance: $\mb{P}(D=1|Z=1, X=x) > \mb{P}(D=1|Z=0, X=x)$, (iii)  Common support: $0 < \mb{P}(Z=1 | X=x) < 1$.
\end{assumption}
We denote the population size of defiers by $\pdf(x)$ given $X=x$. We consider  the sensitivity   parameter $\pdf$ such that for all $x \in \{x_1, \dots, x_K\}$
$$\pdf(x) \leq \pdf .$$
This parameterization implies that the population size of defiers is bounded from above for each value of the covariates. We note that this  parameterization implies without further assumptions conservative bounds as long as  $\pdf(x) \neq \pdf$ for some values of $x$.\footnote{ We consider two alternative parameterization: First, one could argue that $\pdf(x)= \pdf$ for all $x \in \mc X$. The implied bounds would be sharp, but this  assumption is very restrictive.  Second, we could consider a setting of $\pdf(x)= \pdf^x$. In this parameterization, however,  the parameter space might be very large and therefore difficult to interpret. The  parameterization chosen in the text is plausible and interpretable.}

By similar reasoning the heterogeneity in the outcome distribution is restricted by 
$$|F_{Y_d^{CO}|X=x}(y|X=x) -F_{Y_d^{DF}|X=x}(y|X=x)| \leq \delta_x.$$
Based on the pre-intervention covariates one can calculate for each $k$ lower and upper bounds on the population size of defiers $\updf(x_k)$ and $\opdf(x_k)$, respectively. The bounds on the sensitivity parameters can then be calculated based on the definition of the sensitivity parameters by $\updf=\min_{x\in \mc X}(\updf(x))$  and $\opdf=\max_{x \in \mc X}( \opdf(x))$.
Let $\pdf(x_k)= \min\{\opdf(x_k)  \max\{\pdf, \updf(x_k)\}\} $ and $\pdf^x = \sum_{k=1}^K \pdf(x_k)$. We  denote the lower bound by
\begin{align*}
	\Flcod^x(y, \pdf, \delta_x)& = \frac{1}{\pdf^x} \sum_{k=1}^K \mb{P}(X=x_k, \pdf(x_k) ) \; \Flcod^x(y, \pdf(x_k) , \delta_x |X=x_k)
\end{align*}
and the upper bound by 
\begin{align*}
	\Fucod^x(y, \pdf, \delta_x )& = \frac{1}{\pdf^x} \sum_{k=1}^K \mb{P}(X=x_k, \pdf(x_k)) \; \Fucod^x(y, \pdf(x_k) , \delta_x |X=x_k).
\end{align*}
\begin{proposition}\label{prop_extensions_covariates}
	Suppose that Assumption~\ref{assumptions_late_cond} holds,  and the data generating process is compatible with the sensitivity parameters $(\pdf, \delta_x)$. Then,
	for $d \in \{0,1\}$
	$$ \underline F_{Y_d^{CO}} (y, \pdf, \delta_x) \leq   F_{Y_d^{CO}} (y, \pdf, \delta_x,) \leq  \overline  F_{Y_d} (y, \pdf, \delta_x). $$
	Moreover, there exist data generating processes  which are compatible with the model assumptions such that the outcome distribution of compliers equals either $\overline{F}^x_{Y_d^{CO}}(y,\pdf,\delta_x)$, $\underline{F}^x_{Y_d^{CO}}(y,\pdf,\delta_x)$, or  any convex combination of these bounds, if $\pdf^x=\pdf$, and if  for all $x \in \mc X$ and  for $d\in\{0,1\}$, it holds that $$\sup_{y \in \mathbb Y} |F_{Y_d^{CO}|X=x}(y|X=x) -F_{Y_d^{DF}|X=x}(y|X=x)| = \delta.$$ 
\end{proposition}

The derivation of the sensitivity and robust region follows from the same arguments as in Section~\ref{sectionsensitivtyanalysis}.

\subsection{Form of Sensitivity and Robust Region for Binary Outcome Variable}\label{sec_appendix_more_details_bianry_outcome}
Since our inference procedure exploits the shape of the sensitivity and robust region for a particular empirical conclusion about a binary outcome variable, we discuss this form in this section.
We note again that they are determined by the following parameters. 
\begin{align*}
	\phi_b(\theta_b, \pdf) = (\updf, \, - \opdf, \, - \deltamax( \pdf), \, BP(\pdf)), \label{equ_mapping}
\end{align*} 
We discuss the components in turn.	 	
The sensitivity region is determined based on four parameters: the lower and upper bound on the population size of defiers and the lower and upper bound on the sensitivity parameter of outcome heterogeneity. Due to their simple form, we do not have to discuss the lower and upper bound on the population size of defiers further, neither the lower bound on the outcome heterogeneity.
However, the upper bound on the sensitivity parameter of outcome heterogeneity is given by
\begin{gather*}\overline{\delta}_b(\pdf)  = \max_{d \in \{0,1\}}  \max \{ |\underline P_d^{CO}(\pdf , 1) - \underline P_d^{DF}(\pdf , 1)| ,|\overline P_d^{CO}(\pdf , 1) - \overline P_d^{DF}(\pdf , 1)|   \}.
\end{gather*}
Following the discussion about how the bounds are constructed, e.g.,  in Appendix~\ref{sec_illustraion_bounds}, it follows  that the upper bound on outcome heterogeneity has to be decreasing in the population size of defiers as the distribution of both compliers and defiers become more similar.  We now consider the breakdown point as a function of the population size of defiers. We note that it can be rewritten as
\begin{align*}
	& BP(\pdf)= \frac{1}{\pdf} \max \{ BP_0(\pdf) , BP_1(\pdf),  BP_2(\pdf)\},
\end{align*}
where $BP_0(\pdf)$ is decreasing, $BP_1(\pdf)$ and $BP_2(\pdf)$ are potentially  increasing so that
\begin{align*}
	&	BP_0(\pdf)	  \\
	& =   \max \left\lbrace    P_{11}-P_{10} - (\mu+\frac{P_{00}-P_{01}+\pdf}{\pco} ) \pi_{\Delta}    , -( (\mu-\frac{P_{11}  - P_{10} }{\pco} ) \pi_{\Delta}  +P_{00} -  P_{01})    \right.  \\
	& \qquad \left. -  \left( \mu \cdot \pi_{\Delta} +P_{00} - P_{01} \right),   P_{11}-P_{10} - (\mu+1) \pi_{\Delta} ,     \frac{1}{2} ( P_{11} -  P_{10} - P_{00}  + P_{01}   -\mu \cdot \pi_{\Delta}),0 \right\rbrace  \\ 
	&	BP_1(\pdf)	  	=   \max \left\lbrace  0, -( (\mu-\frac{P_{11}  - \pat}{\pco} )\pi_{\Delta}   +P_{00} -  P_{01})   \right\rbrace   \\
	&	BP_2(\pdf)	  	=   \max \left\lbrace  0,  P_{11}-P_{10} - (\mu+\frac{P_{00}}{\pco}) \pi_{\Delta}   \right\rbrace.   
\end{align*}
We therefore denote by
\begin{align} \tilde \phi_b(\theta_b, \pdf) = (\updf, \, - \opdf, \, - \deltamax( \pdf), \, BP_0(\pdf), BP_1(\pdf), BP_2(\pdf)).\label{equ_binary_increasing}
\end{align}
Each component of the mapping  $\tilde \phi_b$ is either nondecreasing or nonincreasing in $\pdf$. We exploit this shape constraint to construct confidence sets for the sensitivity and robust region that are uniformly valid in $\pdf$. 

\section{Proofs of Main Results}\label{appendixproofs}
In this section, we prove the main results of this paper. 

\subsection{Proof of Theorem~\ref{theoremdist}}\label{sec_proof_theoremdist}
As we consider a fix sensitivity parameter pair $(\pdf, \delta)$, we omit the dependence of all functions on the sensitivity  parameter in this section; for instance,  we write $\Flcod(y)$ instead of $\Flcod(y, \pdf, \delta)$.

\newpage
We first consider how to determine whether a distribution function $F_{\wt Y_d^{CO}}$ is indeed a feasible candidate of $\Fcod$. We therefore have to construct a random variable  $\wt W \equiv (\wt Y_0, \wt Y_1, \wt D_0, \wt D_1, \wt Z)$, that is compatible with the model assumptions, the observed probabilities and the sensitivity parameters, and implies $F_{\wt Y_d^{CO}}$ as outcome distribution of compliers. However, based on our model assumptions, e.g. independence and the definition of the groups,  it indeed suffices to construct marginal outcome distribution functions of 
$$ F_{\wt Y_d^{CO}} , \; F_{\wt Y_d^{DF}}, \; F_{\wt Y_d^{AT}}, \; F_{\wt Y_d^{NT}}, $$
which are consistent with the observed probabilities and the sensitivity parameters. 
 Since the data are also noninformative about the distribution functions $F_{Y_0^{AT}}$ and $F_{Y_1^{NT}}$ these distributions are left unrestricted as well.  
  As outlined in the main text, any given candidate distribution function of $\Fcod$ implies functions of $\Fdfd$ and $F_{Y_d^{DT}}$ given our sensitivity parameters and the observed distributions $(Y,D,Z)$. it follows that $F_{\wt Y_d^{CO}}$  is a feasible candidate of $\Fcod$ if the  implied functions of $\Fdfd$ and $F_{Y_d^{dT}}$ are indeed distribution functions.

We argue in the main text that any feasible candidate of the distribution function $\Fcod$ has to satisfy at least that
\begin{align} \underline H_{Y_d^{CO}}(y)\leq 	 F_{Y^{CO}_d}(y) \leq \overline H_{Y_d^{CO}}(y) \label{equ_hconstraint0}
\end{align}
The proof now proceeds in two parts. In part I, we exploit which information can be obtained from the observed probabilities about the compliers outcome distribution to show which additional restriction, besides \eqref{equ_hconstraint0}, any feasible candidate of $F_{Y_d^{CO}}$ has to satisfy. In part II, we then verify that the proposed bounds $\Flcod$ and $\Fucod$   are feasible candidates of the distribution function  $\Fcod$. We show that  these bounds satisfy that  any value outside of these bounds is incompatible with the distribution of $(Y,D,Z)$, and our assumptions. We denote by $\underline y$ and $\overline y$ the respectively left and right limits of $\mb Y$, which might equal $\pm \infty$.	

Let  $\Gdsup(y)=\sup_{ \wh{y} \leq y} G_d(\wh{y})$ that is (i) a nondecreasing function  and   (ii)  satisfying $G_d(y) \leq \Gdsup(y)$ for all $y \in \mb Y$. It further holds that $\Gdsup(y)$  is such that  $\Gdsup(y) \leq \wt G_d(y)$ for all $y \in \mb Y$, where $\wt G_d$  is any  real-valued function $\wt G_d$ satisfying  conditions (i) and (ii).
Similarly, let $\Gdinf(y)=\inf_{ \wh{y} \geq y} G_d(\wh{y})$  that is (i') a nondecreasing function  and   (ii')  satisfying $G_d(y) \geq \Gdinf(y)$ for all $y \in \mb Y$. It further holds that $\Gdinf(y)$  is such that  $\Gdinf(y) \geq \wt G_d(y)$ for all $y \in \mb Y$, where $\wt G_d$  is any  real-valued function $\wt G_d$ satisfying  conditions (i') and (ii').

\subsubsection*{Part I}
Using \eqref{equ_otheroutcomes_compliers_always_takers} and that distribution functions are nondecreasing, any feasible candidate of 	$F_{Y_d^{CO}}$ has to satisfy that, for any $y, y' \in \mathbb Y$ and $y' \leq y$,
\begin{align}	  F_{Y^{CO}_d}(y) - F_{Y^{CO}_d}(y') \leq \frac{Q_{dd}(y)- Q_{dd}(y')}{\pco}. \label{equ_monotonicityat}
\end{align}
Using the same reasoning and \eqref{equ_otheroutcomes_compliers_defiers}, it follows that
\begin{equation*}		
	F_{Y^{CO}_d}(y) - F_{Y^{CO}_d}(y') \geq \frac{\pi_{\Delta}}{\pco} \left( G_d(y)- G_d(y') \right).
\end{equation*}
for any arbitrary $y$ and $y'$. As $G_d(y)$ is not necessarily nondecreasing,  we can similarly conclude that it has to hold that
$$ \mb P(Y_d^{CO} \in B) \geq \frac{\pi_{\Delta}}{\pco} \left(  \mb P(Y \in B, D=d|Z=d )  -  \mb P(Y \in B, D=d|Z=1-d ) \right) $$
for any $B \in \mathscr B$ 
and therefore 
\begin{align}
	F_{Y^{CO}_d}(y) - F_{Y^{CO}_d}(y') \geq \frac{\pi_{\Delta}}{\pco} \left( G_d^+(y)- G_d^+(y') \right). \label{equ_monotonicitydefiers}
\end{align}
Any feasible candidate of 	$F_{Y_d^{CO}}$ has to further satisfy the conditions 
\begin{align}
	(viii) \; \lim_{y \rightarrow \underline y} F_{Y_d^{CO}}(y)=0 \quad  \text{  and } \quad \lim_{y \rightarrow \overline y} F_{Y_d^{CO}}(y)=1.\label{equ_limits}
\end{align}
The distribution functions $ \Fdfd$, and $F_{Y_d^{dT}}$ fulfill then these  limit conditions based on \eqref{equ_otheroutcomes_compliers_always_takers}~and~\eqref{equ_otheroutcomes_compliers_defiers}, as it holds that $\lim_{y \rightarrow  \underline y} G_d(y)= \lim_{y \rightarrow  \underline y} Q_{ds}(y)=0 $   and   $\lim_{y \rightarrow \overline y} G_d(y)= 1$ and $\lim_{y \rightarrow  \overline y} Q_{dd}(y)= \pi_d +\pco $ and   $\lim_{y \rightarrow  \overline y} Q_{d(1-d)}(y)= \pi_d +\pdf $ for any $d,s \in \{0,1\}$. 

Any real-valued function, which is defined on $\mb Y$ and right-continuous,  which left-limits exists and which  satisfy equations \eqref{equ_hconstraint0}~--~\eqref{equ_limits} implies by construction potential outcome distributions for all four groups, which are consistent with the imposed model  assumption,  the sensitivity parameter constraints, and  the observed probability functions. It is thus a feasible candidate of $\Fcod$.   It is clear that the simple additive structure of all imposed conditions implies that if there are two different such feasible candidate functions, any convex combinations of these functions satisfy these conditions as well. 

\subsubsection*{Part II}
We show in the following both that the proposed bounds $\Flcod$ and $\Fucod$ satisfy the constraints in \eqref{equ_hconstraint0}~--~\eqref{equ_limits} and  that any function which takes values outside of these bounds contradicts one of these conditions and is therefore incompatible with the distribution of $(Y,D,Z)$, our assumption and the sensitivity parameters. 
As the considered sensitivity parameters  lie within the sensitivity region by assumption, bounds on the outcome distribution of compliers exist and are therefore non-intersecting by construction.
The condition in \eqref{equ_hconstraint0} is therefore satisfied if our bounds $\Flcod$ and $\Fucod$ satisfy
\begin{align}
	\underline H_{Y_d^{CO}}(y) \leq \Flcod(y) \qquad  \text{ and }\qquad  \Fucod(y) \leq \overline H_{Y_d^{CO}}(y) \label{equ_hconstraint} 
\end{align}
for all $y \in \mathbb Y$.
Additionally, both bounds  preserve the existence of limits and continuity.

\subsubsection*{Part II - Lower Bound}
We consider first
\begin{align}
	\underline {H}_1(y)= \frac{1}{\pco} \left( \pi_{\Delta} G_d^+(y)  -  \inf_{\wt y \leq y  } \left( \pi_{\Delta} G_d^+(\wt y) -   \pco \underline{H}(\wt y) \right) \right).\label{equ_h1}
\end{align}  
It clearly holds  that $\underline {H}_1(y)  \geq   \underline{H}_{Y_d^{CO}}(y) $.
Consider again any $y, y' \in \mb Y$ such that $y' \leq y$.
Based on this reasoning $\underline {H}_1(y) $ satisfies constraint \eqref{equ_monotonicitydefiers} as 
\begin{align*}
	&  	\underline {H}_1(y) - 	\underline {H}_1(y')\\
	&   = \frac{1}{\pco}  \left( \pi_{\Delta} G_d^+(y)-\pi_{\Delta} G_d^+(y')   -  \inf_{\wt y \leq y  } \left( \pi_{\Delta} G_d^+(\wt y) -  	\underline{H}(\wt y)  \right)  -\inf_{\wt y \leq y'  } \left( \pi_{\Delta} G_d^+(\wt y') -  \underline{H}_{Y_d^{CO}}(\wt y')   \right) \right)\\
	& \geq \frac{1}{\pco}  \left( \pi_{\Delta} G_d^+(y)-\pi_{\Delta} G_d^+(y')\right).
\end{align*} 
Any function $F$ such that   $F(y) \leq \underline H_1(y)$ either violates \eqref{equ_monotonicitydefiers} or   \eqref{equ_hconstraint} for some $y \in \mathbb Y $. We conclude that any feasible candidate function of $\Fcod$ has to satisfy 
\begin{align} \underline H_{1}(y)	 \leq   F_{Y^{CO}_d}(y)  \label{equ_h1constraintsmallerequal}
\end{align}
We now consider our final lower bound  
$$\Flcod(y)=\frac{1}{\pco} \left( Q_{dd}(y) -\inf_{ \wt y \geq y } \left(   Q_{dd}( \wt{y}) -  \underline {H}_1(\wt y) \right)\right). $$
It is clear that $\Flcod(y) \geq \underline H_1(y)$ and that $Q_{dd}(y)-Q_{dd}(y') \geq \Flcod(y)  -\Flcod(y')  $. 
As it further  holds that, for any $y, y' \in \mathbb Y$ and $y' \leq y$, 
$$ Q_{dd}(y) - Q_{dd}(y') \geq  G_d^+(y) - G_d^+(y')$$
$\Flcod(y)$  satiates \eqref{equ_monotonicitydefiers},  \eqref{equ_monotonicityat}, and it holds that $\Flcod(y) \geq \underline {H}_1(y)$. Clearly, any function  $F$ such that $F(y) \leq \underline H_1(y)$ is incompatible with the distribution of $(Y,D,Z)$, the sensitivity parameters and our assumptions. 

We now show that   $\Flcod(y)$ satisfies  \eqref{equ_limits} .
By construction,  $\underline{F}_{Y_d^{CO}}\in [0,1]$. We therefore show that  $\lim_{y \rightarrow  \underline y} \underline{F}_{Y_d^{CO}}(y) \leq0$ and  $\lim_{y \rightarrow  \overline y} \underline{F}_{Y_d^{CO}}(y) \geq1$.  It holds that
\begin{align*}
	\lim_{y\rightarrow  \underline y} 	&  \underline{F}_{Y_d^{CO}}(y) = \frac{1}{\pco} \inf_{ \wt{y} \in  \mb{R} } \left(  Q_{dd}( \wt{y})-( \pi_{\Delta} G_d^+( \wt{y})  - \inf_{\wh y \leq \wt y} \left( \pi_{\Delta}  G_d^+(\wh{y}) +\pco  \underline{H}_{Y_d^{CO}}(\wh{y})   \right) ) \right) 
\end{align*}
The  equality follows as $\lim_{y\rightarrow \underline y} Q_{dd} (y)=0$. We note that for all $y,y'\in \mb Y$ and $y'\leq y$
$Q_{dd}(y)-Q_{dd}(y') \geq \pi_{\Delta}  \left( G_d^+(y)-G_d^+(y') \right).$
It follows that 
\begin{align*}
	&	\lim_{y\rightarrow  \underline y} \underline{F}_{Y_d^{CO}}(y)\\ 
	&	\leq \frac{1}{ \pco} \inf_{\wh y \in \mb Y} \big(    \max\{\underbrace{0}_{(1a)},  \underbrace{Q_{dd}(\wh{y})-\pi_{d}}_{(2a)},  \underbrace{\pi_{\Delta} G_d^{\sup}(\wh{y})}_{(3a)},
	\underbrace{\pco G_d^{\sup}(\wh{y})- \pco \frac{\pdf}{\pi_{\Delta}}\delta }_{(4a)}\}-Q_{dd}(\wh{y}) \big).
\end{align*}
We now show that each of the expressions (1a)--(4a) evaluated at any $\wh y \in \mb Y$ is bounded by $Q_{dd}(\wh{y})$ so that it holds that
$	\lim_{y\rightarrow  \underline y} \pco \underline{F}_{Y_d^{CO}}(y) \leq 0.$  
It is obvious that  expressions $(1a)$ and $(2a)$ satisfy this reasoning. Considering $(3a)$, we note that it holds that  
$\pi_{\Delta} G_d^{\sup}(\wh y) \leq Q_{dd}(\wh y).  $
We turn to (4a). It holds that
\begin{align*}
	G_d^{\sup}(\wh{y})- \frac{\pdf}{\pi_{\Delta}}  \delta-\frac{Q_{dd}(\wh{y})}{\pco} 	&  \leq F_{Y^{CO}_{d}}(\wh y) + \frac{\pdf}{\pi_{\Delta}}  \delta- \frac{\pdf}{\pi_{\Delta}} \delta  -\frac{Q_{dd}(\wh{y})}{\pco}   \leq F_{Y^{CO}_{d}}(\wh y)-\frac{Q_{dd}(\wh{y})}{\pco}  \leq 0 
\end{align*}
We consider the right limit. It holds that $\underline{F}_{Y_d^{CO}}
\geq\underline{H}_{Y_d^{CO}}$ and therefore
\begin{align*}
	\lim_{y\rightarrow \overline y} \underline{F}_{Y_d^{CO}}(y)
	& \geq \lim_{y\rightarrow \overline y} \max\{0, G_d^{\sup}(y)-\frac{\pdf}{\pi_{\Delta}} \delta, \frac{\pi_{\Delta}}{\pco} G_d^{\sup}(y)  , \frac{Q_{dd}(y)- \pi_{d}}{\pco}\} \geq 1. 
\end{align*}
The  second inequality follows as $\lim_{y\rightarrow \overline y}  Q_{dd}(y)- \pi_{d} =  \pco + \pi_d - \pi_{d}= \pco.$
This reasoning concludes the  proof of the lower bound.

\subsubsection*{Part II - Upper Bound}
A similar reasoning applies to the upper bound. To briefly sketch this reasoning, let 
$$ \overline H_1(y)=   Q_{dd}(y)-  \sup_{\wh{y} \leq y} \left(  Q_{dd}(\wh{y})- \pco \overline{H}_{Y_d^{CO}}(\wh y)   \right) . $$

It is clear that $\overline H_1(y) \geq \overline H_{Y_d^{CO}} (y)$ and that $Q_{dd}(y)-Q_{dd}(y') \geq \overline H_1(y) -\overline H_1(y')  $. 
It holds that $\overline H_1(y')$ satisfies   \eqref{equ_monotonicityat}. Clearly, any function  $F$ such that $F(y) \leq \underline H_1(y)$ is incompatible with the distribution of $(Y,D,Z)$, the sensitivity parameters and our assumptions. It therefore follows that any function which is a feasible candidate of the distribution function $\Fcod$ has to satisfy
\begin{align}
	\Fcod(y) \leq \overline H_1(y)\label{equ_h1_upperbound}
\end{align}
We now consider our proposed bound.  $$ \Fucod(y)  =   \pi_{\Delta}  G_d^+(y)  - \sup_{\wt{y}\geq y} \left(   \pi_{\Delta}  G_d^+(\wt{y}) -   \overline H_1(y) \right). $$
It follows from the same reasoning as above that $\Fucod(y)$ satisfies \eqref{equ_h1_upperbound},  \eqref{equ_monotonicityat} 
and  \eqref{equ_monotonicitydefiers}. 
Clearly, any function  $F$ such that $F(y) \geq \Fucod(y)$ is incompatible with the distribution of $(Y,D,Z)$, the sensitivity parameters and our assumptions. 

We conclude by showing that $\Fucod(y)$  satisfies \eqref{equ_limits}.
It holds that
\begin{align*}
	\lim_{y\rightarrow \underline y} \overline{F}_{Y_d^{CO}}(y) \leq \lim_{y\rightarrow \underline y} \min\{1, G^{\inf}_d(y)+ \frac{\pdf}{\pi_{\Delta}} \delta,   \frac{\pco}{\pi_{\Delta}} G^{\inf}_d(y)+ \frac{\pco}{\pdf},  \frac{Q_{dd}(y)}{\pco}\} \leq 0,
\end{align*}
where the second inequality  follows by  $	\lim_{y\rightarrow  \underline y} \frac{Q_{dd}(y)}{\pco}=0$. We now consider
\begin{align*}
	&	\lim_{y\rightarrow \overline y}  \overline{F}_{Y_d^{CO}}(y)  \\
	& = 	\frac{\pco+\pi_d}{\pco}   - \sup_{\wh{y}\in \mb{Y}} \big(\frac{Q_{dd}(\wh{y})}{\pco}- \min\{\underbrace{1}_{(1b)}, \underbrace{\frac{Q_{dd}(\wh{y})}{\pco}}_{(2b)},
	\underbrace{\frac{\pi_{\Delta}}{\pco} G^{\inf}_d(\wh{y})+ \frac{\pdf}{\pco}}_{(3b)},
	\underbrace{ G^{\inf}_d(\wh{y})+ \frac{\pdf}{\pi_{\Delta}} \delta}_{(4b)}\}  \big).  
\end{align*}
We show that  (1b)--(4b) are bounded from below by $\frac{Q_{dd}(\wh{y})}{\pco}-\frac{\pi_d}{\pco}$ such that
\begin{align*}
	\lim_{y\rightarrow\overline y} \overline{F}_{Y_d^{CO}}(y) \geq \frac{\pco+\pi_d}{\pco}  - \frac{\pi_d}{\pco} = 1.
\end{align*}
It is clear that (1b)-(2b) satisfies this restriction. Concerning (3b), we note that
\begin{align*}
	\frac{1}{\pco} \left(   \pi_{\Delta} G^{\inf}_d(\wh{y})+  \pdf \right)  &\geq  \frac{1}{\pco} (  Q_{dd}(\wh{y}) + \pdf - \pd -  \pdf)
	= \frac{Q_{dd}(\wh{y})-\pd}{\pco} .
\end{align*}
Concerning (4b), we note that
\begin{align*}
	G^{\inf}_d(\wh{y})+ \frac{\pdf}{\pi_{\Delta}} \delta \geq F_{Y_d^{CO}}(\wh y) - \frac{\pdf}{\pi_{\Delta}}  \delta   + \frac{\pdf}{\pi_{\Delta}} \delta   \geq  \frac{Q_{dd}(\wh{y})}{\pco}-\frac{\pi_d}{\pco}
\end{align*}
This  completes this proof. \qed

\subsection{Proof of Proposition~\ref{theoremlambda}}
We show that the population size of compliers is sharply bounded by $	\underline{\pi}_{CO}  \leq \pco \leq	\overline{\pi}_{CO} $, where for  $ B \in \mathscr{B}$  and $d,z \in \{0,1\}$
\begin{align*}
	\overline{\pi}_{CO} = & \min\{\mb{P} (D=1|Z=1), \mb{P} (D=0|Z=0) \} \\
	\underline{\pi}_{CO}= &  \max_{d \in \{ 0,1 \} } \{ \sup_{B \in \mathscr{B}} \{\mb{P}(Y \in B, D=d |Z=d) - \mb{P}(Y  \in B, D=d |Z=1-d) \} \}.
\end{align*}
The proposition follows from this statement as $\pi_{DF}=\pi_{CO}-\mb{P}(D=1|Z=1) + \mb{P}(D=1|Z=0)$.	Let $\mb{P}(Y_d^t \in B )$ denotes the unobserved  probability  distribution of the  potential outcome of group $t$ with treatment status $d$.\footnote{In principle, Proposition~\ref{theoremlambda} is a Corollary of Theorem~\ref{theoremdist}. Considering the sharp lower bound on the population size of defiers, one could simply use the bounds to solve  for the minimal size of defiers for which there exists one value of outcome heterogeneity $\delta$ such that the bounds are non-intersecting. However, this exercise is tedious, and we propose a simpler and direct proof for this claim in this section.} 

$\underline{\pi}_{CO}$ is a valid lower bound of the population size of compliers as it follows from the definition of groups  that
\begin{align*}
	\overline{\pi}_{CO}= \min \{ \pi_{AT} + \pi_{CO}, \pi_{NT}+ \pi_{CO}  \} \geq \pi_{CO}.
\end{align*}
Similarly, $\underline{\pi}_{CO}$ bounds the population size of compliers from below as  $\underline{\pi}_{CO}$ equals	
\begin{align*}
	\max  & \left\lbrace  \sup_{B \in \mathscr{B}}  \left\lbrace  \mb{P}(Y_1 \in B, AT) + \mb{P}(Y_1 \in B, CO)  - \mb{P}(Y_1 \in B, AT)-\mb{P}(Y_1 \in B, DF)\right\rbrace, \right. \\
	&\quad  \left.  \sup_{B \in \mathscr{B}} \left\lbrace  \mb{P}(Y_0 \in B, NT)+ \mb{P}(Y_0 \in B, CO)   - \mb{P}(Y_0 \in B, NT) - \mb{P}(Y_0 \in B, DF)\right\rbrace \right\rbrace \\
	&	\leq   \max \{ \sup_{B \in \mathscr{B}} \{ \mb{P}(Y_1 \in B, CO)\}, \sup_{B \in \mathscr{B}} \{ \mb{P}(Y_0 \in B, CO) \} \} = \pi_{CO}.  
\end{align*}
The  inequality follows from the independence assumption and the definition of the groups.
It is therefore clear that the population size of compliers  lies within the bounds. It remains to show  that these bounds are sharp. To show this, we consider any fix $\wt{\pi}_{CO} \in [\underline{\pi}_{CO}, \overline{\pi}_{CO}]$.  Let $B \in \mathscr{B}$ and $\mathscr{B}_B=\{ A \cap B | A \in \mathscr{B} \}$.  Using the discussion of the proof of Theorem~\ref{theoremdist} about how to verify that a candidate distribution is a feasible distribution,  we consider the following marginal outcome distributions of the groups.\footnote{Otherwise $\mb{P}(\wt{Y}_d \in B,  T= dT)$ is defined to be zero if $\mb{P}(D=d|Z=d)= \sup_{C \in \mathscr{B}} (\mb{P}(Y \in C, D=d|Z=d)-\mb{P}(Y \in C, D=d|Z=1-d))$. The other probability distributions stay the same.} 
\begin{align*}
	\mb{P}(\wt{Y}_d \in B, T= CO) & = \mb{P}(Y \in B,D=d|Z=d) - \mb{P}(\wt{Y}_d \in B, T= dT),  \\
	\mb{P}(\wt{Y}_d \in B,  T= DF)  &=\mb{P}(Y \in B,D=d|Z=1-d) - \mb{P}(\wt{Y}_d \in B, T=\;  dT), \\
	\mb{P}(\wt{Y}_d \in B,  T= dT) &= L_1 \cdot L_2, 
\end{align*}
where 
\begin{align*} 
	L_1& = 	  \frac{\mb{P}(D=d|Z=d)-\wt{\pi}_{CO} } {\mb{P}(D=d|Z=d)- \sup_{C \in \mathscr{B}} (\mb{P}(Y \in C, D=d|Z=d)-\mb{P}(Y \in C, D=d|Z=1-d))},  \\
	L_2	& =   \mb{P}(Y \in B, D=d|Z=d) \\
	& \qquad  - \sup_{C  \in \mathscr{B}_B} (\mb{P}(Y \in C, D=d|Z=d)-\mb{P}(Y \in C, D=d|Z=1-d)) .
\end{align*}
The outcome distribution of group $dT$ is the product of two terms.  The  term $L_1$ guarantees that the probability distributions  integrate to the corresponding population size. The term  $L_2$ guarantees that the outcome probabilities of  compliers and defiers are  nonnegative. The outcome distributions of the other groups are respectively defined.

By construction, the proposed outcome probability distributions imply the observed outcome probability distributions.\footnote{This means that  $ \forall  B \in \mathscr{B}$ and $\forall d,z \in \{0,1\}$  $\mb{P}(Y \in B, D=d| Z=z) = \mb{P}(\wt{Y} \in B, \wt{D}=d| \wt{Z}=z)$.} We show now that the implied probability distributions are indeed  distributions, which  satisfy $\forall \; T \in \{CO, DF, AT, NT\}$, $d\in \{0,1\}$, and  $B, B' \in \mathscr{B}$, where  $B' \subseteq B $: (i) $\mb{P}(\wt{Y}_d^T \in \mb{Y} )=1$; (ii) $\mb{P}(\wt{Y}_d^T \in B )\geq 0$;    and (iii)    $\mb{P}(\wt{Y}_d^T \in B )\geq\mb{P}(\wt{Y}_d^T \in B' )  $. We consider  any $B \in \mathcal B $ and any $d \in \{0,1\}$ in the following. 

We first consider condition (i). It  clearly holds that $\mb{P}( \wt{Y}_d \in \mb{Y},T=CO)  =\wt{\pi}_{CO}$, and 
\begin{align*}
	&&\mb{P} (\wt{Y}_d \in \mb{Y}, T= dT) &= \mb{P}(D=d|Z=d)-\wt{\pi}_{CO}=
	\wt{\pi}_{dT}\\
	&&	\mb{P}( \wt{Y}_d \in \mb{Y},T=DF) & = \mb{P}(D=d|Z=1-d) - \mb{P}(D=d|Z=d)+\wt{\pi}_{CO}=  \wt{\pi}_{DF}.
\end{align*}

We turn to condition (ii).
First note that it follows from  the bounds  on the population size of compliers $\underline{\pi}_{CO}$ that $0 \leq L_1 \leq 1 $. Second, we note that
\begin{gather*}
	L_2 	\geq  \mb{P}(Y \in B, D=d|Z=d) - \sup_{C \in \mathscr{B}_B} (\mb{P}(Y \in C, D=d|Z=d))  =0.
\end{gather*}
This reasoning implies that  $\mb{P}(\wt{Y}_d \in B,  T= dT)\geq 0 $. Further, note that
\begin{align*}
	&\mb{P}(\wt{Y}_d \in B, T= DF)  = \mb{P}(Y \in B ,D=d|Z=1-d) - \mb{P}(\wt{Y}_d \in B, T= dT) \\
	& = \mb{P}(Y \in B ,D=d|Z=1-d)  - L_1 \mb{P}(Y \in B, D=d|Z=d)  \\
	& 	\qquad + L_1  \sup_{C \in \mathscr{B}_B} (\mb{P}(Y \in C, D=d|Z=d)-\mb{P}(Y \in C, D=d|Z=1-d)) \geq 0, 
\end{align*}
by basic arguments about sets. A similar reasoning applies to the  compliers.

We  consider condition (iii).  Let $B' \subseteq B$.  We note that
\begin{align*}
	&	\mb{P}(\wt{Y}_d \in B,  T= dT) - \mb{P}(\wt{Y}_d \in B',  T= dT) \\
	& \geq   \mb{P}(Y \in B\backslash B', D=d|Z=d)   \\
	&\qquad - \sup_{C  \in \mathscr{B}_{B\backslash B'}} (\mb{P}(Y \in C, D=d|Z=d)-\mb{P}(Y \in C, D=d|Z=1-d))
\geq 0 . 
\end{align*}
Using a simple arguments, it further holds that
$\mathbb{P}(\tilde{Y}_d \in B, T= DF) \geq \mathbb{P}(\tilde{Y}_d \in B', T= DF)$ as
\begin{align*} 
	&		\mathbb{P}(\tilde{Y}_d \in B, T= dT)- \mathbb{P}(\tilde{Y}_d \in B, T= dT) \\
	&
	=  \mathbb{P}(Y \in B, D=d|Z=d)-  \mathbb{P}(Y \in B', D=d|Z=d) \\
	&\qquad   - \sup_{C  \in \mathscr{B}_{B\backslash B'}} (\mathbb{P}(Y \in C, D=d|Z=d)-\mathbb{P}(Y \in C, D=d|Z=1-d))\\
	& \leq  \mathbb{P}(Y \in B, D=d|Z=1-d)-  \mathbb{P}(Y \in B', D=d|Z=1-d).
\end{align*} 
A similar reasoning applies to the compliers, which completes this proof.\qed

\subsection{Proof of Proposition~\ref{prop_extensions_average}}\label{appendixproofextensions_average}
It holds by our assumptions that, for $d \in \{0,1\}$,
$$F_{Y_d}(y)= \pi_{1-d} F_{Y_d^{(1-d)T}} (y) + \pco F_{Y_d^{CO}}(y)  + \pdf F_{Y_d^{DF}} (y)  + \pi_d F_{Y_d^{dT}}(y) ,$$
where  $\pi_{1-d}$ is the population size of always takers if $d=0$ and otherwise of never takers and $F_{Y_d^{(1-d)T}}$ respectively. 
In the absence of treatment, the data generating process does not reveal anything about the distribution of the always takers, and neither in the presence of treatment of the never takers. The proof of Theorem~\ref{theoremdist} implies sharp bounds on the remaining six potential outcome distributions. 
Using \eqref{equ_otheroutcomes_compliers_always_takers}~and~\eqref{equ_otheroutcomes_compliers_defiers}, it follows that
\begin{align*}
	& 	F_{Y_d}(y)  \\
	& = \pi_{1-d} F_{Y_d^{(1-d)T}}(y)  + \pco F_{Y_d^{CO}}(y)  - \pi_{\Delta} G_d(y) + \pco F_{Y_d^{CO}} (y)  + Q_{dd}(y)- \pco F_{Y_d^{CO}}(y) . \\
	& = \pi_{1-d} F_{Y_d^{(1-d)T}}(y)  + \pco F_{Y_d^{CO}}(y)  - \pi_{\Delta} G_d(y) +  Q_{dd}(y) \\
	& = \pi_{1-d} F_{Y_d^{(1-d)T}}(y)  + \pco F_{Y_d^{CO}} (y)  + Q_{d(1-d)}(y)
\end{align*}
Sharp bounds in a first-order stochastic dominance sense of $F_{Y_d}(y)$ are therefore obtained by Theorem~\ref{theoremdist} by taking the distribution functions $\Flcod$ and $\Fucod$ and setting $F_{Y_d^{(1-d)T}}(y) $ to its most extreme values, respectively.   
The statement follows from this reasoning.  
\qed

\subsection{Proof of Corollary~\ref{coro_treatment_effect}}\label{appendixprooftreatment_effect}
The statement follows directly from first-order stochastic dominance of the distribution functions $\Flcod$ and $\Fucod$ by Theorem~\ref{theoremdist} and Lemma 1 in  \cite{stoye2010partial}.
\qed

\subsection{Proof of Corollary~\ref{extensions_corollarybinary}}\label{appendixproofextensions_corollarybinary}
The statement directly follows from Theorem~\ref{theoremdist} by noting how the bounds simplify for a binary variable.  \qed 

\subsection{Proof of Proposition~\ref{prop_extensions_covariates}}\label{appendixproofextensions_covariates}
By the same arguments of the proof of Theorem~\ref{theoremdist}, one can show that these bounds are sharp conditionally on the covariates given the respective assumptions. 
\qed

\subsection{Verification of Expressions used throughout the Paper}
In this section,  we verify a few simple expressions that we use through the text for completeness.

\subsubsection{Verification of Properties of the function $G_d^+(y)$}
\begin{lemma}\label{lemma_gdplus}
	Suppose $Q_{ds}(y)$ is continuously differentiable in $y \in \mathbb Y$ for $d,s \in \{0,1\}$. Then, 
	\begin{equation}
		G_d^+(y)=   \int_{\mathbb Y}  \1{z \leq y} \max \{0, g_d(z)\}dz. \label{equ_g_d_plus}
	\end{equation}
\end{lemma}

\noindent\textit{Proof:}
We note that
\begin{align*}
	G_d^+(y)	&  = \frac{1}{\pi_{\Delta}}
	\sup_{B \in \mathscr{B}} \{ \mb{P}(Y \in B, Y\leq y,  D=d|Z=d) - \mb{P}(Y \in B,  Y\leq y,  D=d |Z=1-d) \} \\
	& = \frac{1}{\pi_{\Delta}} 
	\sup_{B \in \mathscr{B}} \{ \int_{ \mathbb Y } \1{z \in B  } \1{z \leq y}  q_{dd}(z)dz  - \int_{\mathbb Y } \1{z \in B  } \1{z \leq y}   q_{d(1-d)}(z)dz   \} \\
	& =  \int_{\mathbb Y}  \1{z \leq y} \max \{0, g_d(z)\}dz.
\end{align*}
The first inequality follows from the definition of probabilities and our definition of $q_{ds}(z)$. The second equality follows  by continuity of  $q_{d(1-s)}$. \qed

\subsubsection{Outer and Inner Set for Sensitivity and Robust Region}\label{sec_preliminariesforconfidencesets}
We first verify that our expression \eqref{equ_goal_of_inference} follows from expression \eqref{equ_goal_of_indeference_2}.	
Let it holds that $\phi_L(\pdf; \theta,) \leq \phi(\pdf; \theta)$ for each component and for all $\pdf \in [0,0.5)$. 
We denote  the $l-th$ unit vector by $e_l$. We then note that by the definition of $\phi_L(\pdf; \theta)$ and $SR$ of Section~\ref{sectionsensitivtyanalysis} that
\begin{align*}
	SR& = \left\lbrace (\pdf, \delta): 	e_1^\top  \phi(\theta, \pdf) \leq \pdf \leq  	- e_2^\top \phi(\theta, \pdf) \right. \\
	& \hspace*{2.5cm} \left.  	e_3^\top  \phi(\theta, \pdf) \leq \delta \leq  	- e_4^\top \phi(\theta, \pdf) \right\rbrace  \\
	& \subseteq  \left\lbrace (\pdf, \delta): 	e_1^\top  \phi_L(\theta, \pdf) \leq \pdf \leq  	- e_2^\top \phi_L(\theta, \pdf) \right.\\
	& \hspace*{2.5cm} \left.  	e_3^\top  \phi_L(\theta, \pdf)  \leq \delta \leq  	- e_4^\top \phi_L(\theta, \pdf) \right\rbrace = SR_L.
\end{align*}
By a similar argument we note that
\begin{align*}
	RR_{L}(\A_L) &   = \{(\pdf, \delta) \in \A_L: \delta \leq   e_5^\top \phi_L(\theta, \pdf)  \}\\
	&\supseteq  \{(\pdf, \delta) \in \A_L: \delta \leq   e_5^\top \phi(\theta, \pdf)  \} \phantom{H} = RR(\A_L).
\end{align*}	
As we have shown above that $SR \subseteq SR_L$ it follows that $RR_{L}(\A) \supseteq  RR(\A)$. \qed

\section{Additional Materials for  Estimation and Inference}\label{appendixestimationandinference}
In this section, we present more details on the estimation and inference methods proposed in the main text. We first consider a binary and then a continuous outcome variable. For both cases, we provide more details about estimating  the sensitivity and robust regions, we discuss the imposed assumptions and then proceed by showing asymptotic results.
Many of the following results are based on applications and ideas from other papers and we therefore only sketch most of them. 

\subsection{Estimation for a binary outcome variable}\label{appendixestimation}
We consider a binary outcome variable, where the mapping of interest is  $\phi_b(\theta_b, \pdf)$. As shown in Section~\ref{sec:binaryoutcomemodel}, the underlying parameters $\theta_b$ are given by $(P_{11},P_{10},P_{01}, P_{00}, P_0, P_1)$. We estimate the probabilities by their sample counterparts, i.e.
$P_{ds}= \frac{1}{n_s} \sum_{i=1}^{n_s} \1{Y^s_i=1, D^s_i=d}$ and   $P_{s}= \frac{1}{n_s} \sum_{i=1}^{n_s} \1{D^s_i=1}$. We then estimate $\phi_b(\wh \theta_b, \pdf)$ by simple plug-in estimates, where the precise formulas are given in Section~\ref{sec:binaryoutcomemodel} and Appendix~\ref{sec_appendix_more_details_bianry_outcome}.	

\subsection{Assumptions}\label{sec_appendix_assumption}
We consider the following sampling process. 
\begin{assumption} \label{assumption_inference_sampling_distribution}
	For $z\in \{0,1\}$,  $\{(Y^z_{i},D^z_{i})\}_{i=1}^{n_z}$ are identically and independently distributed according to the distribution of $(Y^z,D^z)$ which is drawn conditional on $Z=z$ with support $\mb{Y}\times \{0,1\}$. It holds that $n_0/n$ converges to a nonzero constant as $n \to \infty$.
\end{assumption}	
By Assumption~\ref{assumptionLATE}, the instrument is independent of all potential outcomes, so that  the distribution of the instrument does not contain any further information and we can assume that the sampling is conditionally on the instrument \citep[see, e.g.,][]{Kitagawa2015}.

\subsection{Inference for a binary outcome variable}\label{sec:appendix:inference_binary}
In this section, we present more details on how to construct confidence sets for a binary outcome variable.	Based on the derivation in Section~\ref{sec:binaryoutcomemodel} and in Appendix~\ref{sec_appendix_more_details_bianry_outcome}, it suffices to construct a lower confidence band for $\tilde \phi_b(\hat \theta_b, \pdf)$ given in \eqref{equ_binary_increasing}. To unify the notation, let us denote the $i$-th component of this mapping by $\phi_{b,i}(\theta_b, \pdf)$ for $i \in \{1,\dotsm, 6\}$. 
We note that each of these components can be written as
$$ \phi_{b,i}(\theta_b, \pdf)= \max\{ \psi_{i,j}(\theta_b, \pdf) \}_{j=1}^{J(i)},$$
where $\psi_{i,j}(\theta_b, \pdf)$ are Hadamard-differentiable functions of $(\theta_b, \pdf)$ by the relevance assumption. The mappings $ \phi_{b,i}(\theta_b, \pdf)$ are not Hadamard-differentiable on $(\theta_b, \pdf)$, but they are Hadamard-directionally-differentiable in the direction of $\theta_b$ when evaluated at any finite set of  $\{\pdf^k\}_{k=1}^K $ , where $\pdf^k\in [0, 0.5]$ and $K$ is some finite number. 

Following ideas of \cite{fansantos2016} and \cite{masten2020}, we consider a bootstrap method to construct confidence sets $ \tilde{\phi}_{b,i}(\theta_b, \pdf^k)$ which are uniformly valid across $k$ and $i$. Specifically, the directional derivative of $\tilde{\phi}_{b,i}(\theta_b, \pdf)$ in the direction of $\theta_b$ evaluated at some $\pdf$ is given by  
$$\wt{\phi}'_{i,b, \theta_b}( h, \pdf)= 
\underset{j: \psi_{1,j}(\theta_b, \pdf)\geq  \max_{s \leq J(i)} \{ \psi_{1,s}(\theta_b, \pdf) \}  }{\max}  h_{j},$$
for all $h \in \mathbb R^{J(i)}$.\footnote{See Definition 2.1 in \cite{fansantos2016} for a definition of  Hadamard-directional differentiable mappings. }
Following \cite{fansantos2016}, we consider 	as  an estimator of  this directional derivative,
$$\wh{\wt{\phi}}'_{i,b, \theta_b}(h,\pdf)= 
\underset{j: \psi_{1,j}(\theta_b,  \pdf)\geq  \max_{s \leq J(i)} \{ \psi_{1,s}(\theta_b, \pdf) + \kappa\}  }{\max}  h_{j},$$
where $\kappa>0$ and $\kappa \to 0$ and $\kappa \sqrt{n} \to \infty$   as $n\to \infty$. 

We first get estimates of $\theta_b$ and $\phi_b(\theta_b, \pdf^k)$ from the original sample for all $k \in \{1, \dots, K\}$. We then generate  B bootstrap samples $\{ (Y_{i}^{b,z}, D_{i}^{b,z})\}_{i=1}^{n_z}$, $b=1, \dots, B$ by drawing $n_z$ observations with replacements from the original data $\{Y_{i}^z, D_{i}^z\}_{i=1}^{n_z}$ for $z \in \{0,1\}$ and we calculate $ \wh \phi{'}_{b, \theta}$ for each bootstrap iteration. 
We take
$$\displaystyle \wh{\text{cv}}_{1-\alpha} = \inf( z: \mathbb P(  \max_{k \in 1, \dots, K} \tilde \phi'_{b, \theta}   ((\sqrt{n} (\wh \theta_b^\star - \wh \theta_b); \pdf^k) - z) \leq 0 ) \geq 1- \alpha'),$$
where $\alpha'<\alpha$ but arbitrarily close to $\alpha$.\footnote{
	To simplify the notation, we just consider a fix critical value $\text{cv}_{1-\alpha}$ here. In principle, it might be different for each component and for each point of evaluation $\{\pdf^k\}_{k=1}^K$ and indeed it would be more efficient to do this.}	
We then consider as lower confidence set $\tilde \phi_b(\wh \theta_b, \pdf^k )- \wh{\text{cv}}_{1-\alpha} /\sqrt{n}$ for all $k \in \{1, \dots, K\}$.
These lower confidence sets  are uniformly valid for the mapping $\tilde \phi_b$ when evaluated at $\{\pdf^k\}_{k=1}^K $.

To obtain a lower confidence band of $\tilde \phi_b$, which is valid uniformly in $\pdf$, we exploit the functional form of $\tilde \phi_b$ similarly to \cite{masten2020}.  
The lower bound  for intermediates points, that are not within the set $\{\pdf^k\}_{k=1}^K$, is interpolated based on the left and right nearest neighbor of the point of evaluation. The respectively lowest confidence set is taken.	
By monotonicity of $\tilde \phi_b$, this lower confidence set is then also valid uniformly valid in $\pdf$. 

To construct a valid confidence set for $\phi_b$, we then consider a simple projection argument of $\tilde \phi_b$ by taking the maximum of the last three components of $\tilde \phi_b$ into account. 
We construct our confidence set for our sensitivity and our robust region $\wh{RR}_{b,L}$ and  $\wh{ \A}_{b,L}$  based on our  constructed lower confidence set. 	
\begin{proposition}\label{coro_inference_cv_consistentcy_binary} Suppose that Assumption~\ref{assumptionLATE}~and~\ref{assumption_inference_sampling_distribution} hold and the variance of each component of  $\theta_b$ is bounded away from zero.  It then holds that, 	$$ \underset{n \rightarrow \infty}{\lim} \IP ( \wh{RR}_{b,L} \subseteq RR_b, \;   \A_b  \subseteq  \wh{ \A}_{b, L})\geq 1 - \alpha.$$ 
\end{proposition}
We impose the variance condition to ensure that the underlying parameters converge to a non-degenerated distribution.
\subsection{Estimation for a Continuous Outcome Variable}\label{appendixestimation}
In this section, we give further details on the construction of the estimators for a continuous outcome variable.	We first estimate the underlying parameters $\theta.$	
We estimate the conditional joint densities by standard nonparametric kernel density estimator  
\begin{align*}
	\wh{q}_{dz}(y)= \frac{1}{n_zh} \sum_{i=1}^{n_z} K_h(Y_i^z-y) \cdot \1{D_i^z=d},
\end{align*}
where  $K_h(\cdot)=K(\cdot/h)/h$ and $K(\cdot)$ denotes a density function and $h>0$ a bandwidth.  We show in Lemma~\ref{lemma_gdplus} that  our estimator of $G_d^+(y)= \int_{\mathbb Y} \1{\tilde y \leq y} \max\{0,g_d(\tilde y ) \} d\tilde y$ under our assumptions.
We therefore define $$\wh G_d^+(y)=  \int_{\mathbb Y} \1{\tilde y \leq y} \max\{0,\wh{g}_d(\tilde y ) \} d\tilde y,$$ where  $\wh g_d(y)=(\wh{q}_{dd}(y)- \wh{q}_{d(1-d)}(y))/\wh \pi_\Delta$.   	The conditional probability  functions are further estimated by
$\wh Q_{dz}(y)= \int_{\mathbb Y} \1{ \tilde y \leq y} \wh q_{dz}(\tilde y)d\tilde y.$
Based on these estimators, the parameters of $\theta$ are estimated
and 	we estimate $\phi_b(\wh \theta_b, \pdf)$ by simple plug-in methods, where infimum, supremum and integrals are numerically evaluated.

\subsection{Assumptions for a Continuous Outcome Variable}\label{sec_assumption_inference_continuous}

We first impose the following  regularity assumptions.
\begin{assumption} \label{assumption_inference_continuous}	\begin{enumerate*}[label=(\roman*)]	
		\item $\mb{Y}_d$ is given by $[\underline y_d, \overline y_d]$ for $\infty < \underline y_d < \overline y_d< \infty$ for $d \in \{0,1\}$.	
		\item  $\forall d,z \in \{0,1\}$, the functions $q_{dz}(y)$ are bounded and bounded away from zero, absolutely continuous  and two times continuously differentiable  with uniformly bounded derivatives.
		\item For $d \in \{ 0,1\}$, the  functions $q_{dd}(y)$ and $q_{d(1-d)}(y)$ cross at a finite number of times. \label{assumption_item_gdplus}
	\end{enumerate*}
\end{assumption}
Assumption $(i)$ assumes compact support of the outcome variable as it simplifies the following analysis.   Assumption $(ii)$ imposes smoothness conditions on the joint densities, which are standard in the nonparametric literature.  Assumption $(iii)$ is  imposed for simplicity and substantially simplifies the analysis of the estimator of the function  $G_d^+(y)$.\footnote{ This assumption is satisfied if the weighted densities $\pdf f_{Y_d^{DF}}$ and  $\pco f_{Y_d^{CO}}$ intersect  only finitely many times. Without this assumption,  our proposed estimator of  $\wh G_d^+(y)$ is a biased estimator of  $G_d^+(y)$.  Following the arguments of \cite{anderson2010}, one can construct a debiased estimator of $G_d^+(\overline y)$, which  converges  in $\sqrt{n}$ to a mean-zero normal distribution. Based on similar arguments, one could now construct a debiased estimator of $G_d^+(y)$. As this is a rather tedious exercise and not the purpose of this paper, we impose this stronger assumption.}		
\begin{assumption} \label{assumption_inference_kernel}
	$(i)$ The kernel is a second order kernel function, being symmetric around zero, integrates to one, twice continuously differentiable, of bounded variation and zero-valued off, say $[-0.5, 0.5]$. 
	$(ii)$ The bandwidth satisfies: (a) $nh^{4} \rightarrow 0$, (b) $nh^2\rightarrow \infty$, (c) ${nh/ \log(n) \rightarrow \infty}$.
\end{assumption}
Assumption~\ref{assumption_inference_kernel}~(i)  imposes  conditions on the choice of kernel  which can be satisfied by construction and  Assumption~\ref{assumption_inference_kernel}~(ii) imposes conditions on the bandwidth.

\subsection{Asymptotic Results for a Continuous Outcome Variable}\label{sec_underlying_paramters}
We first note that  we have the following result.
\begin{proposition}\label{proposition_underlyingparameter_convergence}
	Suppose  Assumptions~\ref{assumption_inference_sampling_distribution}--\ref{assumption_inference_kernel} hold. It then follows that 	$$\sqrt{n}( \widehat \theta(y) - \theta(y)) \rightarrow \mathcal Z_1(y),$$
	where $\mathcal Z_1(y)$ is a tight mean-zero Gaussian process in $\ell^{\infty}( \mathbb R,\mathbb R^6)$.\footnote{ Let $A$ be some  arbitrary set and $B$ a Banach space. Then  $\ell^{\infty}( A,B)$ denotes the set of all mappings $f: A\rightarrow B$, which satisfy that $\sup_{a\in A} || f(a)||_B \leq \infty$.  }
\end{proposition}
As explained in the main text, we cannot directly base our inference procedure on the mapping $\phi(\theta, \pdf)$, as this mapping is non-smooth and standard asymptotic theory cannot be applied. 
We, therefore, consider a smoothed version of this mapping in this section. 	
To be more precise, we consider the definition of \cite{masten2020}, which we cite here for completeness.
\begin{definitions}[Definition 1, \cite{masten2020}]\label{definitionsmoothing}
	Let $(\Theta,\| \cdot \|_{\Theta})$ and $(\mathcal H,\| \cdot \|_{\mathcal H})$ be Banach spaces.  Let $\leq $ be a partial order on $\mathcal H$. Let $h: \Theta \rightarrow \mathcal H $ be a function.  Consider a function $H_{\kappa}: \Theta \rightarrow \mathcal H$, where $ \kappa \in \mb R_+^{\dim(\kappa)} $ is a vector of smoothing parameters.  Then $H_{\kappa}$ denotes a \textit{smooth  lower approximation} (SLA) of  $H$  if
	\begin{enumerate}[wide, labelindent=0pt]
		\itemsep0em 
		\item Lower envelope: $ H_{\kappa}(\theta) \leq  H(\theta) $ for all $\theta  \in \Theta$ and  $ \kappa \in \mb R_+^{\dim(\kappa)} $ .
		\item Approximating:  For each $\theta \in \Theta $, $ H_{\kappa}(\theta) \rightarrow H(\theta)$ for  $\kappa \rightarrow \infty$ (pointwise).
		\item Smoothing: $ H_{\kappa}$  is Hadamard-differentiable.
	\end{enumerate} 
\end{definitions}
This definition of a \textit{smooth  upper approximation} (SUA) is analogues. 	We now assume that $\phi_{\kappa}$ is a SLA of $\phi$ componentwise, and we show in the subsequent sections how we can obtain such a smooth mapping.
Let $\wh \theta^{\star}$ denotes a draw from the nonparametric bootstrap. We then choose the critical value such that
\begin{align*}
	&\wh{\text{cv}}_{1-\alpha}=\inf \left\lbrace z \in \mb R: \right. \\
	&  \left. \mb P\left((\sup_{\pdf \in [0, 0.5], l\leq 5} \sqrt{n} e_l^\top (\phi_{\kappa}(\wh \theta^{\star}, \pdf) - \phi_{\kappa}(\wh \theta,\pdf)) \leq z| \{ \{Y_i^z, D_i^z\}_{i=1}^{n_z}\}_{z=0}^1 \right) \geq 1- \alpha    \right.\rbrace
\end{align*}
We can also allow that z is a known function of $\pdf$ and $l$. By doing so, we can exploit the trade-off by constructing the confidence set for the sensitivity and robust region.
We construct our function  $\phi_{\kappa,L}(\theta, \pdf) = \phi_{\kappa}(\wh \theta, \pdf) + \wh{\text{cv}}_{1-\alpha}/\sqrt{n}$ and our confidence sets for the sensitivity and robust region $ \widehat{ \A}_{L}(\kappa)$ and $\widehat{RR}_{L}(\kappa)$ are constructed based on this mapping as explained in Section~\ref{sectionsensitivtyanalysis}. We then have  the following result. 
\begin{proposition}\label{coro_inference_cv_consistentcy}
	Suppose Assumptions~\ref{assumptionLATE}  and Assumptions~\ref{assumption_inference_sampling_distribution}--\ref{assumption_inference_kernel} hold. It then follows that	$ \underset{n \rightarrow \infty}{\lim} \IP ( \widehat{RR}_{L}(\kappa) \subseteq RR, \;   \A  \subseteq  \widehat{ \A}_{L}(\kappa)) \geq 1 - \alpha.$
\end{proposition}

\subsection{Population Smoothing}\label{sec_population_smoothing}
\subsubsection{General Introduction}\label{sec_population_smoothing}
We now show how to construct these smoothed mapping $\phi_{\kappa}$.	
As our mapping $\phi$ is a mapping of many non-differentiable mappings, we prove a chain-rule argument, which allows us to consider simpler mappings. 
\begin{lemma}\label{lemma_chainrule}
	Let $\psi$ and $\phi$ be two positive and nondecreasing mappings and denote by 	$\psi^U(\kappa)$   and by $\phi^U(\kappa)$ there respectively SLA, then
	$\psi^U(f, \kappa)$   and by $\phi^U(\psi^U(f, \kappa), \kappa)$
	is a SLA of $\psi(\phi)$. 	
	Accordingly, $\phi^U(\psi^U(f, \kappa), \kappa)$ denotes the SUA. 
\end{lemma}

Based on these definitions, we argue that the mapping $\phi(\theta, \pdf)$ is a composition of non-smooth random functions, where we replace each of them with a respective SLA and SUA. We first consider these mapping separately, and we  then show how to use them to construct our bounds. 	Let $\kappa>1$ be  the smoothing parameter.	Let $(\Theta,\| \cdot \|_{\Theta})$ and $(\mathcal H,\| \cdot \|_{\mathcal H})$ be Banach spaces, where $\leq $ is a partial order on $\mathcal H$. we consider two mappings $f,g: \Theta \rightarrow \mathcal H $ in the following, which are both Hadamard-differentiable.

\textbf{Maximum and minimum: } 
We first consider the
function $ \psi_{av}(f)=|f|,$ where a SLA and SUA is given by
$ \psi^U_{av}(f; \kappa)= \sqrt{f^2 + 1/\kappa}$ and  $\psi^L_{av}(f; \kappa)=  f^2  / (\sqrt{f^2 + 1/\kappa}).$ 

\begin{lemma}\label{lemma_av}
	$ \psi^L_{av}(f; \kappa)$ is a SLA and $ \psi^U_{av}(f; \kappa)$ a SUA for the mapping $\psi_{av}(f)$ and  .
\end{lemma}

Let  $\psi_{\min}(f,g) =\min(f,g)$ and $\psi_{\max}(f,g)=\max(f,g)$. A SLA of $\psi_{\max}(f,g)$ is clearly given by  $\psi^L_{\max}(f,g; \kappa)= f+g+ \psi^L_{av}(f-g; \kappa)$ and a SUA is given by $\psi^U_{\max}(f,g; \kappa)= f+g+ \psi^U_{av}(f-g; \kappa)$.  It follows from a simple induction argument, that one can generalize this procedure to the maximum of a set of finitely many mappings.

\textbf{Supremum and infimum: } 
In the following, we consider the mapping  $\psi_{\sup, \leq }(f,g)(\cdot)=\sup_{z\leq \cdot } f(z)-g(z)$ and the equally binned set $\mb Y= \bigcup_{k=1}^{\kappa} [\underline{y}+(k-1) d_Y, \underline{y}+k d_Y]$, where $d_Y= \frac{1}{\kappa} (\overline{y}-\underline{y}) $. Let  $k_j= \underline y +j \cdot d_y$, where $j \in \{0, 1,2,\dots, \kappa \}$.
\begin{align*}
	\psi^L_{\sup, \leq}(f,g; \kappa)(\cdot) & =  \psi^L_{\max}( \{g(k_{j})  - f(k_{j}) \}_{j: k_{j}\leq \cdot}   ; \kappa). \\ 
	\psi^U_{\sup,  \leq}(f,g; \kappa)(\cdot) & =  \psi^U_{\max}( \{g(k_{j})  - f(\min(\cdot,k_{j+1})) \}_{j: k_{j}<\cdot} \;; \kappa). 
\end{align*}
We similarly define for the mapping $\psi_{\inf, \leq }(f,g)(\cdot)=\inf_{z\leq \cdot } f(z)-g(z)$ that
\begin{align*}
	\psi^U_{\inf}(f,g; \kappa)(\cdot) & =  \psi^U_{\min}( \{g(k_{j} )  - f(k_{j+1}) )\}_{j: k_{j}\leq\cdot}  ; \kappa). \\
	\psi^L_{\inf}(f,g; \kappa)(\cdot) & =  \psi^L_{\min}( \{g(\min(\cdot,k_{j+1})  - f(k_{j}) )\}_{j: k_{j}<\cdot}  ; \kappa).
\end{align*}

\begin{lemma}\label{lemma_supu}
	If $f$ and $g$ are monotone increasing,
	$\psi^L_{\inf,\leq}(f,g; \kappa)$ is a SLA  and $\psi^U_{\sup,\leq}(f,g; \kappa)$ a SUA  to the function $\psi_{\sup, \leq }(f,g)$, and $\psi^U_{\inf, \leq}(f,g; \kappa)$ a SUA and  $\psi^L_{\inf, \leq}(f,g; \kappa)$ a SLA to the function $\psi_{\inf, \leq}(f,g)$.\footnote{By similar reasoning, the functions $\psi_{\sup,\geq}(f,g)$ and $\psi_{\inf,\geq}(f,g)$ can be smoothly approximated.} 
\end{lemma}

\subsubsection{Smoothing the Sensitivity and Robust Regions} We derive the smoothed mapping
$$\phi_{\kappa}( \theta, \pdf)=   \left( \updf^L( \kappa), \; - \opdf^U(\kappa), \; \deltamin^L(\pdf; \kappa), \; - \deltamax^U( \pdf; \kappa ), \; BP^L(\pdf; \kappa ))    \right). $$	
Since our sharp bounds $\Flcod$ and  $\Fucod$ are the key elements in our construction, we consider them first.  
We show how to smooth the lower bound from above.  we note that $\Gdsup (y)= \sup_{z \leq y} G_d^+(z) - G_d^-(z)$, where  $G_d^-(z)=G_d^+(y)- G_d(y).$ We denote  the upper bound by
\begin{align*}
	G_d^{\sup, U} (y; \kappa)= \psi^U_{\sup, \leq}(G_d^{+}-G_d^{-}; \kappa)(y)
\end{align*}
The bound on the outcome distribution of compliers are therefore bounded by 
\begin{align*}
	\underline{H}^U_{Y_d^{CO}}& (y, \pdf, \delta;  \kappa)=  \\
	& \frac{1}{ \pco } \;   \psi^U_{\max}\left( \left\lbrace 0,   \pi_{\Delta}  G_d^{\sup, U} (y; \kappa), \Qdd (y)- \pi_d , \frac{\pco}{ \pi_{\Delta} } (G_d^{\sup, U} (y; \kappa) -\pdf \delta ) \right\rbrace  ; \kappa\right)
\end{align*}
A SUA of $\underline{F}_{Y_d^{CO}}(y, \pdf,\delta) $ is given by 
\begin{align*}
	&\underline{F}^{U}_{Y_d^{CO}}(y, \pdf,\delta; \kappa) = \frac{1}{\pco} \Qdd(y)    \\
	&  - \frac{1}{\pco}
	\psi^L_{\inf,\geq}\left(Q_{dd}(\wt{y}) - \pi_{\Delta} G_d^+(\wt{y})  -   \psi^L_{\inf,\leq}
	\left( \pi_{\Delta} G_d^+(\wh{y}) -  \pco \underline{H}^U(y, \pdf, \delta_;  \kappa) ; \kappa   \right)( \wh y) \;  ;\kappa \right)(\wt y).
\end{align*}
A SLA of $\underline{F}_{Y_d^{CO}}(y, \pdf,\delta) $ can be similarly constructed as well as a smooth lower and upper approximation of $\overline{F}_{Y_d^{CO}}(y, \pdf,\delta) $.	
We now turn to the sensitivity region. The lower bounds on the sensitivity parameter $\opdf$ can be constructed by
$$ \opdf^L(\kappa)= \psi_{\min}^L( \{ \mb P(D=1|Z=0), \mb P(D=0|Z=1) \}; \kappa )$$
and similarly upper bound on the sensitivity parameter $\updf$ by
$$ \opdf^U(\kappa) =  \frac{\pi_{\Delta}}{\pco} \psi_{\max}^U( \{ G_1^+(\overline y), G_0^+(\overline y) \}; \kappa )-1.$$
One can similarly derive the other parameters.

\subsection{Illustration of  Derivation of Bounds on Outcome Distributions}\label{sec_illustraion_bounds}
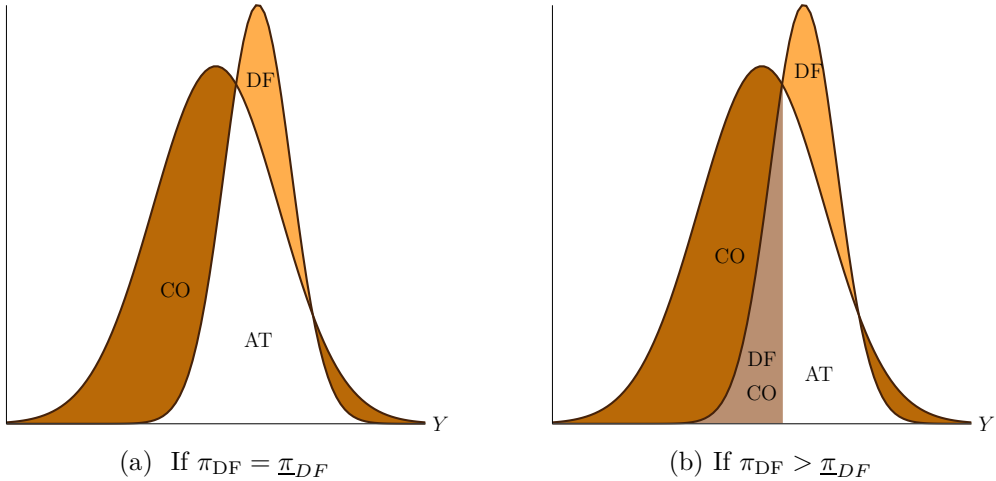
\begin{figure}[!h]
	\centering
	\subfloat[	If $\pdf=\updf$] { 	\resizebox{0.35\linewidth}{!}{\input{Graphics/illustration_bounds_2}}\hfill}%
	$\qquad$
	\subfloat[If $\pdf >\updf$ 	]{		\resizebox{0.35\linewidth}{!}{\input{Graphics/illustration_bounds_3}}}
	\caption{Derivation of the compliers outcome distributions}\label{figureconstruction1} 
\end{figure}
In Figure~\ref{figureconstruction1}, we give some intuition on how the outcome distribution of compliers is constructed. We plot  the functions $q_{11}$ and $q_{10}$. Based on the reasoning of the main text, the function $q_{11}$ is a weighted average of the densities of $f_{Y_1^{CO}}$ and $f_{Y_1^{AT}}$, and the function $q_{10}$  of the densities of $f_{Y_1^{DF}}$ and $f_{Y_1^{AT}}$. It is clear that to be a feasible candidate of $f_{Y_1^{CO}}$, any density has to satisfy that
$$ \frac{1}{\pi_\Delta} \max\{ 0, q_{11}(y)-q_{10}(y) \} \leq f_{Y_1^{CO}}(y) \leq \frac{1}{\pco}  q_{11}(y).$$
In Figure~\ref{figureconstruction1}~(a), the density of $f_{Y_1^{CO}}$	is point identified for the sensitivity parameter $\pdf$ that is the smallest when ignoring the distribution functions in the absence of treatment. However, if $\pdf$ increases the density of $f_{Y_1^{CO}}$ is in general not point identified. The corresponding probability mass of the tails of the function $\min \{q_{11}(y), q_{10}(y)\}$ is then imputed to belong to the compliers and defiers. 	Figure~\ref{figureconstruction1}~(b) gives such an example for a possible candidate of density function of $f_{Y_1^{CO}}$ implying an upper bound on the distribution function of compliers. 
\subsection{Intuition for Lower and Upper Bound on Outcome Heterogeneity}
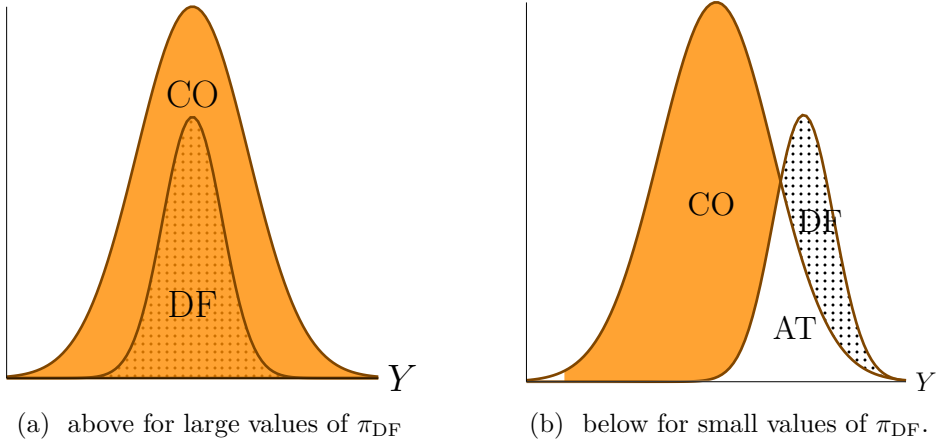
\begin{figure}[!h] 
	\centering
	\subfloat[	above for large values of $\pdf$] { 	\resizebox{0.33\linewidth}{!}{\input{Graphics/sensitivityregion_bound_above}}\hfill}%
		$\qquad$
	\subfloat[	below for small values of $\pdf$.]{		\resizebox{0.33\linewidth}{!}{\input{Graphics/sensitivityregion_bound_below}}}
	\caption{Illustration of sensitivity region. }\label{figureboundsdelta}
\end{figure}
We give some intuition on how the bounds on the sensitivity parameters $\delta$ are derived. Let us first  consider the largest value $\opdf$ ignoring the distribution functions in the absence of treatment. In Figure~\ref{figureboundsdelta}~(a), this value implies that both the outcome distributions of compliers and defiers are point identified, as the population size of always takers would be zero. Thus the  outcome distribution function of defiers  equals $Q_{10}(y)$, and  of compliers equals $Q_{11}(y)$ up to normalization. In this specific example, the outcome heterogeneity would be point identified but especially bounded from above by $0.5$.  In Figure~\ref{figureboundsdelta}~(b),  we consider the smallest possible value of outcome heterogeneity $\pdf=\updf$.  The two outcome distributions are again point identified, and the outcome heterogeneity would be close to one, but especially it would be bounded from below. A similar reasoning then also applies to the absence of treatment.

\singlespacing
\bibliography{bibl}  
\end{document}

%% file: Graphics/average_identification.tex
\begin{tikzpicture}[scale=1, transform shape]
\begin{axis}[
mark = none,
xmin = 0.5, ymin = 0,
xmax = 5,ymax = 8,
axis lines*=middle,
enlargelimits=upper,
clip=false,
xtick={0},
ytick={0}]
\addplot[black, domain=1.02:4.5,restrict y to domain=0:6.8, samples=100]  {4/(x-1)};
\draw [black]  (110,800) -- (110,-1);
\draw [black] (400,-1) -- (400,800);
\coordinate[label=below: $BF$] (G) at (350,220);
\coordinate[label=below:  $\updf$] (A) at (110,-10);
\coordinate[label=below: $\opdf$] (C) at (400,-10);
\coordinate[label=below:  $\delta_a$] (F) at (-30,500);
\draw[fill=gray,nearly transparent]  (1,0) -- (1,800) -- (109,800) -- (109,0) -- cycle;
\draw[fill=gray,nearly transparent]  (400,0) -- (400,800) -- (500,800) -- (500,0) -- cycle;
\coordinate[label=below:Robust Region] (B) at (250,100);
\coordinate[label=below:Nonrobust Region] (D) at (250,650);
\end{axis}
\end{tikzpicture}

%% file: Graphics/quantile_identification.tex
\begin{tikzpicture}[scale=1]%, transform shape]
\begin{axis}[
mark = none,
xmin = 0.5, ymin = 0,
xmax = 5.5,ymax = 5.2,
axis lines*=middle,
enlargelimits=upper,
clip=false,
xtick={-10},
ytick={-20}]
\addplot[name path=IR, black, domain=1.02:4.5,restrict y to domain=0:6, samples=100]  {min(5,3/(x-1)+2.5)};
\addplot[black,name path=BF, domain=2.02:4.5,restrict y to domain=0:6, samples=100]  {min(5,  3/(x-1)+2.5,  2/(x-2)+0.5)};
\draw [black, dashed, name path=one] (0,500) -- (500,500);
\addplot[gray,nearly transparent] fill between [of=IR and one];
\draw [black]  (110,500) -- (110,-1);
\draw [black] (400,-1) -- (400,332);
%\draw [black, dashed] (50,0) -- (50,500);
\coordinate[label=below: BF] (G) at (375,140);
\coordinate[label=below:$\updf$] (A) at (110,-10);
\coordinate[label=below:$\opdf$] (C) at (400,-10);
\coordinate[label=below:$\delta_q$] (F) at (-30,300);
\draw[fill=gray,nearly transparent]  (1,0) -- (1,500) -- (109,500) -- (109,0) -- cycle;
\draw[fill=gray,nearly transparent]  (400,0) -- (400,332) -- (500,500) -- (500,0) -- cycle;
\coordinate[label=below: Robust Region] (B) at (250,100);
\coordinate[label=below:Nonrobust] (D) at (300,350);
\coordinate[label=below: Region] (D2) at (300,300);
\coordinate[label=below: 1] (H) at (-20,520);
\end{axis}
\end{tikzpicture}
 

%% file: Graphics/binary_outcome_angrist_evans_res.tex
% Created by tikzDevice version 0.12.3 on 2020-07-20 20:01:34
% !TEX encoding = UTF-8 Unicode
\begin{tikzpicture}[x=1pt,y=1pt]
\definecolor{fillColor}{RGB}{255,255,255}
\path[use as bounding box,fill=fillColor,fill opacity=0.00] (0,0) rectangle (289.08,216.81);
\begin{scope}
\path[clip] ( 49.20, 61.20) rectangle (263.88,167.61);
\definecolor{drawColor}{RGB}{255,0,0}

\path[draw=drawColor,line width= 0.8pt,line join=round,line cap=round] ( 57.49,216.81) --
	( 59.16,156.03) --
	( 61.17,105.68) --
	( 63.17, 92.17) --
	( 65.18, 85.41) --
	( 67.19, 81.36) --
	( 69.20, 78.66) --
	( 71.21, 76.72) --
	( 73.21, 75.28) --
	( 75.22, 74.15) --
	( 77.23, 73.25) --
	( 79.24, 72.51) --
	( 81.25, 71.90) --
	( 83.25, 71.38) --
	( 85.26, 70.93) --
	( 87.27, 70.55) --
	( 89.28, 70.21) --
	( 91.28, 69.91) --
	( 93.29, 69.65) --
	( 95.30, 69.41) --
	( 97.31, 69.20) --
	( 99.32, 69.00) --
	(101.32, 68.83) --
	(103.33, 68.67) --
	(105.34, 68.52) --
	(107.35, 68.38) --
	(109.36, 68.26) --
	(111.36, 68.14) --
	(113.37, 68.04) --
	(115.38, 67.94) --
	(117.39, 67.84) --
	(119.39, 67.76) --
	(121.40, 67.68) --
	(123.41, 67.60) --
	(125.42, 67.53) --
	(127.43, 67.46) --
	(129.43, 67.39) --
	(131.44, 67.33) --
	(133.45, 67.27) --
	(135.46, 67.22) --
	(137.47, 67.17) --
	(139.47, 67.12) --
	(141.48, 67.07) --
	(143.49, 67.03) --
	(145.50, 66.98) --
	(147.50, 66.94) --
	(149.51, 66.90) --
	(151.52, 66.87) --
	(153.53, 66.83) --
	(155.54, 66.80) --
	(157.54, 66.76) --
	(159.55, 66.73) --
	(161.56, 66.70) --
	(163.57, 66.67) --
	(165.58, 66.64) --
	(167.58, 66.62) --
	(169.59, 66.59) --
	(171.60, 66.56) --
	(173.61, 66.54) --
	(175.61, 66.52) --
	(177.62, 66.49) --
	(179.63, 66.47) --
	(181.64, 66.45) --
	(183.65, 66.43) --
	(185.65, 66.41) --
	(187.66, 66.39) --
	(189.67, 66.37) --
	(191.68, 66.48) --
	(193.69, 65.98) --
	(195.69, 66.14) --
	(197.70, 66.30) --
	(199.71, 66.44) --
	(201.72, 66.58) --
	(203.72, 66.71) --
	(205.73, 66.84) --
	(207.74, 66.96) --
	(209.75, 67.07) --
	(211.76, 67.17) --
	(213.76, 67.27) --
	(215.77, 67.37) --
	(217.78, 67.46) --
	(219.79, 67.54) --
	(221.80, 67.62) --
	(223.80, 67.70) --
	(225.81, 67.77) --
	(227.82, 67.84) --
	(229.83, 67.78) --
	(231.83, 67.65) --
	(233.84, 67.52) --
	(235.85, 67.39) --
	(237.86, 67.27) --
	(239.87, 67.16) --
	(241.87, 67.05) --
	(243.88, 67.10) --
	(245.89, 67.09) --
	(247.90, 67.15) --
	(249.91, 67.20) --
	(251.91, 67.26) --
	(253.92, 67.31) --
	(255.93, 67.36);
\end{scope}
\begin{scope}
\path[clip] ( 49.20, 61.20) rectangle (263.88,167.61);
\definecolor{drawColor}{RGB}{0,0,0}

\path[draw=drawColor,line width= 1.6pt,line join=round,line cap=round] ( 57.49,216.81) --
	( 59.16,156.03) --
	( 61.17,149.42) --
	( 63.17,143.72) --
	( 65.18,138.74) --
	( 67.19,134.36) --
	( 69.20,130.47) --
	( 71.21,127.00) --
	( 73.21,123.89) --
	( 75.22,121.08) --
	( 77.23,118.53) --
	( 79.24,116.20) --
	( 81.25,114.07) --
	( 83.25,112.12) --
	( 85.26,110.31) --
	( 87.27,108.64) --
	( 89.28,107.09) --
	( 91.28,105.65) --
	( 93.29,104.31) --
	( 95.30,103.05) --
	( 97.31,101.87) --
	( 99.32,100.77) --
	(101.32, 99.73) --
	(103.33, 98.75) --
	(105.34, 97.83) --
	(107.35, 96.95) --
	(109.36, 96.13) --
	(111.36, 95.34) --
	(113.37, 94.59) --
	(115.38, 93.89) --
	(117.39, 93.21) --
	(119.39, 92.57) --
	(121.40, 91.95) --
	(123.41, 91.37) --
	(125.42, 90.81) --
	(127.43, 90.27) --
	(129.43, 89.76) --
	(131.44, 89.26) --
	(133.45, 88.79) --
	(135.46, 88.34) --
	(137.47, 87.90) --
	(139.47, 87.48) --
	(141.48, 87.08) --
	(143.49, 86.69) --
	(145.50, 86.26) --
	(147.50, 85.14) --
	(149.51, 84.09) --
	(151.52, 83.11) --
	(153.53, 82.19) --
	(155.54, 81.32) --
	(157.54, 80.51) --
	(159.55, 79.74) --
	(161.56, 79.02) --
	(163.57, 78.33) --
	(165.58, 77.69) --
	(167.58, 77.08) --
	(169.59, 76.50) --
	(171.60, 75.96) --
	(173.61, 75.44) --
	(175.61, 74.95) --
	(177.62, 74.49) --
	(179.63, 74.05) --
	(181.64, 73.63) --
	(183.65, 73.23) --
	(185.65, 72.85) --
	(187.66, 72.49) --
	(189.67, 72.15) --
	(191.68, 71.82) --
	(193.69, 71.51) --
	(195.69, 71.22) --
	(197.70, 70.94) --
	(199.71, 70.67) --
	(201.72, 70.41) --
	(203.72, 70.17) --
	(205.73, 69.93) --
	(207.74, 69.71) --
	(209.75, 69.49) --
	(211.76, 69.29) --
	(213.76, 69.09) --
	(215.77, 68.90) --
	(217.78, 68.72) --
	(219.79, 68.55) --
	(221.80, 68.38) --
	(223.80, 68.22) --
	(225.81, 68.07) --
	(227.82, 67.92) --
	(229.83, 67.78) --
	(231.83, 67.65) --
	(233.84, 67.52) --
	(235.85, 67.39) --
	(237.86, 67.27) --
	(239.87, 67.16) --
	(241.87, 67.05) --
	(243.88, 67.10) --
	(245.89, 67.09) --
	(247.90, 67.15) --
	(249.91, 67.20) --
	(251.91, 67.26) --
	(253.92, 67.31) --
	(255.93, 67.36);

\path[draw=drawColor,line width= 1.6pt,line join=round,line cap=round] ( 57.15, 65.14) --
	( 57.28,  0.00);

\path[draw=drawColor,line width= 1.6pt,line join=round,line cap=round] ( 88.77,  0.00) --
	( 89.28,  1.07) --
	( 91.28,  4.84) --
	( 93.29,  8.19) --
	( 95.30, 11.19) --
	( 97.31, 13.88) --
	( 99.32, 16.32) --
	(101.32, 18.54) --
	(103.33, 20.57) --
	(105.34, 22.43) --
	(107.35, 24.13) --
	(109.36, 25.71) --
	(111.36, 27.17) --
	(113.37, 28.53) --
	(115.38, 29.79) --
	(117.39, 30.97) --
	(119.39, 32.07) --
	(121.40, 33.10) --
	(123.41, 34.08) --
	(125.42, 34.99) --
	(127.43, 35.85) --
	(129.43, 36.66) --
	(131.44, 37.43) --
	(133.45, 38.16) --
	(135.46, 38.85) --
	(137.47, 39.51) --
	(139.47, 40.14) --
	(141.48, 40.73) --
	(143.49, 41.30) --
	(145.50, 41.84) --
	(147.50, 42.36) --
	(149.51, 42.86) --
	(151.52, 43.33) --
	(153.53, 43.78) --
	(155.54, 44.22) --
	(157.54, 44.64) --
	(159.55, 45.04) --
	(161.56, 45.43) --
	(163.57, 45.80) --
	(165.58, 46.16) --
	(167.58, 46.50) --
	(169.59, 46.83) --
	(171.60, 47.16) --
	(173.61, 47.47) --
	(175.61, 47.77) --
	(177.62, 48.06) --
	(179.63, 48.34) --
	(181.64, 48.61) --
	(183.65, 48.87) --
	(185.65, 49.12) --
	(187.66, 49.37) --
	(189.67, 49.61) --
	(191.68, 49.84) --
	(193.69, 50.07) --
	(195.69, 50.28) --
	(197.70, 50.50) --
	(199.71, 50.70) --
	(201.72, 50.90) --
	(203.72, 51.10) --
	(205.73, 51.29) --
	(207.74, 51.47) --
	(209.75, 51.65) --
	(211.76, 51.83) --
	(213.76, 52.00) --
	(215.77, 52.16) --
	(217.78, 52.33) --
	(219.79, 52.48) --
	(221.80, 52.64) --
	(223.80, 52.79) --
	(225.81, 52.94) --
	(227.82, 53.08) --
	(229.83, 53.22) --
	(231.83, 53.36) --
	(233.84, 53.49) --
	(235.85, 53.62) --
	(237.86, 53.75) --
	(239.87, 53.88) --
	(241.87, 54.00) --
	(243.88, 54.12) --
	(245.89, 54.24) --
	(247.90, 54.35) --
	(249.91, 54.46) --
	(251.91, 54.57) --
	(253.92, 54.68) --
	(255.93, 54.79);

\path[draw=drawColor,line width= 1.6pt,line join=round,line cap=round] ( 57.15, 61.20) -- ( 57.15,167.61);

\path[draw=drawColor,line width= 1.6pt,line join=round,line cap=round] (206.73, 61.20) -- (206.73,167.61);
\end{scope}
\begin{scope}
\path[clip] (  0.00,  0.00) rectangle (289.08,216.81);
\definecolor{drawColor}{RGB}{0,0,0}

\path[draw=drawColor,line width= 1.6pt,line join=round,line cap=round] ( 49.20, 61.20) -- (263.88, 61.20);

\path[draw=drawColor,line width= 1.6pt,line join=round,line cap=round] ( 57.15, 61.20) -- ( 57.15, 55.20);

\path[draw=drawColor,line width= 1.6pt,line join=round,line cap=round] ( 96.91, 61.20) -- ( 96.91, 55.20);

\path[draw=drawColor,line width= 1.6pt,line join=round,line cap=round] (136.66, 61.20) -- (136.66, 55.20);

\path[draw=drawColor,line width= 1.6pt,line join=round,line cap=round] (176.42, 61.20) -- (176.42, 55.20);

\path[draw=drawColor,line width= 1.6pt,line join=round,line cap=round] (216.17, 61.20) -- (216.17, 55.20);

\path[draw=drawColor,line width= 1.6pt,line join=round,line cap=round] (255.93, 61.20) -- (255.93, 55.20);

\node[text=drawColor,anchor=base,inner sep=0pt, outer sep=0pt, scale=  1.00] at ( 57.15, 39.60) {0};

\node[text=drawColor,anchor=base,inner sep=0pt, outer sep=0pt, scale=  1.00] at ( 96.91, 39.60) {0.1};

\node[text=drawColor,anchor=base,inner sep=0pt, outer sep=0pt, scale=  1.00] at (136.66, 39.60) {0.2};

\node[text=drawColor,anchor=base,inner sep=0pt, outer sep=0pt, scale=  1.00] at (176.42, 39.60) {0.3};

\node[text=drawColor,anchor=base,inner sep=0pt, outer sep=0pt, scale=  1.00] at (216.17, 39.60) {0.4};

\node[text=drawColor,anchor=base,inner sep=0pt, outer sep=0pt, scale=  1.00] at (255.93, 39.60) {0.5};

\path[draw=drawColor,line width= 1.6pt,line join=round,line cap=round] ( 49.20, 61.20) -- ( 49.20,167.61);

\path[draw=drawColor,line width= 1.6pt,line join=round,line cap=round] ( 49.20, 65.14) -- ( 43.20, 65.14);

\path[draw=drawColor,line width= 1.6pt,line join=round,line cap=round] ( 49.20, 81.56) -- ( 43.20, 81.56);

\path[draw=drawColor,line width= 1.6pt,line join=round,line cap=round] ( 49.20, 97.98) -- ( 43.20, 97.98);

\path[draw=drawColor,line width= 1.6pt,line join=round,line cap=round] ( 49.20,114.41) -- ( 43.20,114.41);

\path[draw=drawColor,line width= 1.6pt,line join=round,line cap=round] ( 49.20,130.83) -- ( 43.20,130.83);

\path[draw=drawColor,line width= 1.6pt,line join=round,line cap=round] ( 49.20,147.25) -- ( 43.20,147.25);

\path[draw=drawColor,line width= 1.6pt,line join=round,line cap=round] ( 49.20,163.67) -- ( 43.20,163.67);

\node[text=drawColor,anchor=base east,inner sep=0pt, outer sep=0pt, scale=  1.00] at ( 37.20, 61.70) {0};

\node[text=drawColor,anchor=base east,inner sep=0pt, outer sep=0pt, scale=  1.00] at ( 37.20, 78.12) {0.1};

\node[text=drawColor,anchor=base east,inner sep=0pt, outer sep=0pt, scale=  1.00] at ( 37.20, 94.54) {0.2};

\node[text=drawColor,anchor=base east,inner sep=0pt, outer sep=0pt, scale=  1.00] at ( 37.20,110.96) {0.3};

\node[text=drawColor,anchor=base east,inner sep=0pt, outer sep=0pt, scale=  1.00] at ( 37.20,127.38) {0.4};

\node[text=drawColor,anchor=base east,inner sep=0pt, outer sep=0pt, scale=  1.00] at ( 37.20,143.80) {0.5};

\node[text=drawColor,anchor=base east,inner sep=0pt, outer sep=0pt, scale=  1.00] at ( 37.20,160.23) {0.6};
\end{scope}
\begin{scope}
\path[clip] ( 49.20, 61.20) rectangle (263.88,167.61);
\definecolor{drawColor}{RGB}{190,190,190}

\path[draw=drawColor,line width= 1.6pt,dash pattern=on 4pt off 4pt ,line join=round,line cap=round] ( 81.00, 61.20) -- ( 81.00,167.61);
\end{scope}
\begin{scope}
	\path[clip] (  0.00,  0.00) rectangle (361.35,216.81);
	\definecolor{drawColor}{RGB}{0,0,0}
	
	\node[text=drawColor,anchor=base,inner sep=0pt, outer sep=0pt, scale=  1.00] at (150, 15.60) {$\pdf$};
	
	\node[text=drawColor, anchor=base,inner sep=0pt, outer sep=0pt, scale=  1.00] at ( 10.80,114.41) {$\delta_b$};
\end{scope}
\end{tikzpicture}

%% file: Graphics/application_pic_continuous.tex
% Created by tikzDevice version 0.12.3.1 on 2020-09-24 00:04:35
% !TEX encoding = UTF-8 Unicode
\begin{tikzpicture}[x=1pt,y=1pt]
\definecolor{fillColor}{RGB}{255,255,255}
\path[use as bounding box,fill=fillColor,fill opacity=0.00] (0,0) rectangle (361.35,216.81);
\begin{scope}
\path[clip] ( 49.20, 61.20) rectangle (336.15,167.61);
\definecolor{drawColor}{RGB}{255,0,0}

\path[draw=drawColor,line width= 1.6pt,line join=round,line cap=round] ( 62.48,163.67) --
	( 65.14,163.67) --
	( 67.80,163.67) --
	( 70.46,163.67) --
	( 73.11,163.67) --
	( 75.77,163.67) --
	( 78.43,163.67) --
	( 81.08,163.67) --
	( 83.74,163.67) --
	( 86.40,163.67) --
	( 89.05,163.67) --
	( 91.71,163.67) --
	( 94.37,163.67) --
	( 97.02, 86.40) --
	( 99.68, 84.18) --
	(102.34, 81.85) --
	(105.00, 80.94) --
	(107.65, 79.19) --
	(110.31, 77.66) --
	(112.97, 77.02) --
	(115.62, 75.69) --
	(118.28, 75.24) --
	(120.94, 74.52) --
	(123.59, 73.50) --
	(126.25, 73.35) --
	(128.91, 72.59) --
	(131.57, 71.92) --
	(134.22, 71.89) --
	(136.88, 71.28) --
	(139.54, 70.75) --
	(142.19, 70.72) --
	(144.85, 70.28) --
	(147.51, 70.10) --
	(150.16, 69.88) --
	(152.82, 69.37) --
	(155.48, 69.37) --
	(158.13, 69.21) --
	(160.79, 68.76) --
	(163.45, 68.88) --
	(166.11, 68.54) --
	(168.76, 68.28) --
	(171.42, 68.44) --
	(174.08, 68.11) --
	(176.73, 67.88) --
	(179.39, 67.99) --
	(182.05, 67.66) --
	(184.70, 67.73) --
	(187.36, 67.67) --
	(190.02, 67.37) --
	(192.68, 67.51) --
	(195.33, 67.28) --
	(197.99, 67.13) --
	(200.65, 67.27) --
	(203.30, 67.06) --
	(205.96, 66.92) --
	(208.62, 67.04) --
	(211.27, 66.80) --
	(213.93, 66.89) --
	(216.59, 66.86) --
	(219.24, 66.64) --
	(221.90, 66.79) --
	(224.56, 66.59) --
	(227.22, 66.49) --
	(229.87, 66.52) --
	(232.53, 66.39) --
	(235.19, 66.40) --
	(237.84, 66.51) --
	(240.50, 66.46) --
	(243.16, 66.44) --
	(245.81, 66.51) --
	(248.47, 66.53) --
	(251.13, 66.47) --
	(253.78, 66.59) --
	(256.44, 66.52) --
	(259.10, 66.69) --
	(261.76, 66.64) --
	(264.41, 66.57) --
	(267.07, 66.67) --
	(269.73, 66.69) --
	(272.38, 66.62) --
	(275.04, 66.71) --
	(277.70, 66.72) --
	(280.35, 66.66);
\end{scope}
\begin{scope}
\path[clip] (  0.00,  0.00) rectangle (361.35,216.81);
\definecolor{drawColor}{RGB}{0,0,0}

\node[text=drawColor,anchor=base,inner sep=0pt, outer sep=0pt, scale=  1.00] at (192.68, 15.60) {$\pdf$};

\node[text=drawColor, anchor=base,inner sep=0pt, outer sep=0pt, scale=  1.00] at ( 10.80,114.41) {$\delta$};
\end{scope}
\begin{scope}
\path[clip] (  0.00,  0.00) rectangle (361.35,216.81);
\definecolor{drawColor}{RGB}{0,0,0}

\path[draw=drawColor,line width= 1.6pt,line join=round,line cap=round] ( 49.20, 61.20) -- (336.15, 61.20);

\path[draw=drawColor,line width= 1.6pt,line join=round,line cap=round] ( 59.83, 61.20) -- ( 59.83, 55.20);

\path[draw=drawColor,line width= 1.6pt,line join=round,line cap=round] (118.87, 61.20) -- (118.87, 55.20);

\path[draw=drawColor,line width= 1.6pt,line join=round,line cap=round] (177.91, 61.20) -- (177.91, 55.20);

\path[draw=drawColor,line width= 1.6pt,line join=round,line cap=round] (236.96, 61.20) -- (236.96, 55.20);

\path[draw=drawColor,line width= 1.6pt,line join=round,line cap=round] (296.00, 61.20) -- (296.00, 55.20);

\node[text=drawColor,anchor=base,inner sep=0pt, outer sep=0pt, scale=  1.00] at ( 59.83, 39.60) {0.0};

\node[text=drawColor,anchor=base,inner sep=0pt, outer sep=0pt, scale=  1.00] at (118.87, 39.60) {0.1};

\node[text=drawColor,anchor=base,inner sep=0pt, outer sep=0pt, scale=  1.00] at (177.91, 39.60) {0.2};

\node[text=drawColor,anchor=base,inner sep=0pt, outer sep=0pt, scale=  1.00] at (236.96, 39.60) {0.3};

\node[text=drawColor,anchor=base,inner sep=0pt, outer sep=0pt, scale=  1.00] at (296.00, 39.60) {0.4};

\path[draw=drawColor,line width= 1.6pt,line join=round,line cap=round] ( 49.20, 61.20) -- ( 49.20,167.61);

\path[draw=drawColor,line width= 1.6pt,line join=round,line cap=round] ( 49.20, 65.14) -- ( 43.20, 65.14);

\path[draw=drawColor,line width= 1.6pt,line join=round,line cap=round] ( 49.20, 84.85) -- ( 43.20, 84.85);

\path[draw=drawColor,line width= 1.6pt,line join=round,line cap=round] ( 49.20,104.55) -- ( 43.20,104.55);

\path[draw=drawColor,line width= 1.6pt,line join=round,line cap=round] ( 49.20,124.26) -- ( 43.20,124.26);

\path[draw=drawColor,line width= 1.6pt,line join=round,line cap=round] ( 49.20,143.96) -- ( 43.20,143.96);

\path[draw=drawColor,line width= 1.6pt,line join=round,line cap=round] ( 49.20,163.67) -- ( 43.20,163.67);

\node[text=drawColor,anchor=base east,inner sep=0pt, outer sep=0pt, scale=  1.00] at ( 37.20, 61.70) {0.0};

\node[text=drawColor,anchor=base east,inner sep=0pt, outer sep=0pt, scale=  1.00] at ( 37.20, 81.40) {0.2};

\node[text=drawColor,anchor=base east,inner sep=0pt, outer sep=0pt, scale=  1.00] at ( 37.20,101.11) {0.4};

\node[text=drawColor,anchor=base east,inner sep=0pt, outer sep=0pt, scale=  1.00] at ( 37.20,120.81) {0.6};

\node[text=drawColor,anchor=base east,inner sep=0pt, outer sep=0pt, scale=  1.00] at ( 37.20,140.52) {0.8};

\node[text=drawColor,anchor=base east,inner sep=0pt, outer sep=0pt, scale=  1.00] at ( 37.20,160.23) {1.0};
\end{scope}
\begin{scope}
\path[clip] ( 49.20, 61.20) rectangle (336.15,167.61);
\definecolor{drawColor}{RGB}{0,0,0}

\path[draw=drawColor,line width= 1.6pt,line join=round,line cap=round] ( 62.48,163.67) --
	( 65.14,163.67) --
	( 67.80,163.67) --
	( 70.46,163.67) --
	( 73.11,163.67) --
	( 75.77,163.67) --
	( 78.43,163.67) --
	( 81.08,163.67) --
	( 83.74,163.67) --
	( 86.40,163.67) --
	( 89.05,163.67) --
	( 91.71,163.67) --
	( 94.37,163.67) --
	( 97.02,163.67) --
	( 99.68,163.67) --
	(102.34,163.67) --
	(105.00,163.67) --
	(107.65,163.67) --
	(110.31,163.67) --
	(112.97,162.35) --
	(115.62,153.22) --
	(118.28,145.72) --
	(120.94,141.09) --
	(123.59,134.54) --
	(126.25,129.26) --
	(128.91,125.89) --
	(131.57,121.03) --
	(134.22,117.11) --
	(136.88,114.66) --
	(139.54,110.96) --
	(142.19,108.73) --
	(144.85,105.61) --
	(147.51,102.87) --
	(150.16,101.55) --
	(152.82, 99.10) --
	(155.48, 96.93) --
	(158.13, 95.92) --
	(160.79, 93.96) --
	(163.45, 92.38) --
	(166.11, 91.45) --
	(168.76, 89.87) --
	(171.42, 88.62) --
	(174.08, 87.49) --
	(176.73, 86.37) --
	(179.39, 85.86) --
	(182.05, 84.75) --
	(184.70, 83.83) --
	(187.36, 83.39) --
	(190.02, 82.48) --
	(192.68, 81.80) --
	(195.33, 81.35) --
	(197.99, 80.59) --
	(200.65, 79.95) --
	(203.30, 79.50) --
	(205.96, 78.92) --
	(208.62, 78.38) --
	(211.27, 77.94) --
	(213.93, 77.48) --
	(216.59, 77.15) --
	(219.24, 76.88) --
	(221.90, 76.29) --
	(224.56, 76.31) --
	(227.22, 76.03) --
	(229.87, 75.62) --
	(232.53, 75.52) --
	(235.19, 75.30) --
	(237.84, 75.04) --
	(240.50, 74.92) --
	(243.16, 74.75) --
	(245.81, 74.56) --
	(248.47, 74.42) --
	(251.13, 74.11) --
	(253.78, 74.10) --
	(256.44, 73.95) --
	(259.10, 73.80) --
	(261.76, 73.65) --
	(264.41, 73.44) --
	(267.07, 73.21) --
	(269.73, 72.95) --
	(272.38, 72.65) --
	(275.04, 72.45) --
	(277.70, 72.21) --
	(280.35, 71.94);

\path[draw=drawColor,line width= 1.6pt,line join=round,line cap=round] ( 60.52, 61.20) -- ( 60.52,167.61);

\path[draw=drawColor,line width= 1.6pt,line join=round,line cap=round] (281.94, 61.20) -- (281.94,167.61);
\end{scope}
\end{tikzpicture}

%% file: Graphics/illustration_bounds_2.tex
	\begin{tikzpicture}[scale=1]
	\begin{axis}[
	no markers, domain=0:10, domain y=0:10, samples=100,
	axis lines*=left, xlabel=$ $, ylabel=$ $,
	every axis y label/.style={at=(current axis.above origin),anchor=south},
	height=10cm, width=10cm,
	xtick\empty, ytick=\empty,
	enlargelimits=false, clip=false, axis on top,
	grid = major
	]
	\addplot [name path=Afun, very thick,darkbrown!50!black] {0.5*gauss(6,0.8)};
	\addplot [name path=Bfun, very thick, darkbrown!50!black] {0.8*gauss(5,1.5)};
	\addplot [name path=Cfun, very thin, white] {0.00001*gauss(5,1.5)};
	\node [right] at (350,80) {\normalsize{CO}};
	\node [right] at (550,50) {\normalsize{AT}};
		\node [right] at (555,205) {\normalsize{DF}};
	\node [right] at (1000,0) {\normalsize{$Y$}};
	\addplot fill between[of = Afun and Bfun,
		soft clip={domain=0:10,domain y=0:10},
	 every even segment/.style={black,opacity=.9}];
	 \addplot fill between[of = Afun and Cfun,
	 soft clip={domain=0:10,domain y=0:10},
	 every even segment/.style={white,opacity=1}];
	\addplot fill between[of = Afun and Bfun,
soft clip={domain=0:10,domain y=0:10},
every even segment/.style={darkorange,opacity=.7}];
	\end{axis}
	\end{tikzpicture}

%% file: Graphics/illustration_bounds_3.tex
	\begin{tikzpicture}[scale=1,
	hatch distance/.store in=\hatchdistance,
	hatch distance=10pt,
	hatch thickness/.store in=\hatchthickness,
	hatch thickness=2pt
	]
	\makeatletter
%	\pgfdeclarepatternformonly[\hatchdistance,\hatchthickness]{flexible hatch}
%	{\pgfqpoint{0pt}{0pt}}
%	{\pgfqpoint{\hatchdistance}{\hatchdistance}}
%	{\pgfpoint{\hatchdistance-1pt}{\hatchdistance-1pt}}%
%	{
%		\pgfsetcolor{\tikz@pattern@color}
%		\pgfsetlinewidth{\hatchthickness}
		\pgfpathmoveto{\pgfqpoint{0pt}{0pt}}
%		\pgfpathlineto{\pgfqpoint{\hatchdistance}{\hatchdistance}}
%		\pgfusepath{stroke}
%	}]
\begin{axis}[
no markers, domain=0:10, domain y=0:10, samples=100,
axis lines*=left, xlabel=$ $, ylabel=$ $,
every axis y label/.style={at=(current axis.above origin),anchor=south},
height=10cm, width=10cm,
xtick\empty, ytick=\empty,
enlargelimits=false, clip=false, axis on top,
grid = major
]
\addplot [name path=Afun, very thick,darkbrown!50!black] {0.5*gauss(6,0.8)};
\addplot [name path=Bfun, very thick, darkbrown!50!black] {0.8*gauss(5,1.5)};
\addplot [name path=Cfun, very thin, white] {0.00001*gauss(5,1.5)};
\addplot [name path=Dfun, very thin, white] {0.00001*gauss(5,1.5)};
\node [right] at (560,210) {\normalsize{DF}};
\node [right] at (370,100) {\normalsize{CO}};
\node [right] at (585,30) {\normalsize{AT}};
\node [above] at (500,10) {\normalsize{CO}};
\node [above] at (500,30) {\normalsize{DF}};
\node [right] at (1000,0) {\normalsize{$Y$}};
\addplot fill between[of = Afun and Bfun,
soft clip={domain=0:10,domain y=0:10},
every even segment/.style={black,opacity=.9}];
\addplot fill between[of = Afun and Cfun,
soft clip={domain=0:10,domain y=0:10},
every even segment/.style={white,opacity=1}];
\addplot fill between[of = Afun and Bfun,
soft clip={domain=0:10,domain y=0:10},
every even segment/.style={darkorange,opacity=.7}];
\addplot fill between[of = Afun and Dfun,
soft clip={domain=0:5.5},
every even segment/.style={darkbrown,opacity=.6}];
\end{axis}
\end{tikzpicture}

%% file: Graphics/sensitivityregion_bound_above.tex
\begin{tikzpicture}[scale=1]
				\begin{axis}[
					no markers, domain=0:10, domain y=1:10, samples=100,
					axis lines*=left, xlabel=$ $, ylabel=$ $,
					every axis y label/.style={at=(current axis.above origin),anchor=south},
					%every axis x label/.style={at=(current axis.right of origin),anchor=west},
					height=7.5cm, width=7.5cm,
					xtick\empty, ytick=\empty,
					enlargelimits=false, clip=false, axis on top,
					grid = major
					]
					\addplot [name path=Afun, very thick, darkorange!50!black] {0.3*gauss(5,0.8)};
					\addplot [name path=Bfun, very thick,darkorange!50!black] {0.8*gauss(5,1.5)};
					\addplot [name path=Cfun, very thick,darkorange!50!black] {0.0001*gauss(5,1.5)};
					\node [above] at (500,150) {\Large{CO}};
					\node [above] at (500,30)  {\Large{DF}};
					\node [right] at (1000,0) {\Large{$Y$}};
				\addplot fill between[of = Afun and Bfun,
				soft clip={domain=0:1,domain y=0:10},
				every even segment/.style={darkorange, opacity=0.8}];
				\addplot fill between[of = Afun and Cfun,
				soft clip={domain=0:1,domain y=0:10},
				every even segment/.style={white,opacity=1}];
					\addplot fill between[of=Afun and Cfun,soft clip={domain=1:10},	split,
					every segment no 1/.style={darkorange,opacity=.4},
					every segment no 0/.style={pattern=dots,pattern color=black},];
					\addplot fill between[of=Afun and Cfun,soft clip={domain=1:10},	split,
					every segment no 1/.style={darkorange,opacity=.4},
					every segment no 0/.style={darkorange,opacity=.8},];							
				\end{axis}
		\end{tikzpicture}
 

%% file: Graphics/sensitivityregion_bound_below.tex
\begin{tikzpicture}[scale=1]
		\begin{axis}[
			no markers, domain=0:10, domain y=0:10, samples=100,
			axis lines*=left, xlabel=$ $, ylabel=$ $,
			every axis y label/.style={at=(current axis.above origin),anchor=south},
			height=7.5cm, width=7.5cm,
			xtick\empty, ytick=\empty,
			enlargelimits=false, clip=false, axis on top,
			grid = major
			]
			\addplot+[name path=Afun, very thick,darkorange!50!black] {0.3*gauss(7.3,0.8)};
			\addplot+[name path=Bfun, very thick, darkorange!50!black] {0.8*gauss(5,1.5)};
			\addplot+[name path=Cfun, very thin, white] {0.001*gauss(5,1.5)};
			\addplot fill between[of=Afun and Bfun,soft clip={domain=1:10},
		split,
		every segment no 0/.style={darkorange,opacity=.8},
		every segment no 1/.style={pattern=dots,pattern color=black},];	
			\node [right] at (400,100) {\large{CO}};
			\node [right] at (690,90) {\large{DF}};
			\node [right] at (625,30) {\large{AT}};
			\node [right] at (1000,0) {$Y$};		
		\end{axis}
	\end{tikzpicture}